\documentclass[apj,tighten,numberedappendix,appendixfloats,iop]{emulateapj}

\usepackage%
[
pagebackref, 
colorlinks=true,
linkcolor=red,
filecolor=red,
citecolor=blue,
urlcolor=cyan,
pdftitle={Three-dimensional Disk-Planet Torques},
pdfauthor={D'Angelo, Lubow},
pdfsubject={Three-dimensional Disk-Planet Torques},
pdfcreator={KAZEditor},
pdfproducer={Kazzimiei},
pdfkeywords={disk-planet interactions; accretion disks; hydrodynamics; 
methods: numerical; planetary systems: formation; planetary systems: protoplanetary disks},
bookmarks=true,
breaklinks=true,
hypertexnames=false,
pdfstartview=FitV,
hyperfootnotes=true,
bookmarksopen=false,
pdfpagemode=None]{hyperref}
\usepackage{breakurl}

\newcommand{\MEarth}{\mbox{$M_{\mathrm{E}}$}}        
\newcommand{\Mp}{\mbox{$M_{p}$}}                     
\newcommand{\Ms}{\mbox{$M_{s}$}}                     
\newcommand{\Rhill}{\mbox{$R_\mathrm{H}$}}           
\newcommand{\viscu}{\mbox{$a^{2}\,\Omega_{\mathrm{K}}(a)$}}
\newcommand{\AU}{\mbox{AU}}                          
\shorttitle{THREE-DIMENSIONAL DISK-PLANET TORQUES}
\shortauthors{D'Angelo \& Lubow}
\begin{document}
\title{THREE-DIMENSIONAL DISK-PLANET TORQUES IN A LOCALLY ISOTHERMAL DISK\altaffilmark{\dag}}
\author{Gennaro D'Angelo\altaffilmark{1,2} and 
        Stephen H. Lubow\altaffilmark{3,4}}
\slugcomment{\today}
\altaffiltext{$^\dag$}{%
                      To appear in %
                      \textsc{The Astrophysical Journal}
                      (November 10, 2010 issue).%
                               }
\altaffiltext{1}{NASA Ames Research Center, MS 245-3, Moffett Field, 
                 CA 94035, USA; \texttt{\email{gennaro.dangelo@nasa.gov}}.}
\altaffiltext{2}{UCO/Lick Observatory, Department of Astronomy and
                 Astrophysics, University of California, Santa Cruz,
                 CA 95064, USA.}
\altaffiltext{3}{Space Telescope Science Institute, 3700 San Martin
       Drive, Baltimore, MD 21218, USA; \texttt{\email{lubow@stsci.edu}}.}
\altaffiltext{4}{Institute of Astronomy, Madingley Road, Cambridge, 
       CB3 0HA, UK.}
\begin{abstract}
We determine an expression for the Type~I planet migration torque 
involving a locally isothermal disk, with moderate turbulent viscosity
($5\times 10^{-4} \lesssim \alpha \lesssim 0.05$), based on three-dimensional nonlinear 
hydrodynamical simulations. The radial gradients (in a dimensionless
logarithmic form) of density and temperature are assumed to be constant 
near the planet. We find that the torque is roughly equally sensitive to 
the surface density and temperature radial gradients. 
Both gradients contribute to inward migration when they are negative.
Our results indicate that two-dimensional calculations with a smoothed 
planet potential, used to account for the effects of the 
third dimension, do not accurately determine the effects of density 
and temperature gradients on the three-dimensional torque.
The results suggest that substantially slowing or stopping planet 
migration by means of changes in disk opacity or shadowing is 
difficult and appears unlikely for a disk that is locally isothermal.
The scalings of the torque and torque density with planet mass 
and gas sound speed follow the expectations of linear theory. 
We also determine an improved formula for the torque density 
distribution that can be used in one-dimensional long-term evolution studies
of planets embedded in locally isothermal disks. 
This formula can be also applied in the presence of mildly varying 
radial gradients and of planets that open gaps.
We illustrate its use in the case of migrating super-Earths and 
determine some conditions sufficient for survival.
\end{abstract}

\section{INTRODUCTION}
\defcitealias{menou2004}{MG04}
\defcitealias{tanaka2002}{TTW02}
\defcitealias{gennaro2008}{DL08}

Important interactions occur between a young planet and its surrounding 
gaseous disk. Such interactions are generally strong and can lead to 
planet migration on relatively short timescales 
\citep[see also review by \citealp{lubow_chapter}]{goldreich1980,lin1986b,ward1997}.
The shortness of the migration timescales may be a problem, since
they can be shorter than planet formation timescales ($\gtrsim 10^{6}$
years) predicted by the core nucleated accretion models of giant planet 
formation \citep[e.g.,][and references therein]{lissauer2009}.

There are distinct regimes of planet migration. The Type~I regime applies 
to low-mass planets whose tidal torques are too weak to open a gap in the 
disk. In this case, the planet 
is fully embedded in the disk that is nearly unperturbed by the presence 
of the planet. Several calculations have been carried out to determine a 
Type~I torque formula for disk-planet interactions, some of which are
described below. The tidal torque on 
the planet involves opposing contributions from the material outside and 
inside the orbit of the planet. As a result, the net torque on the planet
depends on both the values of the gas properties (such as density and 
temperature) and their radial gradients near the orbit of the planet.
Several studies have investigated the possibility of stopping migration 
by trapping the planet in regions where the radial gradients of disk 
properties may be strong enough to counteract the general tendency 
to inward migration. Such disk regions may occur where the density 
and temperature vary rapidly in radius due to changes in disk opacity 
or turbulent viscosity (e.g., \citealp{menou2004}, hereafter 
\citetalias{menou2004}; \citealp{matsumura2007}).
Disk temperature gradients are modified near a planet due to enhanced 
stellar exposure and shadowing effects caused by the variations in the 
vertical thickness of the disk. The modified temperature structure due 
to this effect may also act to somewhat slow migration  
\citep{jang-condell2005}.

Most torque calculations to date have assumed that the disk is locally 
isothermal and the disk viscosity is not very low ($\alpha \gtrsim 0.001$). 
Under the locally isothermal assumption, the gas temperature is a fixed 
imposed function of distance from the central star that is independent 
of height above the disk midplane. This approximation is formally justified
in disk regions where the thermal timescale is shorter than the local
orbital period. This approximation represents a limitation of our analysis, 
but it allows us to directly connect the results with those of previous studies 
that made the same approximation.

Some studies have suggested that important modifications to the migration 
rate can occur if the disk coorbital region responds nearly adiabatically 
\citep[e.g.,][]{baruteau2008,paardekooper2008} 
or if the disk turbulent viscosity is very low 
\citep{ward1997,rafikov2002,li2009}.  
In the nearly adiabatic case, Type~I outward migration appears possible if 
(1) the turbulent viscosity is sufficiently high to allow the corotation torques 
to be effective, 
(2) the disk thermal timescale near the planet is longer than the time for the 
gas to make U-turns in its horseshoe orbit near the planet (of order the 
planet's orbital period), 
(3) the disk thermal timescale is shorter than the gas libration timescale in 
the coorbital region, 
and (4) the disk radial entropy gradient is negative. 
The radial entropy gradient is zero if the surface density is $\propto r^{-3/2}$, 
as in the standard Minimum-Mass Solar Nebula \citep{hayashi1981}, and the 
temperature is $\propto r^{-1}$, as it would be the case in a disk with constant 
opening angle. 

A weak turbulent viscosity regime ($\alpha \la 10^{-4}$) may occur in 
``dead zones'' of disks where gas ionization levels may be too low to sustain 
magnetically driven turbulence below the disk's outer layers 
\citep[e.g.,][]{gammie1996}.
Under such conditions, two-dimensional simulations suggest that the migration of a
$\approx 10\,\MEarth$ planet can be halted by the strongly nonlinear effects 
of waves launched at Lindblad resonances that result in shocks 
\citep{li2009,yu2010}. At sufficiently low levels of turbulent viscosity
($\alpha \la 10^{-5}$), vortices can form and collide with the planet, 
producing a jumpy and chaotic form of migration.

We restrict attention in this paper to locally isothermal disk regions with 
moderate levels of disk turbulent viscosity ($5\times 10^{-4} \la \alpha \la 0.05$). 
For such cases, the well known disk torque formula of 
\citet[][hereafter \citetalias{tanaka2002}]{tanaka2002} describes the 
situation of a disk that is spatially isothermal (the temperature is 
radially and vertically constant) by means of a linear analysis of the three-dimensional 
dynamical equations. Various three-dimensional nonlinear simulations have obtained 
good agreement with these migration rates 
\citep[e.g.,][]{gennaro2003a,bate2003}.
However, there has not been a systematic analysis, based on three-dimensional 
simulations, to study the effects of temperature and density gradients 
on migration, which we provide here.

Others, such as \citetalias{menou2004},  generalized the three-dimensional analytic
disk torque calculation of \citetalias{tanaka2002} to locally isothermal 
cases where radial disk temperature variations occur. The analysis of
\citetalias{menou2004} was done using two-dimensional resonant torque
expressions, together with a softened potential to mimic the 
effects of the third dimension (perpendicular to the disk midplane). 
This calculation included only the effects of Lindblad torques and 
neglected the corotation torques.
However, upon closer inspection of the three-dimensional Lindblad torque contribution 
of \citetalias{tanaka2002}, we find that it disagrees with the results 
of \citetalias{menou2004}, applied to a fully isothermal disk, in 
important ways. This discrepancy suggests that the torque 
dependence on disk gradients and the trapping mechanisms in 
general require further examination.
One goal of this paper is to determine the torque formula for a planet 
embedded in a three-dimensional, locally isothermal disk by means of numerical 
simulations that includes effects of density and temperature gradients,
as well as the coorbital torque (see \S~\ref{sec:SimulationDescription}).
This analysis is described in \S~\ref{sec:TypeIFormula}. 

The distribution of torques per unit disk mass, as a function of 
radius, provides a description of where the disk contributions to 
the planet torque occur 
\citep[e.g.,][hereafter \citetalias{gennaro2008}]{gennaro2008}.
Such torque distributions are useful in the application to one-dimensional disk 
evolution models that account for planets, including cases where 
the planets open gaps in the disk, the so-called Type~II regime 
of planetary migration \citep[see, e.g.,][and references therein]{ward1997}. 
Current multi-dimensional hydrodynamical 
codes cannot practically follow the disk evolution over long times, 
such as disk viscous times, while one-dimensional codes can. 
The torque density distribution is typically estimated by means of 
the impulse approximation \citep{lin1986b}. But that distribution 
contains a long tail that is unphysical, particularly near the star 
\citep[e.g.,][]{veras2004}. 
Another goal of the paper is to provide a more accurate formula for 
the torque density distribution that can be applied to such and similar 
problems, 
again based on three-dimensional simulations. These distributions are described in 
\S~\ref{sec:dTdM}. Some applications of these formulae to migrating 
super-Earths are discussed in \S~\ref{sec:tracks}. 
Extensions of the torque density formula to situations of mildly varying 
disk gradients and gap-opening planets are presented in 
\S~\ref{sec:dTdMvars} and \S~\ref{sec:dTdMgap}, respectively. 
Section~\ref{sec:summary} contains the summary of the results.

\section{DESCRIPTION OF SIMULATIONS}
\label{sec:SimulationDescription}

\subsection{Disk Model}
\label{sec:DiskModel}

The simulations are carried out along the lines of \citetalias{gennaro2008}. 
We use a spherical polar coordinates system $\{O; R, \theta, \phi\}$, whose 
origin $O$ is located at the star-planet center of mass and which rotates 
at the same angular velocity as the planet orbits about the star.
The planet's orbit is circular and has zero inclination relative to the
disk midplane $\theta=\pi/2$. Since the disk is assumed to be symmetric 
relative to its midplane, only the disk's northern hemisphere, 
$\theta\le\pi/2$, is explicitly simulated.

We assume that the disk's gas is locally isothermal and that the pressure, 
$p$, can be written as
\begin{equation}
  p(R, \theta, \phi) = \rho(R, \theta, \phi)\,c^{2}_{s}(r),
  \label{eq:p}
\end{equation}
where $\rho(R, \theta, \phi)$ is the mass density. The sound speed of the 
gas, $c_{s}(r)$, is assumed to be a function of cylindrical radius 
$r=R \sin{\theta}$. From the equation of state for a perfect gas and 
Equation~(\ref{eq:p}), it follows that the temperature 
$T\propto p/\rho=c^{2}_{s}$. Therefore, to determine the effects of 
temperature gradients on disk torques, we vary $T$ in radius by varying
the gas sound speed, which is taken to be a power law in $r$. The
hydrostatic disk scale-height is defined as
\begin{equation}
  H(r)=\frac{c_{s}}{\Omega_{\mathrm{K}}},
  \label{eq:H}
\end{equation}
where $\Omega_{\mathrm{K}}$ is the Keplerian angular velocity. 
Notice that, since the vertical motion and structure of the disk is 
directly computed from the three-dimensional Navier-Stokes equations, the 
actual scale-height of the disk may deviate from the quantity
$H(r)$. For brevity, however, we will refer to $H(r)$ simply as disk 
scale-height.

The disk extends in azimuth over $2 \pi$ radians around the 
star. In models with planet-to-star mass ratio $\Mp/\Ms < 3\times 10^{-4}$,
the radial disk region extends from $0.4\,a$ to $2.5\,a$ or from 
$0.5\,a$ to $1.55\,a$, where $a$ is the orbital radius of the planet.  
In models with planet-to-star mass ratio $\Mp/\Ms\ge 3\times 10^{-4}$,  
the disk extends from $0.4\,a$ to $4.0\,a$ in radius. 
In the $\theta$-direction, the disk domain contains from a few to many
disk scale-heights above the midplane, depending on both $r$ and $H(r)$.

The initial mass density, $\rho$, is independent of the azimuth, $\phi$, 
and has a Gaussian profile in the meridional direction, $\theta$. 
In the radial direction, $\rho\propto \Sigma/H$ and the initial surface 
density, $\Sigma$, is usually taken to be proportional to a power of $r$.
In \S~\ref{sec:dTdMvars}, we consider a more general case. 
The initial flow field has radial and meridional components
$u_{R}=-3\,\nu/(2R)$, where $\nu$ is the kinematic viscosity of the gas, 
and $u_{\theta}=0$. The initial azimuthal velocity is
$u_{\phi}=r\,[\Omega-\Omega_{\mathrm{K}}(a)]$. The initial (unperturbed) 
rotation rate $\Omega$ accounts for the effects of pressure support 
produced by both density and temperature gradients. For a pressure
supported disk
\begin{equation}
 \Omega^{2}=\Omega^{2}_{\mathrm{K}}%
           +\frac{1}{r \rho}\left(\frac{\partial p}{\partial r}\right).
 \label{eq:OmegaEq}
\end{equation}
Although $\Omega$ has a dependence on distance from the midplane
(see, e.g.,  Equation~(4) of  \citetalias{tanaka2002}), we adopt as an initial 
condition its value at the midplane that is given by
\begin{equation}
  \Omega^{2}=\Omega^{2}_{\mathrm{K}}\left[1-%
  \left(\beta+\frac{\zeta}{2}+\frac{3}{2}\right)%
  \left(\frac{H}{r}\right)^{2}\right].
  \label{eq:Omega}
\end{equation}
In the above equation, we set $\beta=-d\ln{\Sigma}/d\ln{r}$ 
and $\zeta=-d\ln{T}/d\ln{r}$.
Notice that, in an infinitesimally thin (i.e., two-dimensional) disk,
Equation~(\ref{eq:OmegaEq}) involves the surface density and a vertically 
integrated pressure and thus Equation~(\ref{eq:Omega}) becomes 
$\Omega^{2}=\Omega^{2}_{\mathrm{K}}[1-(\beta+\zeta)(H/r)^2]$.

We typically impose a kinematic viscosity proportional to 
the inverse of the initial surface density distribution, so that
the mass accretion rate through the disk ($\propto \nu\,\Sigma$) is 
nearly constant in radius and the disk can achieve a steady state. 
In such cases, the initial imposed value of the parameter $\beta$ 
is therefore preserved in time.
In the calculations discussed in \S~\ref{sec:TypeIFormula}, 
$\nu(a)=10^{-5}\,\viscu$, which corresponds to a turbulent viscosity 
parameter $\alpha=0.004$ at the orbital radius of the planet when 
$H(a)/a=0.05$. However, we also report on models with lower 
(as low as $\alpha\sim 5\times 10^{-4}$) and higher kinematic 
viscosity, up to $\nu(a)\sim 10^{-4}\,\viscu$ or $\alpha\sim 0.05$ 
(see also \S~\ref{sec:dTdM}).

We consider several values of the planet-to-star mass ratio, $\Mp/\Ms$. 
For the Type~I cases discussed in \S~\ref{sec:TypeIFormula}, 
we mainly concentrate on $\Mp/\Ms=3\times 10^{-6}$ 
(or $1\,\MEarth$ planet if $\Ms=1\,M_{\odot}$). 
For Type~II cases discussed in \S~\ref{sec:dTdMgap}, we consider 
planets up to $2$ Jupiter masses ($\Mp=2\times 10^{-3}\,\Ms$).
The gravitational potential of the planet, $\Phi_{p}$, is smoothed
over a fraction $\varepsilon$ of the Hill radius of the planet 
$\Rhill=a \left[\Mp/(3\Ms)\right]^{1/3}$ and is given by
\begin{equation}
 \Phi_{p}=-\frac{G\Mp}{\sqrt{S^2+\varepsilon^2  R_{\mathrm{H}}^2}}, 
 \label{eq:Phip}
\end{equation}
where $S$ is the distance from the planet. We generally apply a 
smoothing (or softening) length with parameter $\varepsilon=0.1$. 
In the low-mass regime,
another relevant length scale is the Bondi radius 
$R_{\mathrm{B}}=G\Mp/c^2_{s}$ (e.g., \citealp{masset2006a};
\citetalias{gennaro2008}), which can be shorter than $\Rhill$.
In these situations, the smoothing length we apply is of order
$R_{\mathrm{B}}$ or smaller.

The softening length plays a much weaker role in three dimensions than it does 
in two-dimensional, since the vertical thickness of the disk effectively smooths 
the potential of the planet. 
We find that our results are insensitive to changes in values of the 
softening length.
For cases with an $\Mp= 5\times 10^{-6}\,\Ms$ planet, changes in 
the softening length by a factor of $10$ produced a change in 
the net torque much less than $1$\%. 
For cases with an $\Mp= 10^{-3}\,\Ms$ planet, changing
$\varepsilon$ by a factor of $2$ resulted in a net torque variation
of about $1$\%. Further details are given in the Appendix.

\subsection{Numerical Method}
\label{sec:NumericalMethod}

The Navier-Stokes equations that describe the motion of the disk's gas
are written in terms of specific momenta, $\rho\,u_{R}$, 
$\rho\,R\,u_{\theta}$, and 
$\rho\,r\,\left[u_{\phi}+r\,\Omega_{\mathrm{K}}(a)\right]$
\citep[see, e.g.,][]{gennaro2005}, and are solved by means of a 
finite-difference code that computes separately the spatial integration 
of advection and source terms (apparent forces, pressure and gravity
gradients, and viscous stresses) using an operator-splitting 
scheme \citep[e.g.,][]{ziegler1997}. 
The algorithm employed in the code is second-order accurate in space 
and semi-second-order accurate in time.
Comparisons on disk-planet interaction problems of this code with 
various other codes have been carried out in several studies 
\citep{kley2001,devalborro2006,masset2006a,devalborro2007}.

The code uses a grid with constant spacing between grid points, 
in each coordinate direction, to discretize the momentum equations
and features grid-refinement capabilities via multiple nested grids
\citep{gennaro2002,gennaro2003a}. This technique allows us to
increase the volume resolution by a factor $2^3$ for each added 
level of grid nesting. Since we are particularly interested in the flow
dynamics in a radial region spanning several times $H$ on either
side of the planet's orbit, to increase the numerical resolution
we use nested subgrids that extend $2 \pi$ radians in azimuth and 
cover the whole domain in the $\theta$-direction.
We employ a variety of grid systems that generate a linear resolution 
$\Delta R\approx a\,\Delta\theta\approx a\,\Delta\phi$, in the disk 
region $|R-a|\le 4\,H(a)$, ranging from $2\times 10^{-3}\,a$ to 
$5\times 10^{-3}\,a$. In the cases discussed in \S~\ref{sec:dTdM}
that consider disk aspect-ratios $H/r\lesssim 0.01$, the grid resolution 
in the region of interest is always such that $H(a)/\Delta R \gtrsim 10$.

To quantify resolution effects on the radial distributions of torques, 
we perform a convergence study that is presented in the Appendix.
The results indicate that discrepancies are at a level of $1$\% or better.

Either nonreflecting \citep{godon1996} or wave damping 
\citep{devalborro2006} boundary conditions are imposed at the
inner and outer radii of the computational domain.
Reflective and symmetry boundary conditions are applied, respectively, 
at the disk surface $\theta=\theta_{\mathrm{min}}$ and at the disk midplane 
$\theta=\pi/2$ \citep[see also][]{masset2006a}.

The calculations discussed in the paper do not involve removal of mass 
from near a low-mass planet. Accreting boundary conditions are instead 
applied for the larger planet masses that we consider, $\Mp\gtrsim 10^{-4}\,\Ms$. 
Accretion is performed within a radius $\sim 0.1\,\Rhill$ of the planet. 
The procedure for mass removal is described in some detail in \S~3 
of \citetalias{gennaro2008}. When accreting boundary conditions are 
applied, the removed mass is not added to the planet's mass in order 
to keep it fixed.
Except for the calculation discussed in \S~\ref{sec:3Dtest}, the planet is 
kept on a fixed orbit.

The disk's evolution is typically followed for about $80$ to about 
$160$ orbital periods of the planet. But a number of models are
evolved for much longer, up to around $2000$ orbits, in order to check 
for possible torque saturation effects. In models that deal with 
a planet mass $\Mp\gtrsim 10^{-4}\,\Ms$, the initial density 
distribution includes a gap along the planet's orbit to account 
for an approximate balance between viscous and tidal torques, 
which reduces the relaxation time toward steady state. 
Torques and torque distributions are evaluated when the flow achieves 
a nearly steady state.

If we denote with $d\mathbf{f}_{g}$ the gravitational force exerted
on the planet by the mass $\rho\,dV$ within a grid element of volume 
$dV$, the associated torque acting on the planet is 
$\mathbf{R}_{p}\,\mathbf{\times}\,d\mathbf{f}_{g}$,
where $\mathbf{R}_{p}$ is the position vector of the planet.
Direct summation of these elemental torques gives the total torque.
Radial torque distributions are obtained by summing elemental torques
over the meridional and azimuthal directions, as outlined in 
\citetalias{gennaro2008}. Both total torques and 
torque distributions are averaged in time over one orbital period of 
the planet.

\section{TYPE I TORQUE FORMULA}
\label{sec:TypeIFormula}

\subsection{Comparison of Three-dimensional and Softened Two-dimensional Torques in a Fully Isothermal Disk}
\label{sec:3Dvs2Dsoft}

Three-dimensional linear analytic calculations of the Type~I migration
rates were carried out by \citetalias{tanaka2002}.
They provide torque formulae for a disk where the gas sound speed is strictly 
constant, i.e., for an isothermal disk.
The torque exerted on a planet, including contributions from Lindblad 
resonances and the corotation resonance, is given by 
\begin{eqnarray}
 \mathcal{T}_{\mbox{\tiny{TTW}}}%
                         &=&-\left(1.36 + 0.54\,\beta\right)\nonumber\\%
                          & &\times\Sigma(a) \, \Omega^{2}(a) \, a^{4}
    \left(\frac{\Mp}{\Ms}\right)^{2} \left(\frac{a}{H}\right)^{2},
\label{eq:Tt}
\end{eqnarray}
where the (azimuthally averaged) gas surface density $\Sigma$, the disk 
rotation rate $\Omega$, and the disk vertical scale-height $H$ are 
evaluated at the orbital radius of the planet, $a$. We recall that 
$\beta=-d\ln{\Sigma}/d\ln{r}$. Equation~(\ref{eq:Tt}) does 
not include the effects of temperature gradients, since the disk is 
assumed to be fully isothermal.

The corotation resonance contains the complication that it can be 
saturated (have zero strength). In the absence of irreversible 
processes, the flow in the coorbital region follows closed horseshoe 
and tadpole orbits in the corotating frame of the planet.
The steady torque associated with such closed orbits is zero (saturated).
Turbulent viscosity introduces an irreversible behavior that results 
in a nonzero torque.
The effects of turbulent viscosity are stronger for lower mass planets
($\Rhill<H$), since the radial width of the horseshoe orbit region, 
$w\approx 2 a\sqrt{(\Mp/\Ms)(a/H)}$ \citep[see][]{masset2006a}, is 
narrower and the turbulence diffusion timescale across this region 
is therefore shorter \citep{ward1992}. 
 
The tables of \citetalias{tanaka2002} describe the various contributions
to the torque in the case that both corotation and Lindblad resonances
contribute in a fully isothermal disk, based on linear theory.  We infer from 
these tables that the three-dimensional torque contribution due to Lindblad resonances is
\begin{eqnarray}
 \mathcal{T}^{\mathrm{L}}_{\mbox{\tiny{TTW}}}%
                         &=&-\left(2.34 - 0.10\,\beta\right)\nonumber\\%
                         & &\times\Sigma(a) \, \Omega^{2}(a) \, a^{4}
    \left(\frac{\Mp}{\Ms}\right)^{2} \left(\frac{a}{H}\right)^{2}.
\label{eq:Ttsat}
\end{eqnarray}

\citet{menou2004} determined the three-dimensional torque for a locally isothermal 
disk (i.e., the temperature depends on $r$) on the basis of two-dimensional linear 
calculations with a softened potential (of the type introduced in 
Equation~(\ref{eq:Phip})) to model the effects of the third (vertical) 
dimension. Only Lindblad resonances are included.
The softening length was chosen to be $H$. As they note, it is not 
clear what value is most appropriate within a range of that order. 
From Figure~3 of \citetalias{menou2004}, we infer that the 
Lindblad-only torque is given by
\begin{eqnarray}
 \mathcal{T}^{\mathrm{L}}_{\mbox{\tiny{SLC}}}%
                         &=&-\left(0.80 -0.77\,\beta + 1.12 \,\zeta\right)\nonumber\\%
                          & &\times\Sigma(a) \, \Omega^{2}(a) \, a^{4}
    \left(\frac{\Mp}{\Ms}\right)^{2} \left(\frac{a}{H}\right)^{2}.
 \label{eq:Tmg}
\end{eqnarray}
The subscript emphasizes that this result derives from linear 
calculations that employ a softened gravitational potential.
We recall that $\zeta=-d\ln{T}/d\ln{r}$.
As is apparent, there is substantial disagreement between 
Equations~(\ref{eq:Ttsat}) and (\ref{eq:Tmg}) for $\zeta=0$, 
where they should agree, in particular for the coefficient of $\beta$. 
Such gradient terms can play an important role in describing the 
possible trapping of a planet in regions of density and temperature
radial variations.

The effects of the density gradient (the $\beta$ term) on the torque
enter in two ways. With a negative radial gradient of the surface 
density ($\beta>0$), the density interior to the orbit of the planet is 
higher than the density exterior to it. As a result, the inner Lindblad 
torques are favored over outer ones, giving rise to a outward 
(positive) torque contribution. On the other hand, the negative 
density gradient slows the disk rotation, due to outward pressure 
forces (see Equation~(\ref{eq:OmegaEq})). This effect acts in the 
same sense as a drag term would in causing inward migration. 
The degree of cancellation of these two effects is critical in 
determining the net torque contributions by a density gradient.
In particular, \citetalias{menou2004} concluded that the cancellation
is nearly complete in two dimensions (i.e., when an unsoftened potential is 
employed), but not in three dimensions. However, \citetalias{tanaka2002} found 
nearly exact cancellation in three dimensions (see Equation~(\ref{eq:Ttsat})).

There are at least three possible explanations for the discrepancy between
the Lindblad-only torque of Equations~(\ref{eq:Ttsat}) and 
(\ref{eq:Tmg}) with $\zeta=0$. One possibility is that the softened 
potential applied to a two-dimensional calculation, Equation~(\ref{eq:Tmg}), does not 
adequately represent the full effects of three-dimensional torques, Equation~(\ref{eq:Ttsat}).
The analytic calculation of \citetalias{tanaka2002} suggests that a more 
complicated disk response may occur in a three-dimensional disk, especially in the presence 
of gradients, and that this response may not be expressible through the use 
of a softened two-dimensional potential.
A second possibility is that the local two-dimensional torque formula used by 
\citetalias{menou2004}, from \citet{ward1997}, is not sufficiently accurate to 
determine the effects of gradients on the torque. This may be the case since 
the two-dimensional (no softened potential) Lindblad-only torque formula of 
\citetalias{tanaka2002} does not agree with that of \citet{ward1997} for a fully 
isothermal disk. The former has a coefficient of $\beta$ equal to $\approx 1.5$,
while the latter has a much smaller value (nearly zero).
A third possibility is that the three-dimensional Lindblad torque of Equation~(\ref{eq:Ttsat}) 
does not provide an accurate description of the Lindblad torque that would 
occur in the absence of a corotation torque. Equation~(\ref{eq:Ttsat}) does 
describe the Lindblad torque contribution in the presence of a coorbital torque.
The Lindblad and corotation resonances can overlap and may affect each other, 
even in the linear theory. However, for $\beta=3/2$, the corotation torque nearly 
vanishes in the three-dimensional calculation and the effects of resonance overlap can be 
largely ignored. (For $\beta=3/2$, there is a small, $\approx 1$\%, nonvanishing 
corotation torque contribution caused by three-dimensional effects.) For $\beta=3/2$, the 
torque predicted by the three-dimensional isothermal analysis, Equation~(\ref{eq:Ttsat}),
has a dimensionless torque that evaluates to $-2.2$, while the 
corresponding result for the softened two-dimensional potential in an isothermal disk
in Equation~(\ref{eq:Tmg}) evaluates to $+0.36$. This large difference 
in predicted torques (and directions) leads us to conclude that the resonance 
overlap, which should be absent in this comparison, is unlikely to explain 
the discrepancy between the calculations.

\subsection{Three-dimensional Torque in a Locally Isothermal Disk}
\label{sec:3Disot}
%
\begin{figure*}
\centering%
\resizebox{\linewidth}{!}{%
\includegraphics{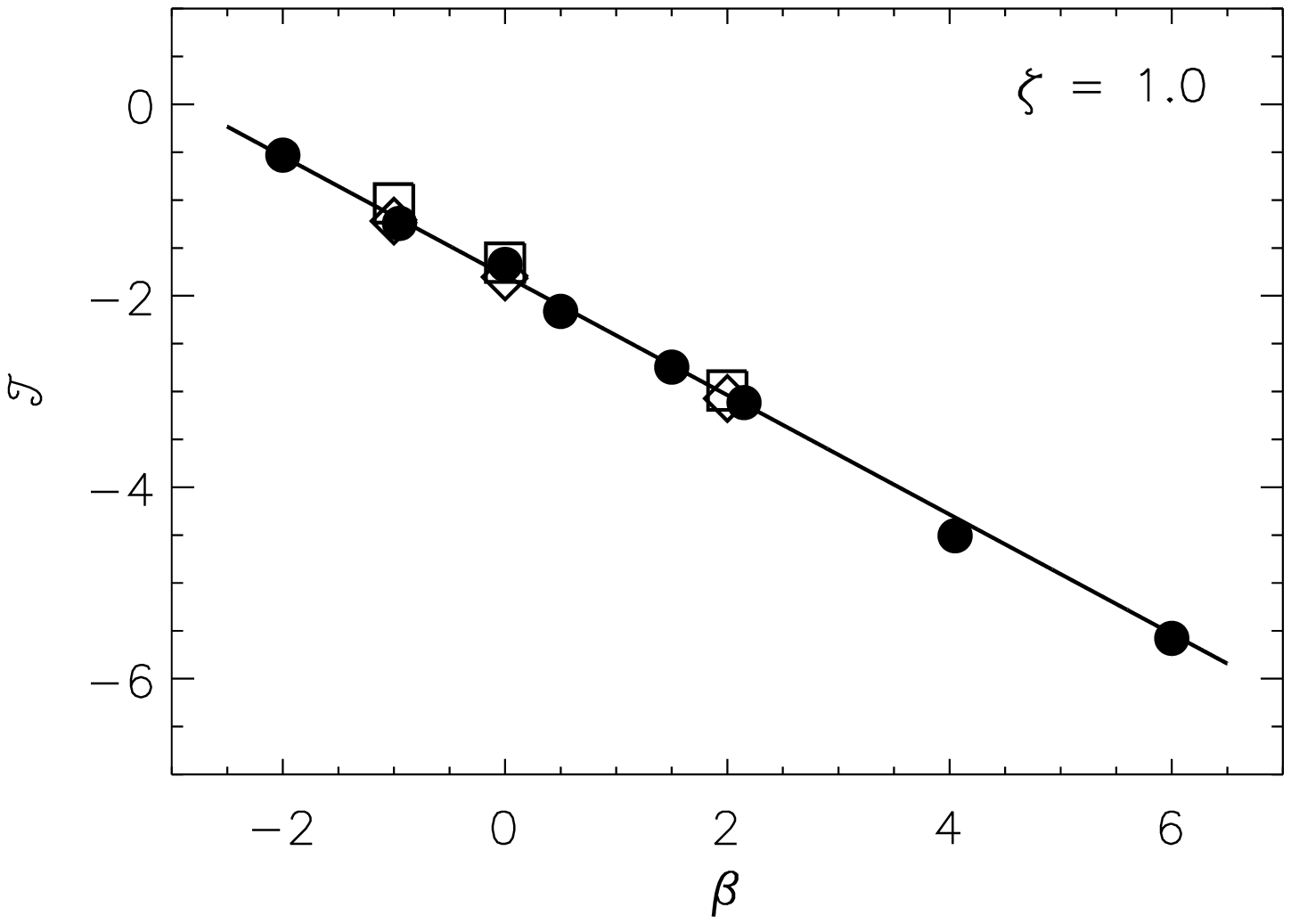}%
\includegraphics{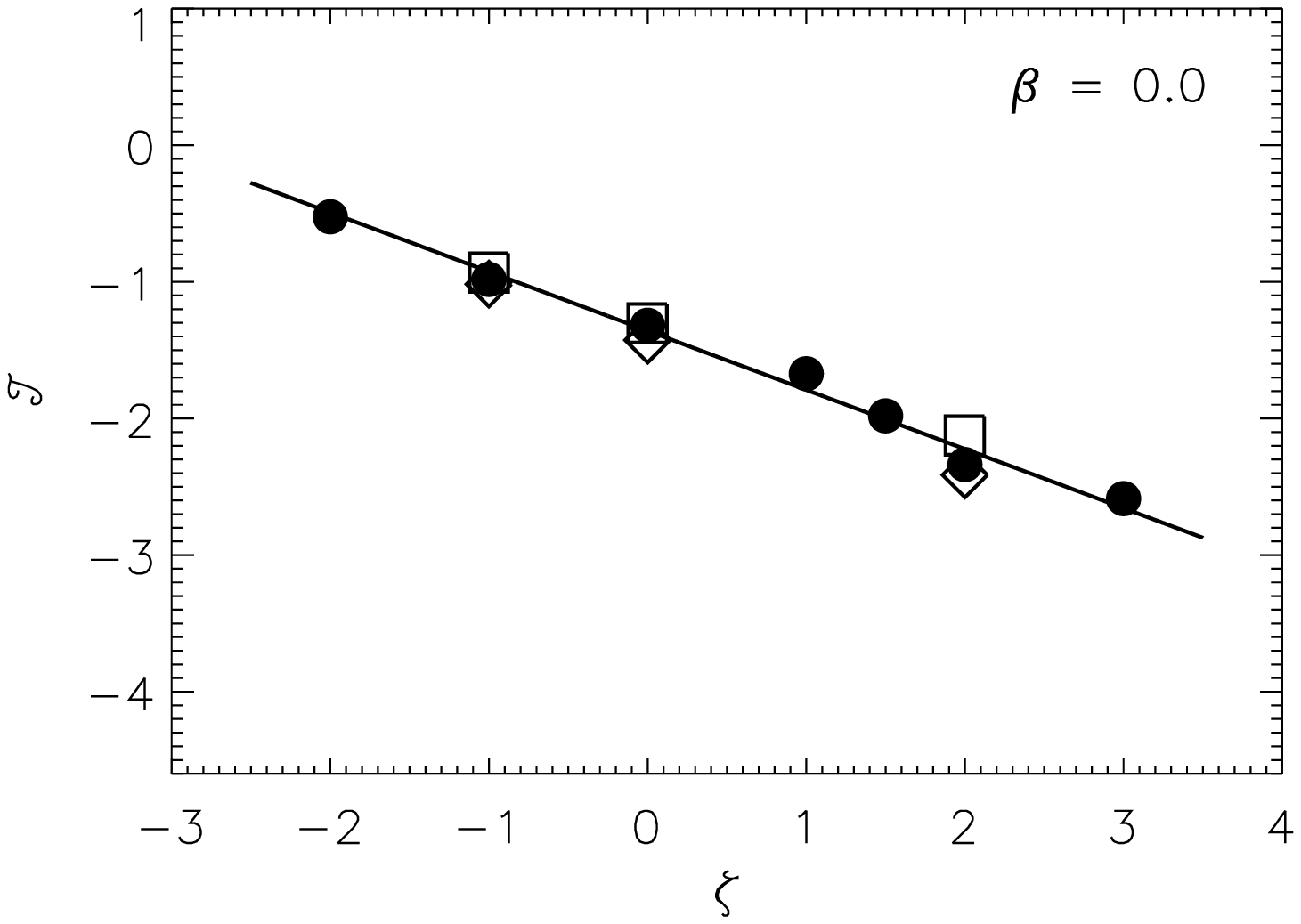}}
\caption{Total torque exerted on a low-mass planet by a locally 
         isothermal disk as a function of surface density 
         (\textit{left panel}) 
         and temperature (\textit{right panel}) gradients, obtained 
         from three-dimensional simulations. We define $\beta=-d\ln{\Sigma}/d\ln{r}$ 
         and $\zeta=-d\ln{T}/d\ln{r}$. The quantity $\Sigma(r)$
         is the azimuthally averaged surface density of the gas.
         The filled circles represent simulation results of total
         torques normalized to
         $\Sigma\,\Omega^{2}\,a^{4}\,(\Mp/\Ms)^{2}\,(a/H)^{2}$,
         in which $\Sigma$, $\Omega$, and $H$ are evaluated at the
         planet orbital radius, $a$. The solid line represents
         a three-parameter, bi-linear fit to data (the leading term in 
         parenthesis in Equation~(\ref{eq:Tdl})), including data not 
         displayed in this figure.
         These results correspond to a disk kinematic viscosity
         equivalent to $\alpha=0.004$. 
         Results from models with lower ($\alpha=0.002$)
         and higher ($\alpha=0.02$) viscosity 
         are also plotted as empty squares and 
         diamonds, respectively, for $\beta$ (\textit{left panel}) and
         $\zeta$ (\textit{right panel}) equal to $-1$, $0$, and $2$.
         }
\label{fig:Tbc}
\end{figure*}
In Figure~\ref{fig:Tbc}, we plot the total torque on the planet as 
a function of density (\textit{left panel}) and temperature 
(\textit{right panel}) gradients ($\beta$ and $\zeta$, respectively) 
obtained from a subset of our simulations (\textit{filled circles}, see figure
caption for details) of an $\Mp=3\times 10^{-6}\,\Ms$ planet ($1\,\MEarth$ 
for a solar mass star) in a disk of fixed temperature and density 
at the orbital radius of the planet. In general, we consider disks 
with values of $\beta$ in the range from $-2$ to $4$ and values of 
$\zeta$ in the range from $-2$ to $2$. But a few models had gradients 
outside of these ranges, as also indicated in Figure~\ref{fig:Tbc}. 
The nonlinear simulation results we obtain are well fit to the following 
torque equation:
\begin{eqnarray}
 \mathcal{T}&=&%
            -\left(1.36 + 0.62\,\beta + 0.43\,\zeta\right)\nonumber\\%
                   & &\times\Sigma(a) \, \Omega^{2}(a) \, a^{4}
    \left(\frac{\Mp}{\Ms}\right)^{2} \left(\frac{a}{H}\right)^{2}.
 \label{eq:Tdl}
\end{eqnarray}
The dependencies on the planet-to-star mass ratio, $\Mp/\Ms$, and 
on disk scale-height, $H$, at the planet's orbit 
(i.e., on the temperature at the orbit of the planet) are tested in 
\S~\ref{sec:dTdM} for fixed values of $\beta$ and $\zeta$. 
We find them to agree very well with the scalings in 
Equation~(\ref{eq:Tdl}) over a broad range of variations.

If we subtract out the corotation contribution in the case of an 
isothermal disk, adopting the corotation torque expression given 
in \citetalias{tanaka2002}, we obtain a Lindblad-only torque of
\begin{eqnarray}
 \mathcal{T}_{\mbox{\tiny{ISO}}}^{\mathrm{L}}&=&%
                         -\left(2.34 - 0.02\,\beta\right)\nonumber\\%
                          & &\times\Sigma(a) \, \Omega^{2}(a) \, a^{4}
    \left(\frac{\Mp}{\Ms} \right)^{2} \left(\frac{a}{H}\right)^{2}.
 \label{eq:Tsat}
\end{eqnarray}
In this case, we infer a nearly exact cancellation between the two 
torque contributions by the density gradient, as described above. 
Not surprisingly, this result is close to the corresponding expression 
by \citetalias{tanaka2002}, Equation~(\ref{eq:Ttsat}).

The criterion for torque saturation of \citet{ward1992} is that the 
libration timescale $\tau_{\mathrm{lib}}$ (over which the specific 
vorticity gradient is removed) is shorter than the viscous diffusion 
timescale across the coorbital zone 
$\tau_{\mathrm{vd}}\approx w^{2}/\nu$  (over which time the specific 
vorticity gradient is reestablished). The relative velocity of the fluid 
on the librating orbit farthest from the planet's is $\approx |3w\Omega/4|$, 
and thus $\tau_{\mathrm{lib}}\approx 16\pi a/(3 w \Omega)$, where 
$\Omega=\Omega(a)$.
The condition $\tau_{\mathrm{lib}}\lesssim \tau_{\mathrm{vd}}$
requires a viscosity such that
\begin{equation}
 \alpha \la \left(\frac{a}{H}\right)^{7/2}%
                 \left(\frac{\Mp}{\Ms} \right)^{3/2}.
 \label{eq:alcr}
\end{equation}
On this basis, we expect that the torques in Figure~\ref{fig:Tbc} 
involve unsaturated corotation torque contributions for the 
viscosity and sound speeds that we impose
($\alpha=0.004$ and $H/a=0.05$ at $r=a$) on a 
$\Mp=3\times 10^{-6}\,\Ms$ planet, since the characteristic 
$\alpha$ required for saturation is only $\lesssim 10^{-4}$. 
In deriving the saturation condition above, we neglect a 
numerical factor of order unity multiplying the right-hand side 
of Equation~(\ref{eq:alcr}) that depends on the precise value of $w$.

If the coorbital torque were partially saturated, it would introduce 
an additional dependence of the overall torque on planet mass. 
This dependence on planet mass would be inconsistent with the 
quadratic dependence predicted by linear theory. The torque 
we obtain in Equation~(\ref{eq:Tdl}) has a  quadratic dependence
on planet mass (see \S~\ref{sec:dTdM}), as expected by linear theory. 
Therefore, our expectation that the torque is not saturated is 
consistent with its quadratic dependence on planet mass.  
We provide additional evidence for the lack of saturation below.

\begin{figure}
\centering%
\resizebox{\linewidth}{!}{%
\includegraphics{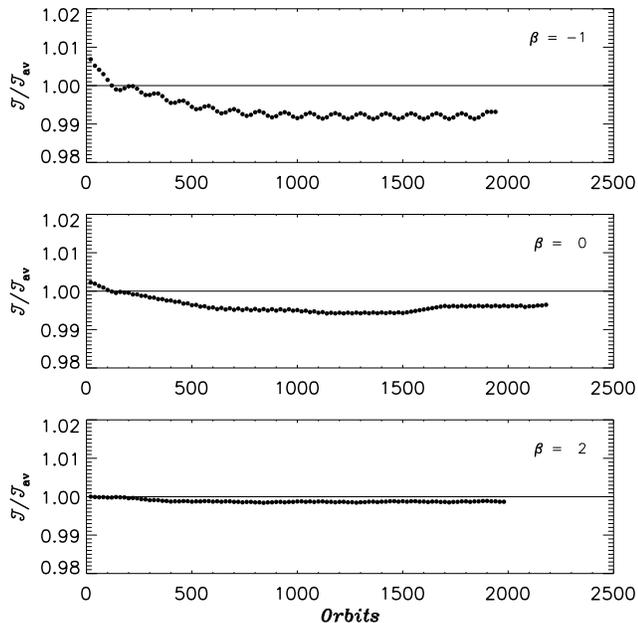}}
\caption{Time behavior of the total torque exerted on a low-mass 
         planet by a locally isothermal disk, corresponding to the cases
         with low viscosity ($\alpha=0.002$) in the left panel of 
         Figure~\ref{fig:Tbc}. The slope of the surface density is indicated
         by the parameter $\beta$ in the top-right corner of each panel. 
         The temperature distribution is such that $\zeta=1$. The torque
         is normalized to $\mathcal{T}_{\mathrm{av}}$, its average value
         between $100$ and $200$ orbits. The libration timescale in these
         models is $\tau_{\mathrm{lib}}\approx 170$ orbits (see the text).
         There is nearly no evolution of the torque (variations are below $1$\%) , 
         as is consistent with it being unsaturated. The torque remains 
         unsaturated because $\tau_{\mathrm{vd}}<\tau_{\mathrm{lib}}$ 
         (see the text for details).
         }
\label{fig:satu}
\end{figure}
To test for saturation further, we run models with lower 
and higher viscosity, for a total variation of $\alpha$ of a 
factor $10$. If saturation of the corotation resonance 
occurred, we would expect to observe a change in the 
total torque equal to $70$\% of the unsaturated torque, 
in a disk with zero surface density and temperature gradients 
(see Equations~(\ref{eq:Tt}) and (\ref{eq:Ttsat})). 
For such cases, however, we find only small 
differences (generally a few per cent) in total torques from 
those cases based on the standard viscosity that we apply.
The results for somewhat lower and much higher viscosity 
are also plotted in Figure~\ref{fig:Tbc} (as 
\textit{empty squares} and \textit{empty diamonds}, respectively).
The small differences in the figure (among different 
symbols) are not due to significant saturation effects, as 
indicated by the time behavior of the total torque illustrated in 
Figure~\ref{fig:satu}, which is nearly constant over time, 
within a margin of $1$\% or less.

If the coorbital torque were to saturate, it would generally decline 
(with some oscillations) from its unsaturated initial value to zero
over several libration timescales of the coorbital region.
Figure~\ref{fig:satu} plots three cases with turbulence parameter 
$\alpha=0.002$, represented as empty squares in the left panel of
Figure~\ref{fig:Tbc}. The torque is normalized to its average value 
between $100$ and $200$ orbits.
Since $\tau_{\mathrm{lib}}\approx 170$ orbital periods in these
models, the torque evolution illustrated in Figure~\ref{fig:satu} spans 
a period of $12$ libration timescales, or longer. There is no evidence 
of a substantial decline in the total torque and therefore no evidence 
of saturation.

In \S~\ref{sec:dTdM}, we provide evidence that the torque behavior
at even lower ($\alpha=5\times 10^{-4}$) and higher viscosity (up to 
values of $\alpha=0.05$) is intrinsic to the torque density distribution, 
whose shape is subject to rather small changes with decreasing or 
increasing viscosity inside this range of the turbulence parameter, 
$\alpha$.
Therefore, we believe that our results apply to a disk where the 
corotation torques are unsaturated (i.e., fully effective).

The torques shown in Figure~\ref{fig:Tbc} disagree with the predictions, 
based on two-dimensional models, of \citet{paardekooper2009} and 
\citet{paardekooper2010}. As we argue in \S~\ref{sec:2dformula},
torques computed in a two dimensional disk are susceptible to smoothing 
length issues and may not always reproduce three-dimensional torques.
On the other hand, \citet{masset2010} obtained a level of agreement with
Equation~(\ref{eq:Tt}) that is similar to ours and a similar variation ($\la 10$\%) 
of torques, as reported in our Figure~\ref{fig:Tbc}, over the range of viscosities 
we consider and for zero disk gradients.

Equation~(\ref{eq:Tdl}) that we obtain agrees fairly well with the 
unsaturated torque expression in Equation~(\ref{eq:Tt}) of 
\citetalias{tanaka2002}, with $\zeta$ set to zero to match the radial 
isothermal assumption in the latter equation.
\citet{tanaka2002} provide the dependence of the torque contributions 
on the radial gradients of the surface density, disk thickness, and 
pressure in various tables. One might think that these tables can 
be used to infer a dependence of the linear torque on the radial 
temperature gradient. If these tables are used in this way, the 
resulting torque expression is in good agreement with 
Equation~(\ref{eq:Tdl}). 
However, it is not clear if the result is meaningful. 
The reason is as follows. Equation~(11) of
\citetalias{tanaka2002} is appropriate for a locally isothermal disk 
with radial temperature variations. But since there is a singularity 
in the term involving $\partial \Omega/ \partial z$ that they want 
to avoid, they set the radial temperature gradient to zero (so that 
$\partial \Omega/ \partial z=0$) in all subsequent equations, including 
the torque derivations. In doing so, they not only discard the 
$\partial \Omega/ \partial z$ term but also other terms, including 
a term that contributes to the coorbital torque. This coorbital torque 
term provides a contribution that is independent of the vortensity 
gradient (the usual coorbital torque contribution) and is due to the 
radial temperature gradient \citep[see also][]{casoli2009}. 
Therefore, the effects of the temperature gradient on the linear three-dimensional 
torque cannot be readily deduced from \citetalias{tanaka2002}.

The coefficient of the temperature gradient that we obtain in
Equation~(\ref{eq:Tdl})  is significantly smaller than the corresponding
coefficient in Equation~(\ref{eq:Tmg}), obtained by using a two-dimensional 
smoothed potential. Furthermore, the contrast between the results is 
even stronger when comparing the ratio of the gradient coefficients to 
the term that is independent of $\beta$ and $\zeta$ in the respective 
equations (i.e., $1.12/0.8$ versus $0.43/1.36$).
The slowing of migration (or trapping of planets) by means of the 
opacity variations, envisaged in \citetalias{menou2004}, relies on having 
a sufficiently steeply declining density profile and a sufficiently shallow 
temperature profile (see Equation~(\ref{eq:Tmg})). However, the three-dimensional 
results (Equation~(\ref{eq:Tdl}) or (\ref{eq:Tsat})) suggest that this 
trapping mechanism does not occur under such conditions.  

Equations~(\ref{eq:Tdl}) and (\ref{eq:Tsat}) are derived under the 
requirement that surface density and temperature gradients  (or, 
more precisely, the derivatives $d\ln\Sigma/d\ln r$ and $d\ln T/d\ln r$) 
are roughly constant over a radial region of about $3 H$ interior 
and exterior to the planet's orbit, where the torque is exerted (see 
\S~\ref{sec:dTdM}).
However, in \S~\ref{sec:dTdM} we will develop a formalism for
computing torques that can be applied in the presence of mildly
varying radial gradients (see \S~\ref{sec:dTdMvars}) and of
gap-opening planets (see \S~\ref{sec:dTdMgap}).

\subsection{A Two-dimensional Torque Formula for a Locally Isothermal Disk}
\label{sec:2dformula}

Two-dimensional ($r$--$\phi$) studies of disk-planet interactions 
often smooth the gravitational potential of the planet, $\Phi_{p}$, 
over a length of order the disk thickness, $H$. This procedure is 
meant to account for the vertical stratification of disk material in 
proximity of the planet. However, as noted by \citetalias{menou2004}, 
it is unclear how well this approximation works since three-dimensional resonant 
torques depend on the vertical distribution of density in the disk 
\citepalias{tanaka2002}. Moreover, two-dimensional simulations of the Type~I 
regime indicate that torques depend on the value of the softening 
length \citep[e.g.,][]{masset2002,rnelson2004}.

We perform two-dimensional simulations using settings and parameters 
described in \S~\ref{sec:SimulationDescription} and applying
a planet potential with a smoothing length equal to $H$, the same 
value of the smoothing length adopted by \citetalias{menou2004}.
The potential is then $\Phi_{p}=-G M_{p}/\sqrt{S^{2}+H^{2}}$, 
where $S$ is the distance from the planet and 
$M_{p}=3\times 10^{-6}\,M_{s}$. Although the softening length
is introduced as a proxy  for the vertical disk thickness, 
it obviously affects the horizontal gradient of the gravitational 
potential as well. The torque density distribution 
of non-gap-opening planets peaks at a radial distance of about $H$ 
from the planet's orbit (see \S~\ref{sec:dTdM}). At those locations, 
the \textit{smoothed} gravitational force, $|\partial \Phi_{p}/\partial S|$, 
is $\sqrt{2}/4$ of that resulting from an unsoftened, point-mass potential.

The grid resolution in these calculations is 
$\Delta r= a\,\Delta\phi=1.5\times 10^{-3}$.  We consider values of
the disk gradients, $\beta$ and $\zeta$, in the range from $-2$ to $2$. 
By fitting the results from the two-dimensional nonlinear models, we obtain a total 
torque that  can be written as
\begin{eqnarray}
 \mathcal{T}_{\mathrm{2D}}&=&%
            -\left(1.17 + 0.15\,\beta + 0.22\,\zeta\right)\nonumber\\%
                                            & &\times\Sigma(a) \, \Omega^{2}(a) \, a^{4}
    \left(\frac{\Mp}{\Ms}\right)^{2} \left(\frac{a}{H}\right)^{2}.
 \label{eq:T2d}
\end{eqnarray}
We check the scaling of Equation~(\ref{eq:T2d}) with disk
thickness, $H$, by performing calculations for given values of $\beta$ 
and $\zeta$ and varying the sound speed at $r=a$. 
The scaling with planet mass was already confirmed by previous 
two-dimensional studies \citep[see, e.g.,][]{rnelson2004,masset2006a}.

Clearly, there are important differences between our three-dimensional and two-dimensional 
results with smoothed potential, Equations~(\ref{eq:Tdl}) and 
(\ref{eq:T2d}) respectively. In fact, although there is not a wide 
discrepancy in total torque for the case of a spatially isothermal 
disk ($\zeta=0$) with constant surface density ($\beta=0$), there 
are factors of $4$ and $2$ difference in the coefficients of 
parameters $\beta$ and $\zeta$, respectively.

We did not check how sensitive Equation~(\ref{eq:T2d}) is to changes
of the softening length and whether it is possible to reproduce more 
closely the three-dimensional formula (Equation~(\ref{eq:Tdl})) with a different a value 
of this parameter. However, there are indications that the coefficients of 
$\beta$ and $\zeta$ in Equation~(\ref{eq:T2d}) are highly dependent 
on the softening length (S.\ Li \ \& H.\ Li 2010, private communication). 

The two-dimensional analytic linear torque of \citetalias{tanaka2002} 
produces a coefficient of $\beta$ equal to $2.83$
\citep[see also the two-dimensional linear calculations by][]{korycansky1993}, which is quite 
different from our value, although the torques are nearly identical for 
$\beta=\zeta=0$. However, \citetalias{tanaka2002} did not apply any 
smoothing in obtaining their result. Therefore, the likely explanation 
for this discrepancy is the smoothing length $H$ that we adopt.

\section{TORQUE DENSITY DISTRIBUTIONS}
\label{sec:dTdM}

Waves carry torques until they are damped and deposit angular momentum 
in the disk. In order to gain some understanding of the radial distribution of 
torques, which is not provided by the integrated torque presented in
\S~\ref{sec:TypeIFormula}, we analyze the azimuthally averaged torque 
density as a function of radius.
This function measures the local strength of disk torques and should be 
insensitive to the details of wave damping that determine where the torque 
is exerted on the disk. 

The torque distribution per unit disk mass as a function of radius,
$d\mathcal{T}(r)/dM$, is defined such that 
\begin{equation}
 \mathcal{T}=2 \pi\!\!\int^{\infty}_{0}{\frac{d \mathcal{T}}{d M}(r)%
                                      \,\Sigma(r)\, r\, dr},
 \label{eq:dTdMdef}
\end{equation}
where $\Sigma(r)$ is the axisymmetric (i.e., azimuthally averaged) 
gas surface density.
This quantity provides an important diagnostic for the nature of the 
disk-planet torques (\citetalias{gennaro2008}; \citealp{li2009}). 
The theory of disk resonances  \citep[e.g.,][]{meyervernet1987,ward1997}
suggests that the torque distribution for a disk with smoothly
varying properties should approximately follow 
\begin{equation}
 \frac{d \mathcal{T}}{d M}(r)=%
 \mathcal{F}\!\left(\frac{r-a}{H}, \beta, \zeta \right)%
 \Omega^{2}(a) \, a^{2}\!%
 \left(\frac{\Mp}{\Ms}\right)^{2}\! \left(\frac{a}{H}\right)^{4}\!,
 \label{eq:dTdM}
\end{equation}
where $\mathcal{F}$ is a dimensionless function and $H$ is evaluated 
at the planet's orbital radius, $a$. 
(There are some modifications to this expression due to resonance widths 
that we ignore.) Notice that, by using Equations~(\ref{eq:dTdMdef}) and
(\ref{eq:dTdM}) along with the approximation $\Sigma(r)\approx\Sigma(a)$
and performing the transformation $r\rightarrow r/H$, one correctly finds 
from the integration that the total torque scales as $(a/H)^{2}$.
In the past, function $\mathcal{F}$ has in effect been 
taken to be an inverse power law with distance from the planet that is 
modified close to the planet. The expression is based on the impulse 
approximation \citep{lin1986b}. We provide here a more accurate
determination of $\mathcal{F}$.

In this section, we describe results for testing the universality of 
function $\mathcal{F}$ over a range of disk parameters by means of 
three-dimensional simulations. For this purpose, we fix surface density and temperature 
gradients and consider the case of a disk with $\beta=0.5$ and $\zeta=1$.
This situation corresponds to a steady-state viscous disk (accretion rate 
independent of radius) with constant aspect-ratio, $H/r$, and constant 
viscosity parameter, $\alpha$. 
Function $\mathcal{F}$ should then depend only on $(r-a)/H$, with these 
assumed values of $\beta$ and $\zeta$.

\begin{figure}
\centering%
\resizebox{\linewidth}{!}{%
\includegraphics{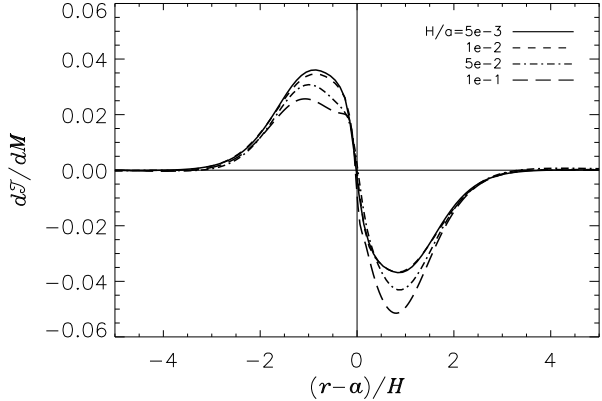}}
\caption{Torque density distributions (torque per unit disk mass)
         in a locally isothermal disk, interacting with a low-mass 
         planet, for different values of the disk scale-height at 
         the orbital radius of the planet. The ratio $H(a)/a$ is 
         indicated in the upper-right corner. Surface density and
         temperature radial gradients are taken to be $\beta=0.5$
         and $\zeta=1$ (see \S~\ref{sec:DiskModel}).
         The torque is scaled by the quantity
         $\Omega^{2}\,a^{2}\,(\Mp/\Ms)^{2}\,(a/H)^{4}$, in which
         $\Omega=\Omega(a)$ and $H=H(a)$.
         }
\label{fig:dtdm_h}
\end{figure}
\begin{figure}
\centering%
\resizebox{\linewidth}{!}{%
\includegraphics{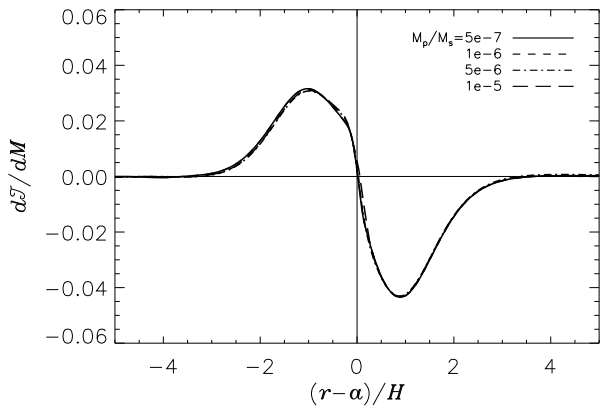}}
\caption{Torque density distributions in a locally isothermal disk, 
         interacting with a low-mass planet, for different values 
         of the planet-to-star mass ratio $\Mp/\Ms$, indicated in 
         the upper-right corner. Density and temperature radial
         gradients in the disk are the same as those in 
         Figure~\ref{fig:dtdm_h}. Again, the torque is scaled by
         $\Omega^{2}\,a^{2}\,(\Mp/\Ms)^{2}\,(a/H)^{4}$, where
         $\Omega=\Omega(a)$ and $H=H(a)$.
         }
\label{fig:dtdm_q}
\end{figure}
\begin{figure}
\centering%
\resizebox{\linewidth}{!}{%
\includegraphics{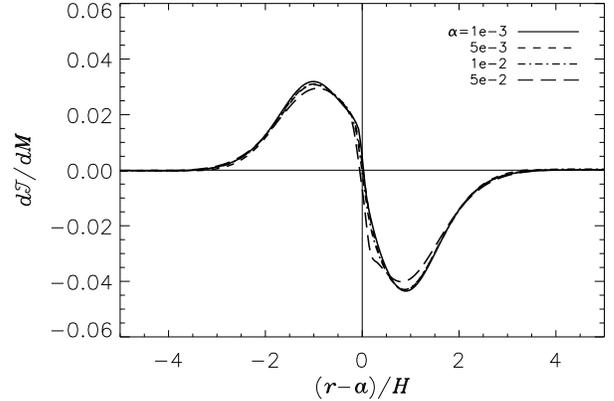}}
\caption{Torque density distributions in a locally isothermal disk,
         interacting with a low-mass planet, for different values
         of the viscosity parameter $\alpha$, indicated in the 
         upper-right corner. The torque is scaled as in 
         Figures~\ref{fig:dtdm_h} and \ref{fig:dtdm_q}. 
         Surface density and temperature radial gradients are given 
         by $\beta=-d\ln{\Sigma}/\ln{r}=0.5$ and 
         $\zeta=-d\ln{T}/\ln{r}=1$.
         }
\label{fig:dtdm_alpha}
\end{figure}
\begin{figure*}
\centering%
\resizebox{\linewidth}{!}{%
\includegraphics{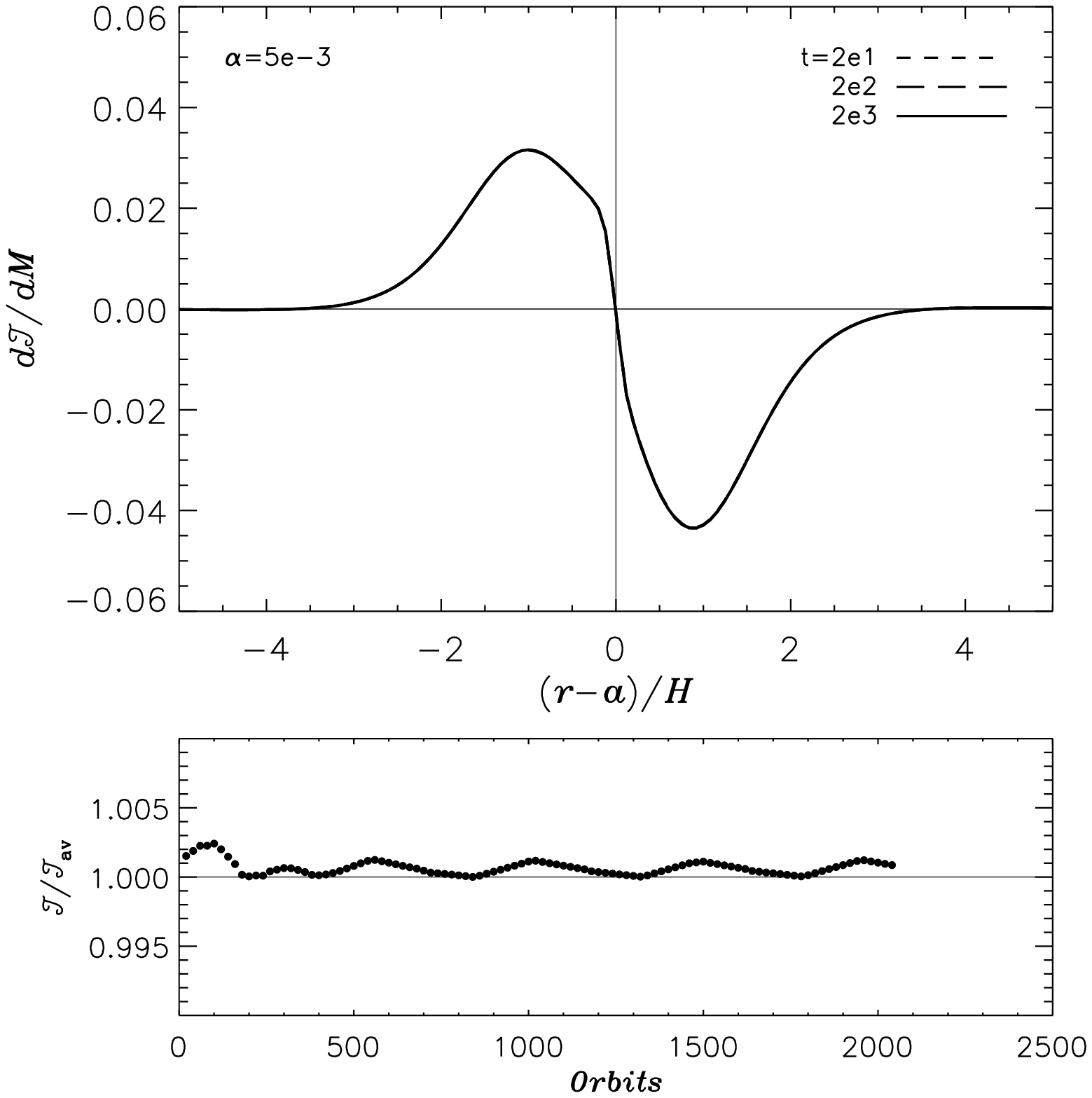}%
\includegraphics{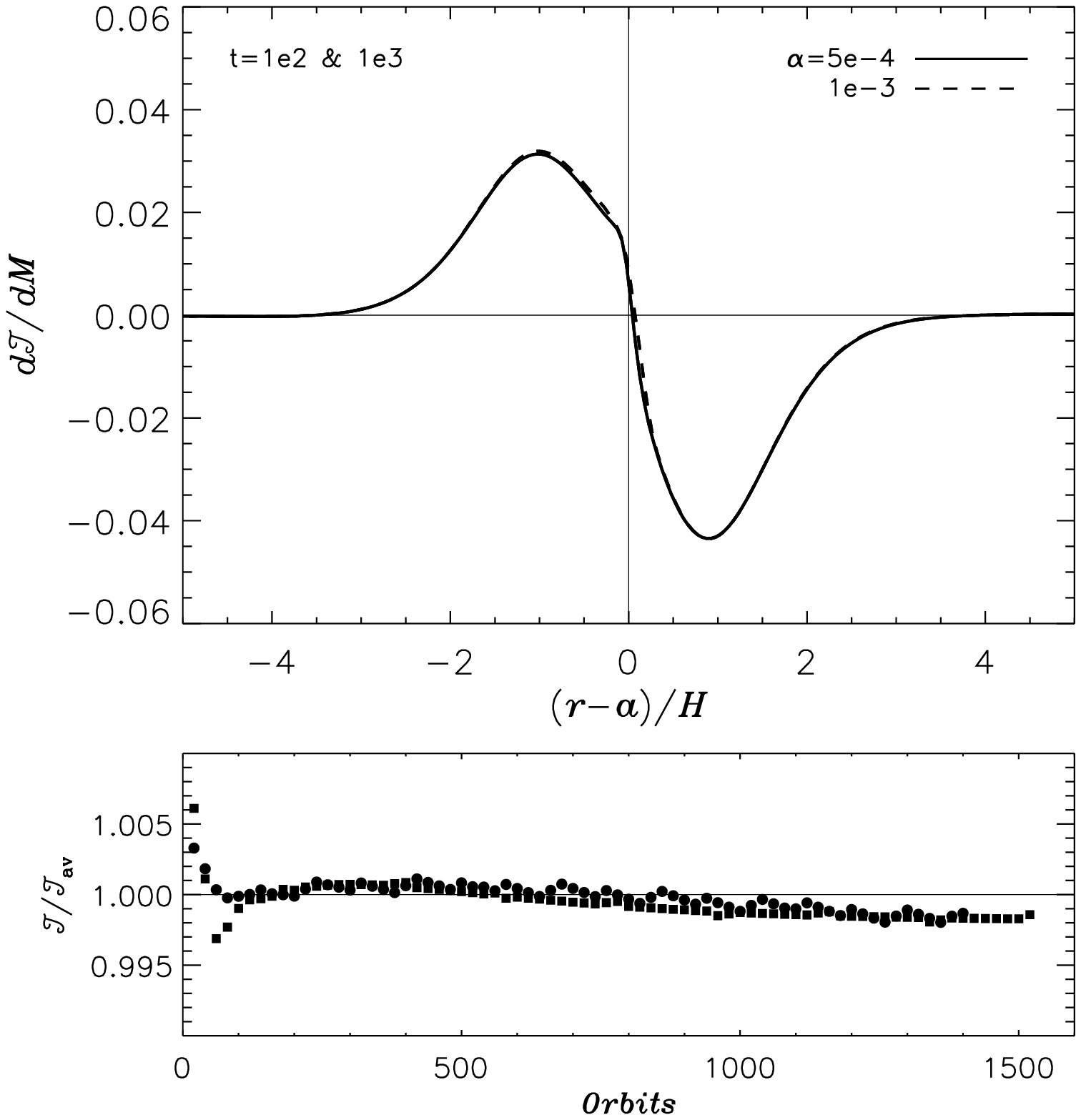}}
\caption{\textit{Top}: time behavior of the torque density distribution 
         in a locally isothermal disk with turbulence parameter 
         $\alpha=0.005$, $0.001$, and $5\times 10^{-4}$, 
         interacting with a low-mass planet. 
         The time in units of orbital periods is indicated in the 
         upper-right (\textit{left panels}) and upper-left corner
         (\textit{right panels}). 
         The torque is scaled as in 
         Figures~\ref{fig:dtdm_h},  \ref{fig:dtdm_q}, and 
         \ref{fig:dtdm_alpha}. Disk's surface density and temperature 
         distributions have radial gradients given by $\beta=0.5$ and 
         $\zeta=1$.
         \textit{Bottom}: total torque as a function of time, for the same 
         models as in the top panels, normalized to the average torque value,
         $\mathcal{T}_{\mathrm{av}}$, between $100$ and $200$ orbits.
         In the right panel, filled squares refer to case with 
         $\alpha=0.001$. As in Figure~\ref{fig:satu} , the lack of significant 
         time evolution of the torque is consistent with coorbital torques being 
         unsaturated.
         }
\label{fig:dtdm_t}
\end{figure*}
We test Equation~(\ref{eq:dTdM}) by considering simulations involving 
four different values for $H/a$ and $\Mp/\Ms$ and levels of viscosity in 
Figures~\ref{fig:dtdm_h}, \ref{fig:dtdm_q}, and \ref{fig:dtdm_alpha}, 
respectively. These figures show the results of $d\mathcal{T}(r)/dM$ 
obtained from simulations that are scaled such that the curves in the 
three figures should lie on top of each other, if the functional form of 
Equation~(\ref{eq:dTdM}) is correct.
The figures demonstrate fairly good agreement with Equation~(\ref{eq:dTdM}) 
for variations of $H/r$. There are $\sim 30$\% variations (measured at the 
curve peaks) in the scaled values of $d\mathcal{T}(r)/dM$ (see
Figure~\ref{fig:dtdm_h}), while the unscaled torque densities differ by 
a factor of $\sim 20^{4}$. There is less than a $2\%$ variation in the 
scaled $d\mathcal{T}(r)/dM$ for a factor of $20^{2}$ change in torque 
density due to planet mass changes (see Figure~\ref{fig:dtdm_q}). 
There is less than $5\%$ change in the scaled $d\mathcal{T}(r)/dM$ 
for a factor of $100$ change in $\alpha$ (see Figures~\ref{fig:dtdm_alpha}
and \ref{fig:dtdm_t}). These results also add support to our claim in 
\S~\ref{sec:TypeIFormula} that the coorbital disk torques are unsaturated.
The reason is that there would be a decrease in the magnitude of 
net torque at about the $30$\% level at higher viscosity, if the
coorbital torques were to switch from being saturated to being 
unsaturated (see Equations~(\ref{eq:Tdl}) and (\ref{eq:Tsat})).
In particular, Figure~\ref{fig:dtdm_t} indicates that there is no significant
change in total torque, over many libration timescales, since fractional 
variations stay at levels $< 0.5$\%.
The slightly lower mass planet, $\Mp=3\times 10^{-6}\,\Ms$ ($1\,\MEarth$), 
used in \S~\ref{sec:TypeIFormula} (versus $\Mp=5\times 10^{-6}\,\Ms$ 
adopted here) is even less prone to saturation according to the criterion 
in Equation~(\ref{eq:alcr}).

\begin{figure*}
\centering%
\resizebox{\linewidth}{!}{%
\includegraphics[clip=true]{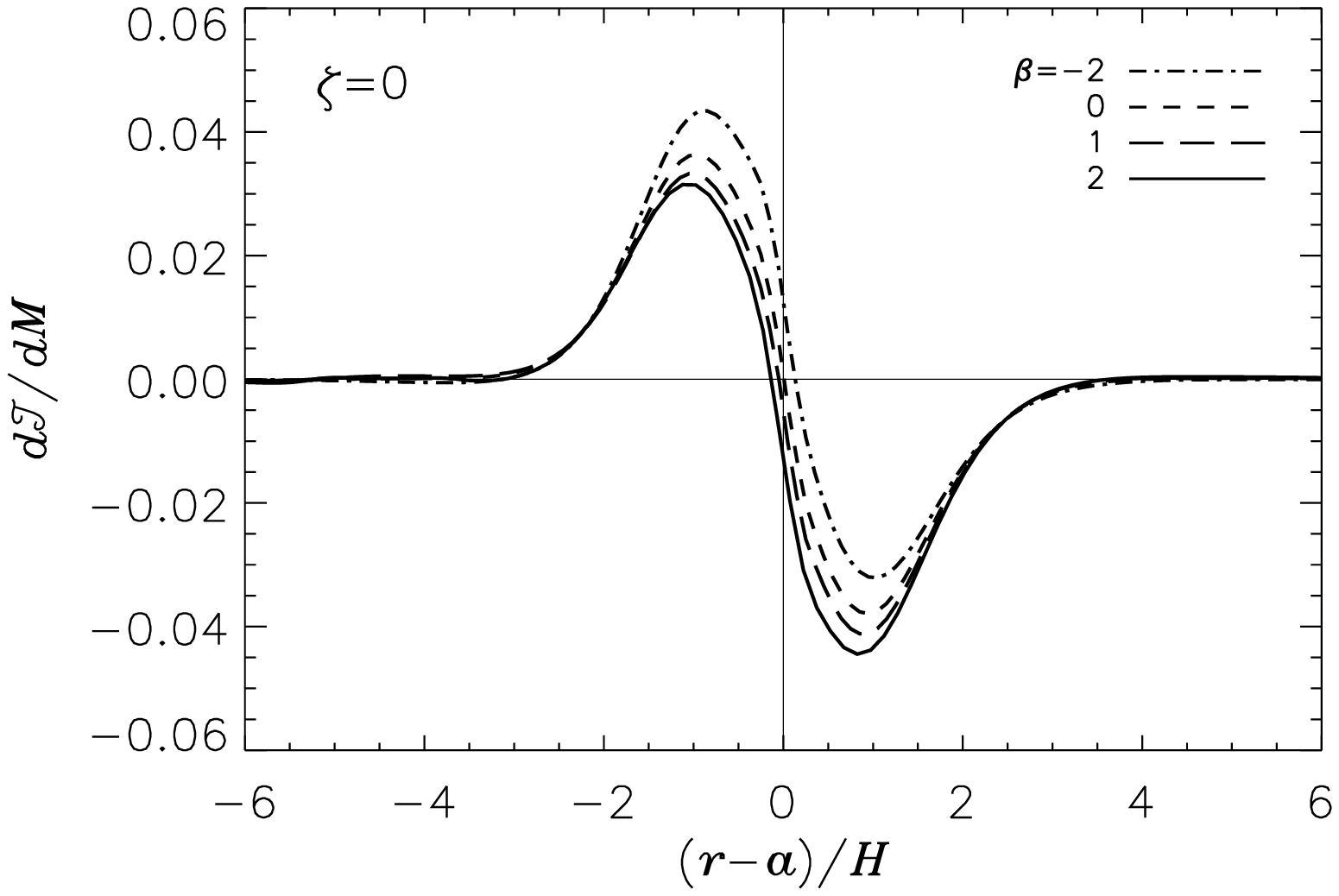}%
\includegraphics[clip=true]{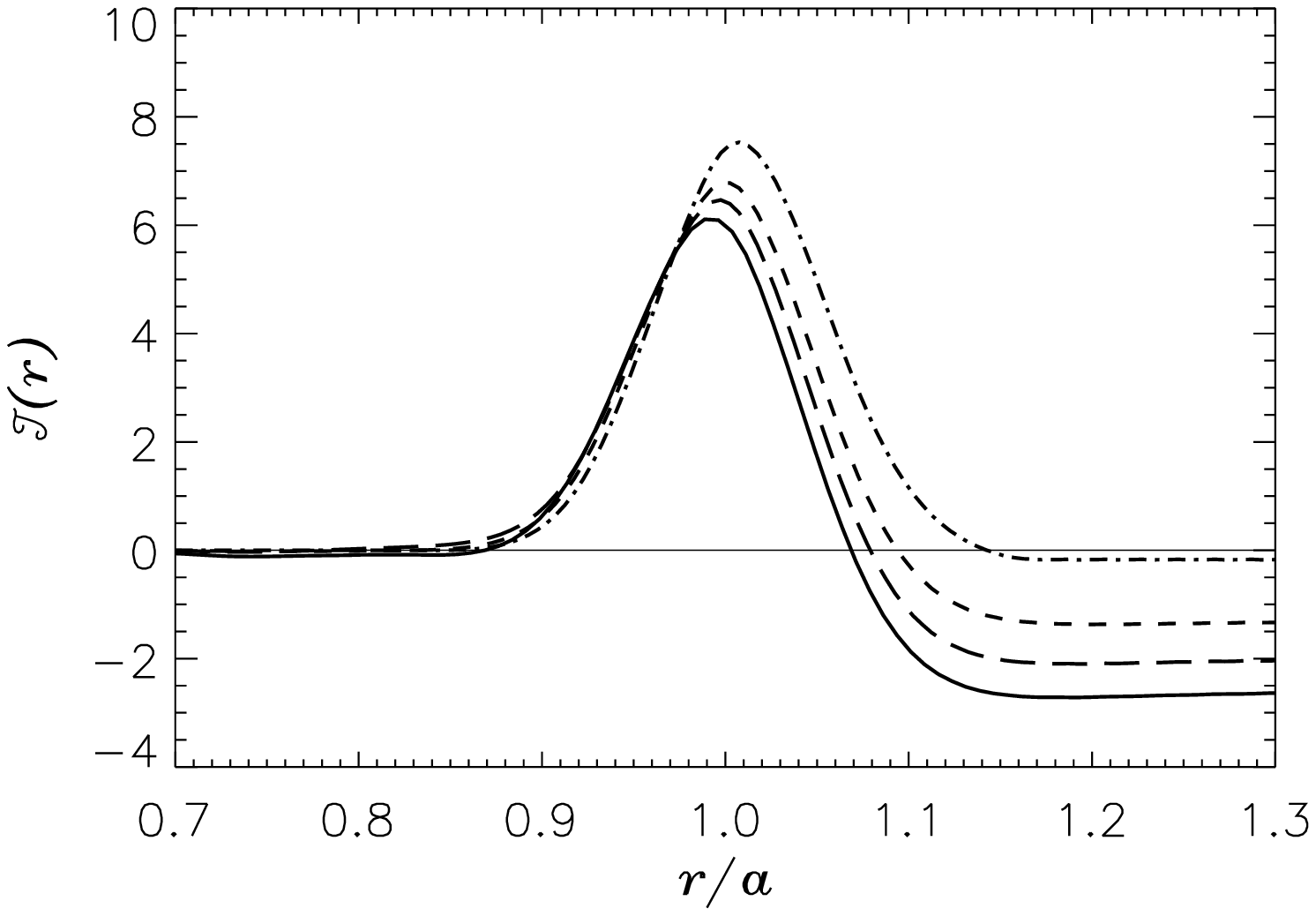}}
\resizebox{\linewidth}{!}{%
\includegraphics[clip=true]{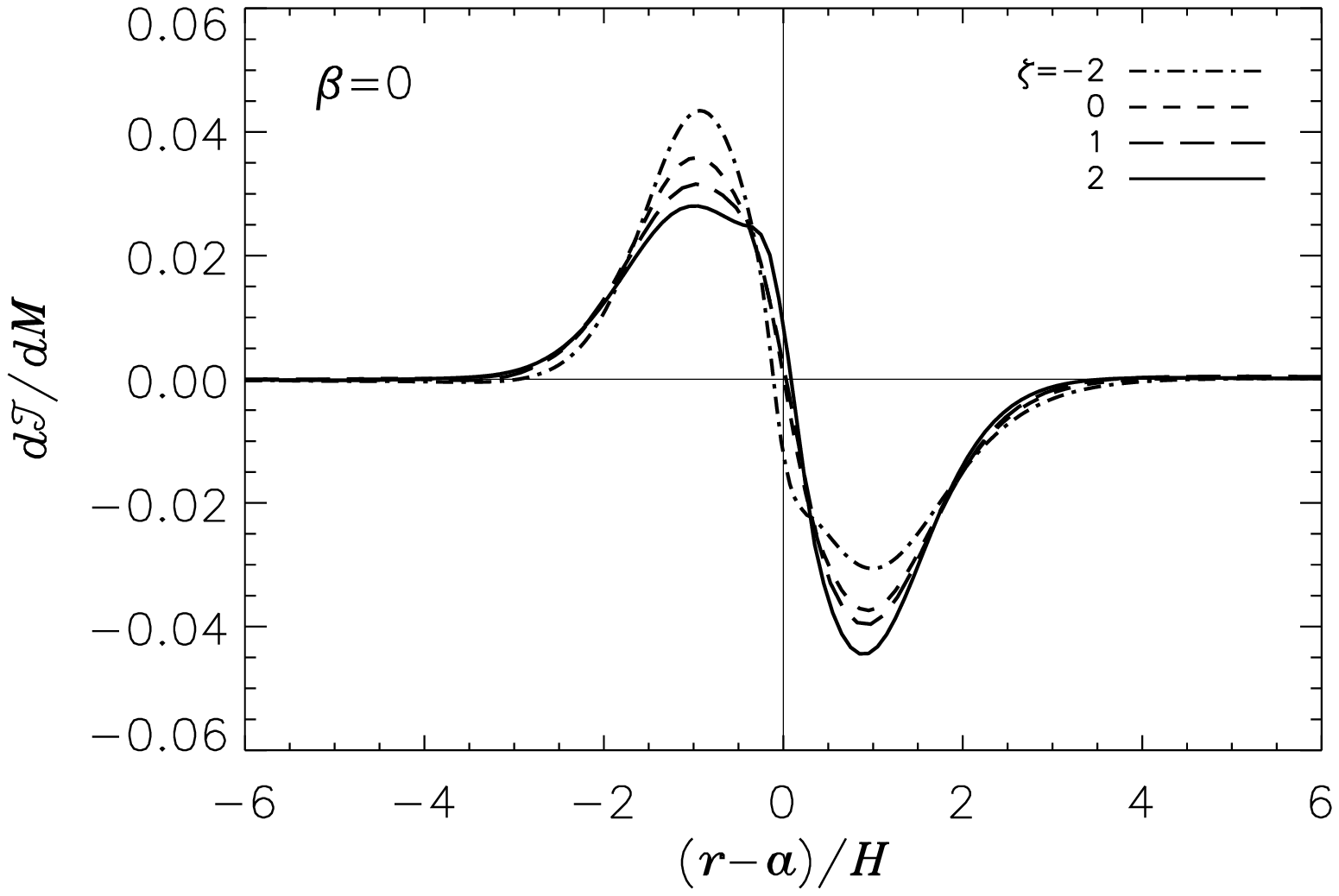}%
\includegraphics[clip=true]{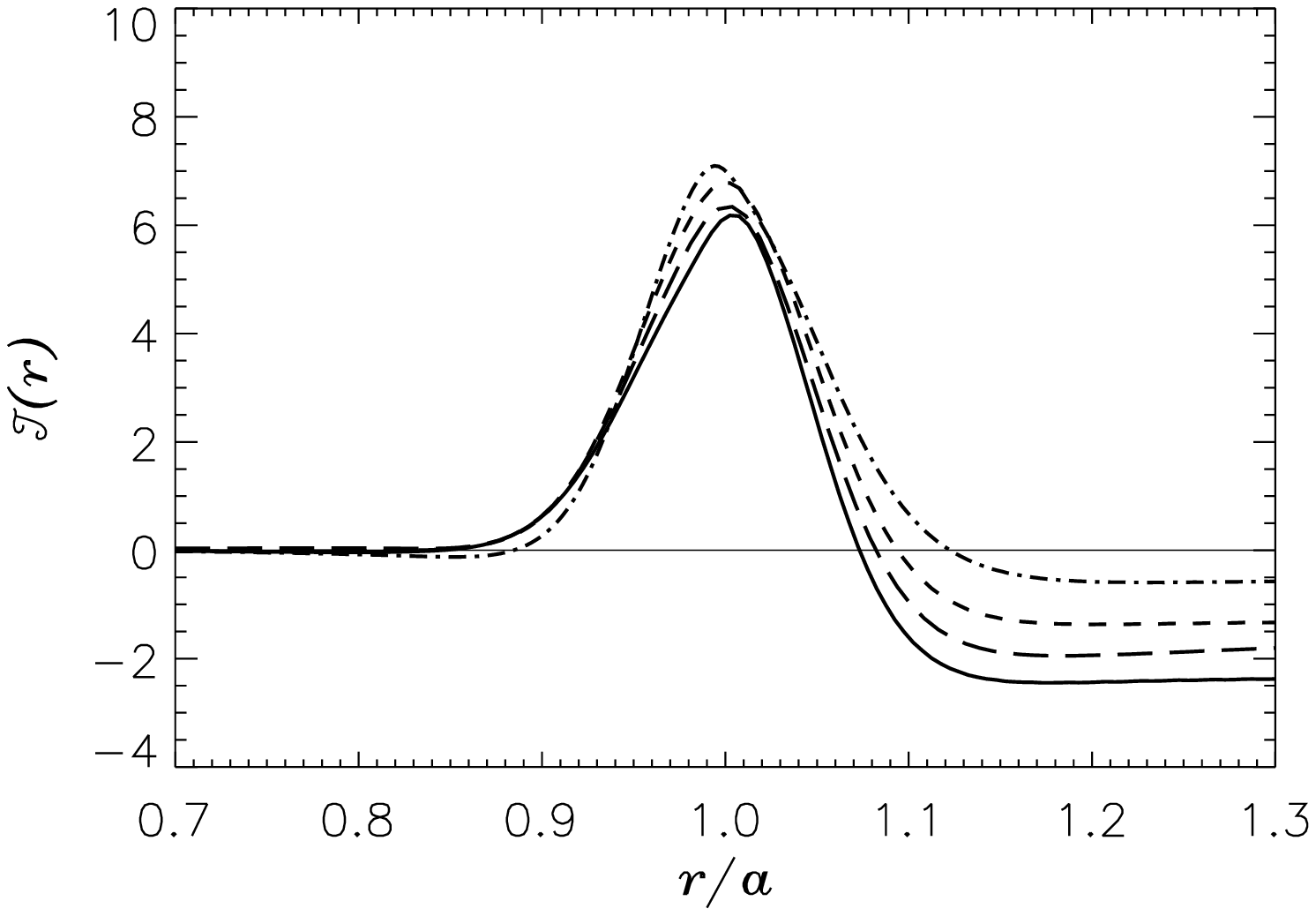}}
\caption{Torque density distributions (\textit{left panels}) and 
         cumulative torques (\textit{right panels}) in a locally
         isothermal disk, interacting with a low-mass planet, for 
         different values of the radial gradient of the surface density
         $\beta$ (\textit{top panels}) and of the radial gradient of 
         temperature $\zeta$ (\textit{bottom panels}). The values 
         of the parameters $\beta$ and $\zeta$ are indicated in 
         the upper right corners of the left panels. Models in the 
         top panels have a fixed temperature gradient $\zeta=0$ 
         (i.e., the disk is isothermal), whereas in the bottom panels 
         the gradient of $\Sigma$ is $\beta=0$. Distributions
         $d\mathcal{T}(r)/dM$ are normalized to
         $\Omega^{2}\,a^{2}\,(\Mp/\Ms)^{2}\,(a/H)^{4}$,
         whereas cumulative torques are normalized to
         $\Sigma\,\Omega^{2}\,a^{4}\,(\Mp/\Ms)^{2}\,(a/H)^{2}$,
         in which $\Sigma$, $\Omega$, and $H$ are all evaluated at
         the orbital radius of the planet, $a$.
         }
\label{fig:Tvsbz}
\end{figure*}
The torque density distributions in, e.g., Figure~\ref{fig:dtdm_q}
would produce a positive total torque, through 
Equation~(\ref{eq:dTdMdef}), for a sufficiently negative density 
gradient ($\beta>0$) since the multiplication by $\Sigma$ would 
make the positive peak of the profile more positive and the negative 
peak less negative. 
However, as indicated in Equation~(\ref{eq:dTdM}), function 
$\mathcal{F}$ depends on the radial gradients of both density and 
temperature. For increasing values of $|\beta|$ and $|\zeta|$,
the asymmetry between the positive and negative peaks becomes larger.
As $\beta$ and $\zeta$ increase, the positive peak becomes shallower 
whereas the negative peak deepens.
This effect outweighs that produced by the surface density. 
This is illustrated in Figure~\ref{fig:Tvsbz}, where 
$d\mathcal{T}(r)/dM$ is shown for various values of $\beta$ 
(\textit{top-left panel}) and of $\zeta$ 
(\textit{bottom-left panel}). In the right panels of the figure
we plot the cumulative torque, $\mathcal{T}(r)$, defined as the integral
on the right-hand side of Equation~(\ref{eq:dTdMdef}) with integration 
limits from $0$ to $r$ (and $\Sigma\propto 1/r^{\beta}$). 
As $\beta$ and $\zeta$ increase, the magnitude of the total torque 
increases. The trend can persist for steeper surface densities, up to 
$\beta\approx 6$, the largest radial gradient we consider.
The total torque tends to decrease (i.e., it becomes less negative) for 
positive radial gradients of surface density and temperature, but 
relatively steep functions seem to be required to result in a positive 
torque (see bottom panels of Figures~\ref{fig:Tvsbz} and \ref{fig:Tfit}).

\subsection{Analytic Approximations of Torque Density Distributions}
\label{sec:AATDD}

%
\begin{figure*}
\centering%
\resizebox{\linewidth}{!}{%
\includegraphics{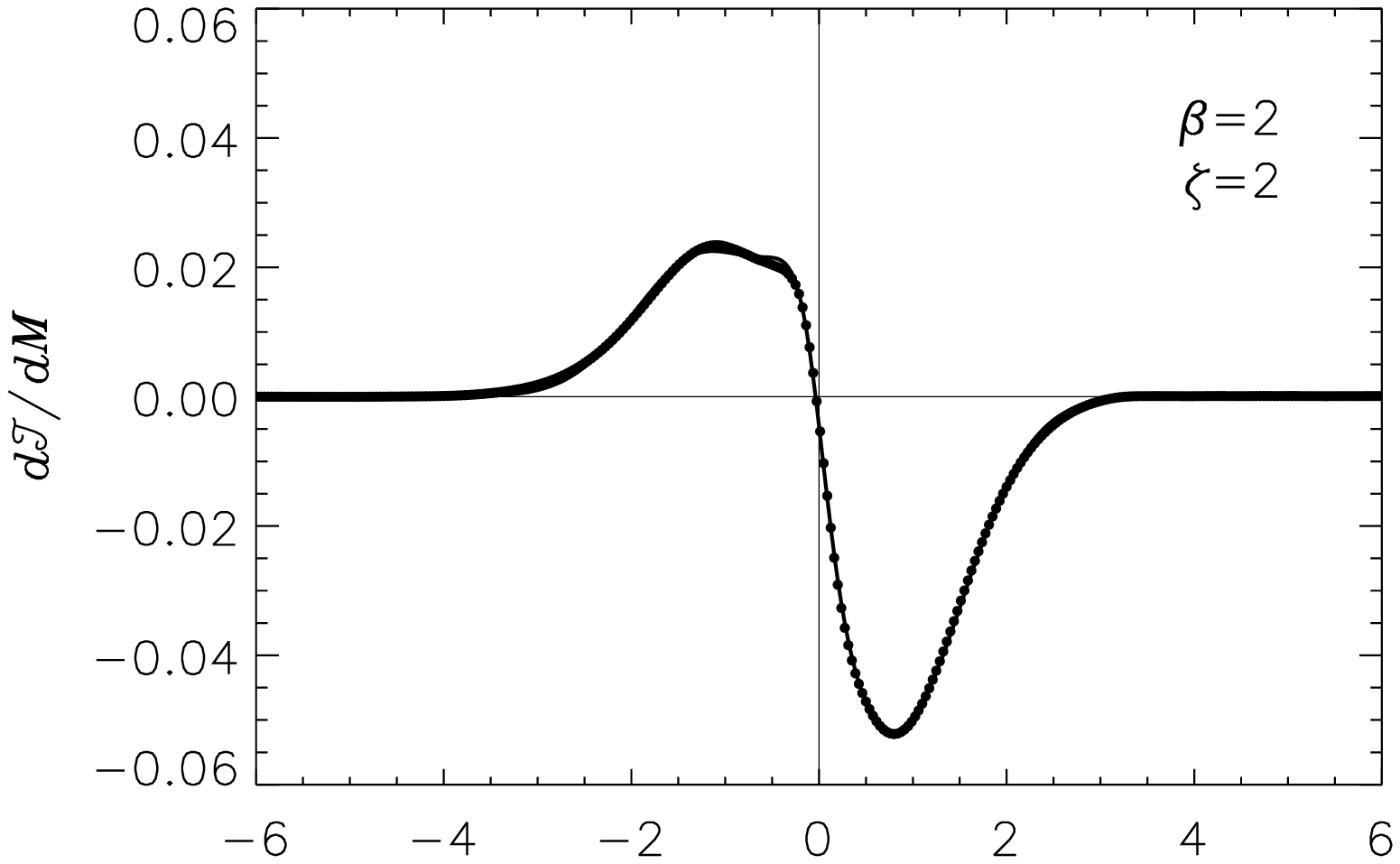}%
\includegraphics{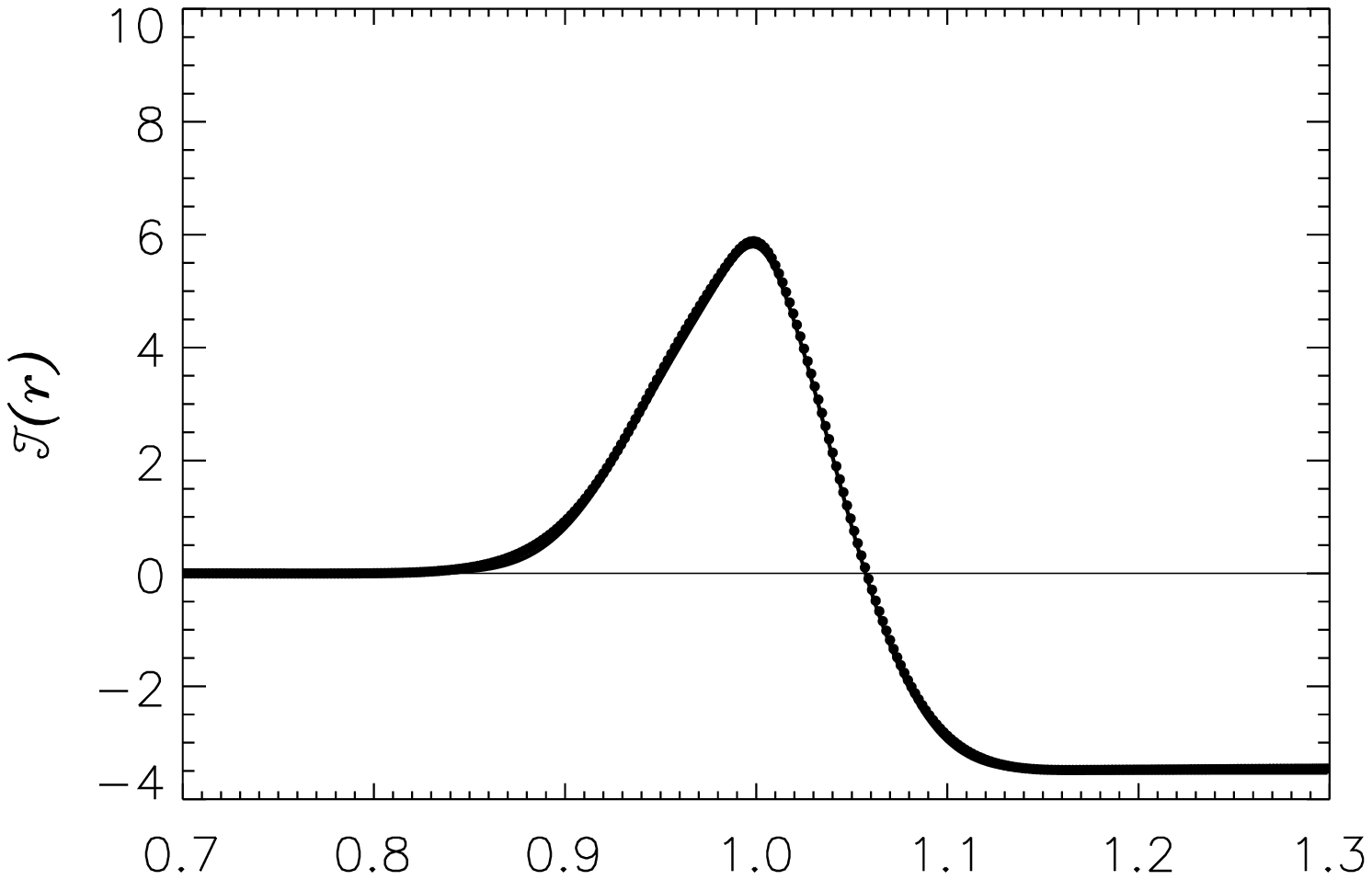}}
\resizebox{\linewidth}{!}{%
\includegraphics{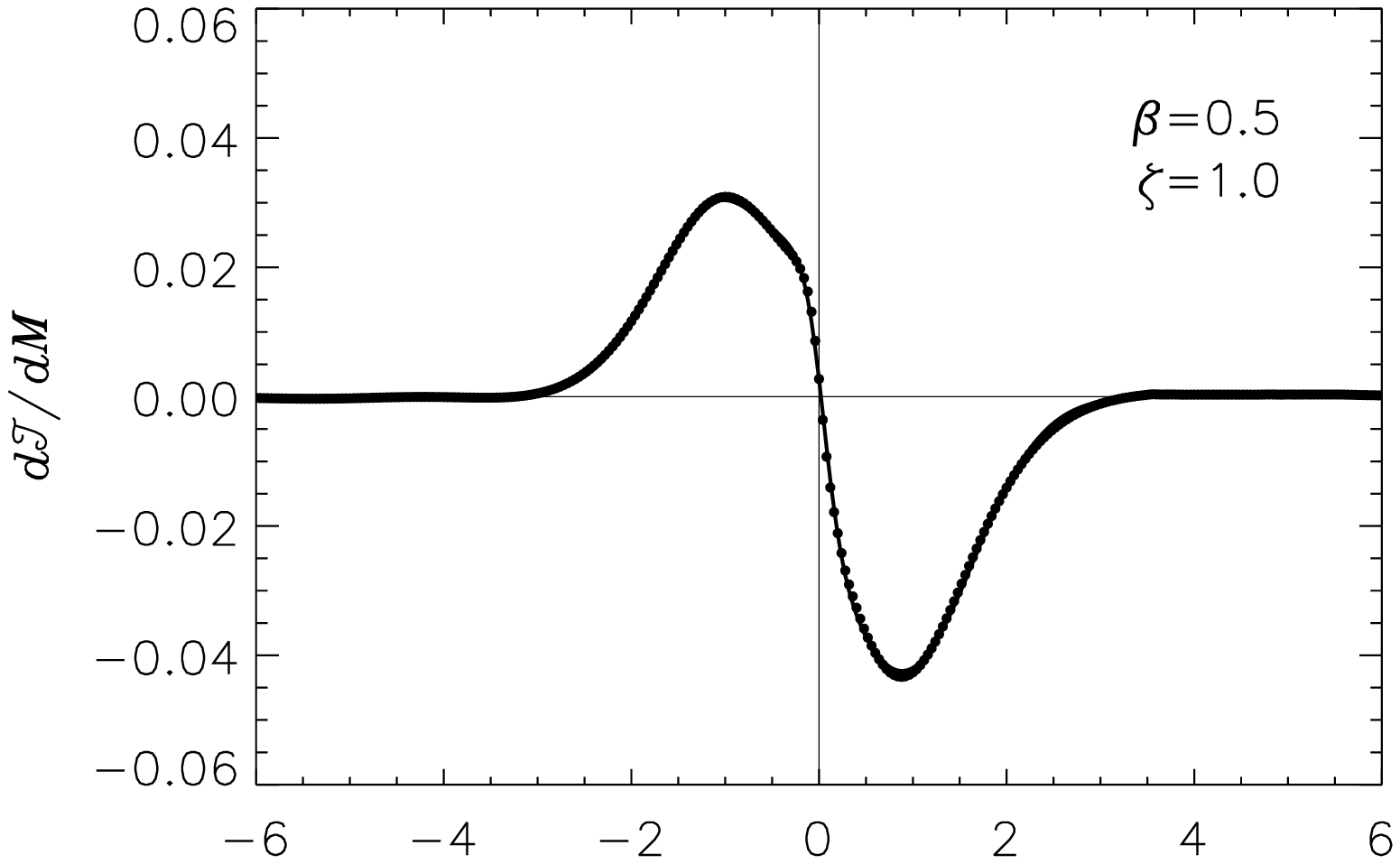}%
\includegraphics{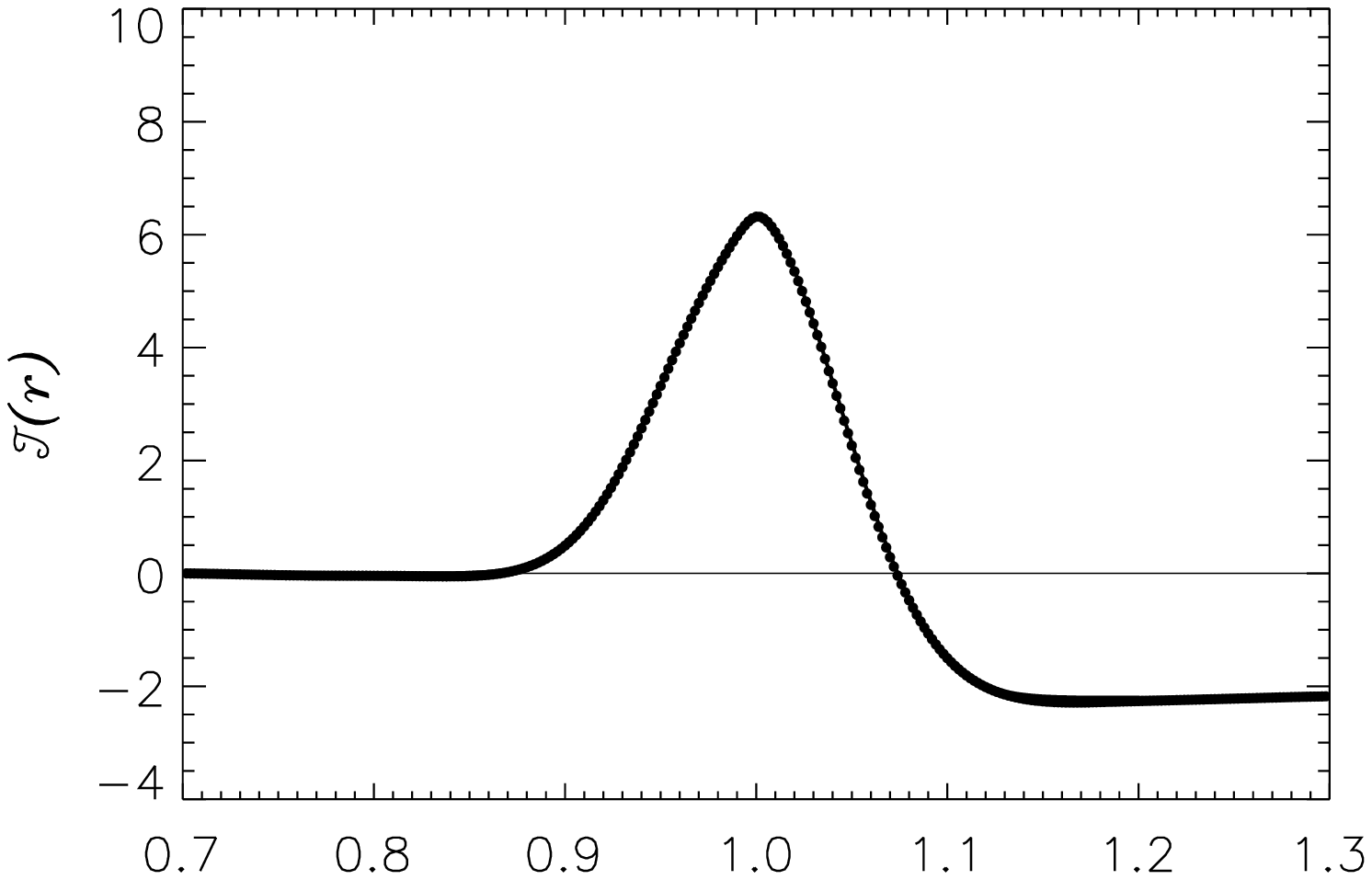}}
\resizebox{\linewidth}{!}{%
\includegraphics{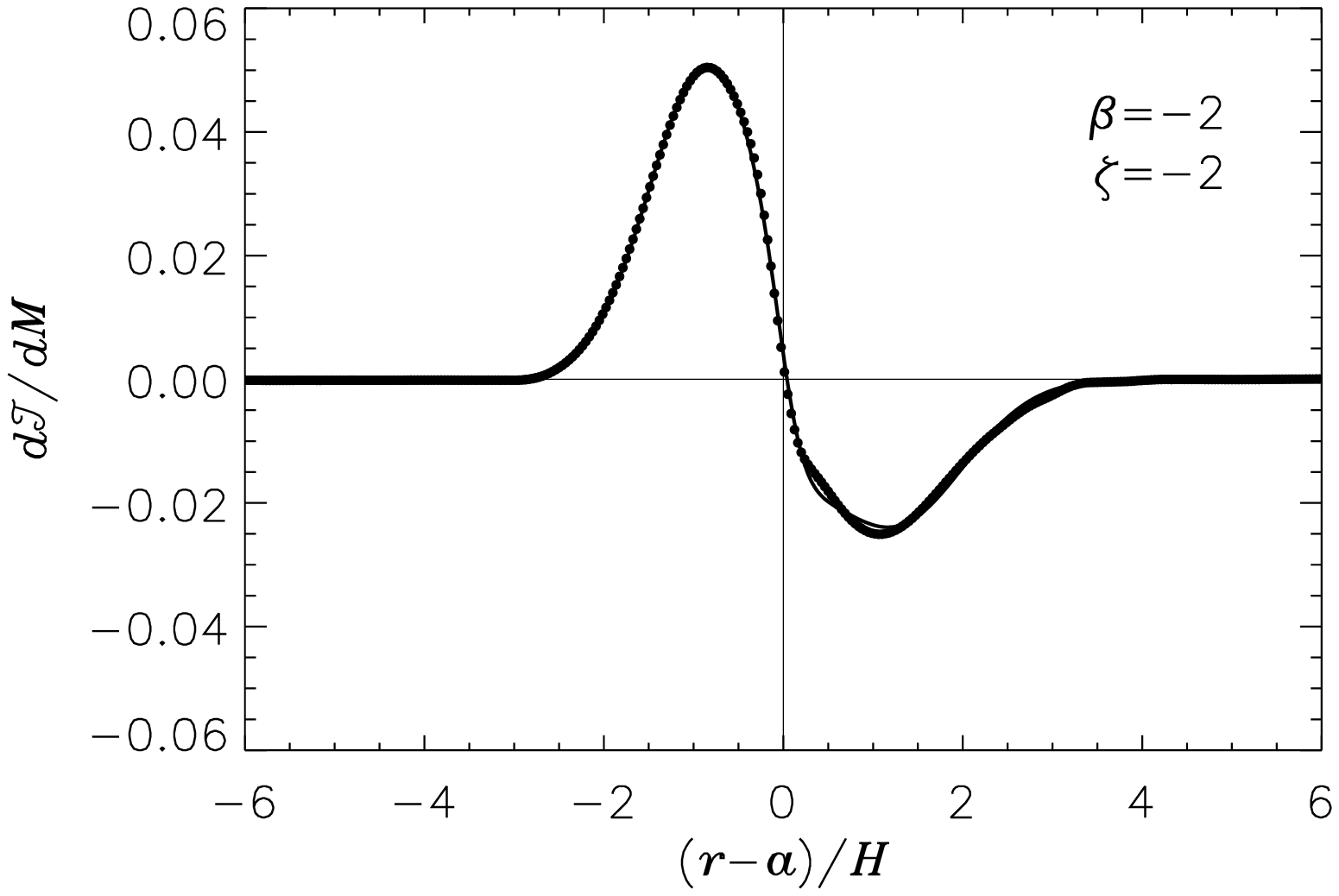}%
\includegraphics{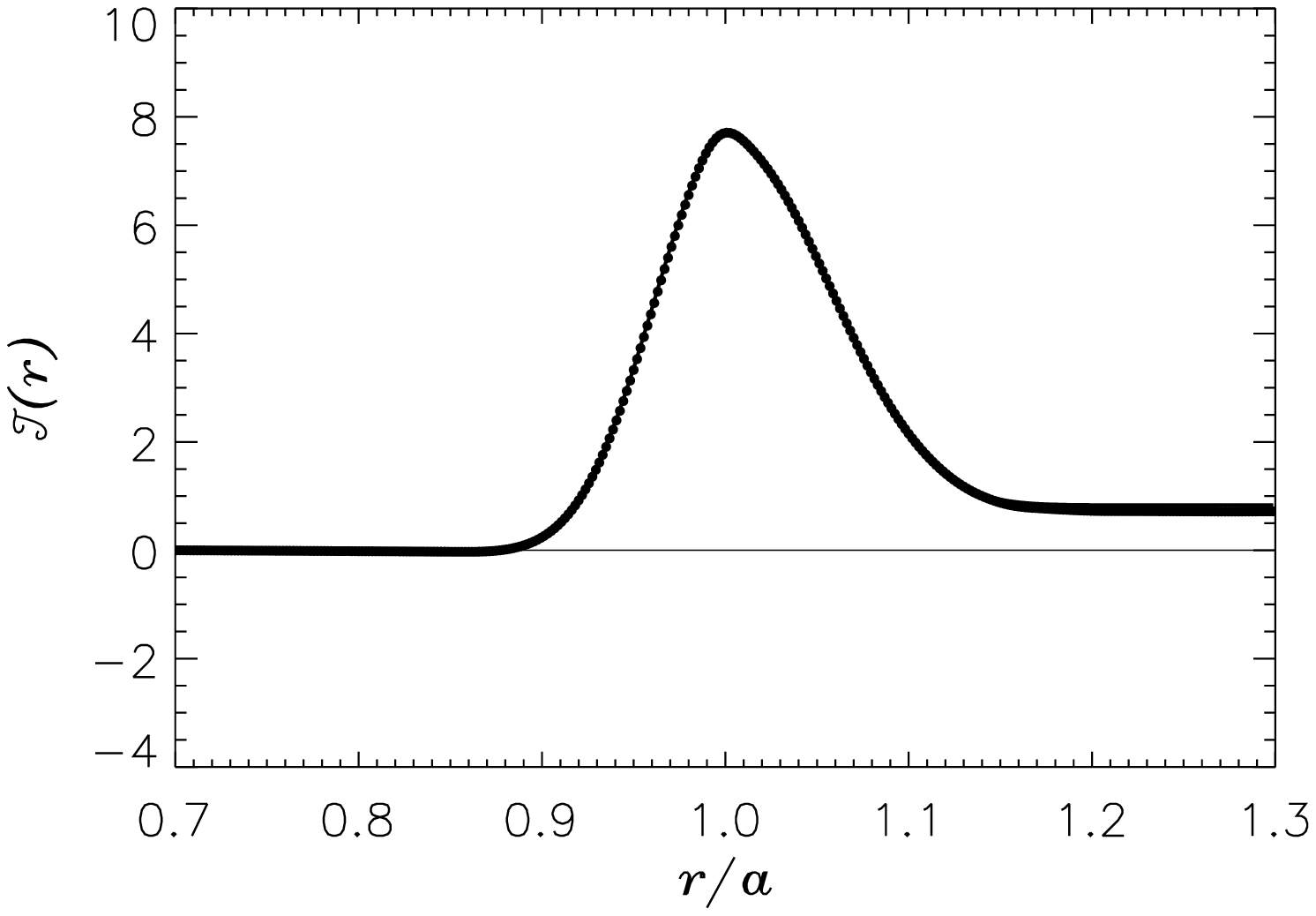}}
\caption{%
         \textit{Left panels}:
         torque density distributions in a locally isothermal
         disk whose surface density and temperature gradients are
         specified by the pair $(\beta,\zeta)$ indicated in the top-right
         corner of each panel.
         The dots indicate the three-dimensional simulation results that are 
         rescaled by
         $\Omega^{2}\,a^{2}\,(\Mp/\Ms)^{2}\,(a/H)^{4}$.
         The solid lines represent fits of
         Equation~(\ref{eq:F}) to the data points, with
         parameters $p_{i}$ listed in Table~\ref{tbl:ppar}.
         \textit{Right panels}:
         cumulative torques, rescaled by
         $\Sigma\,\Omega^{2}\,a^{4}\,(\Mp/\Ms)^{2}\,(a/H)^{2}$,
         resulting from $d\mathcal{T}(r)/dM$ shown in the
         left panels for both simulation data (\textit{dots})
         and fitting functions (\textit{solid line}).
         }
\label{fig:Tfit}
\end{figure*}
Analytic expressions of function $\mathcal{F}$ in Equation~(\ref{eq:dTdM}) 
can be found by fitting data. For this purpose, we assume that 
$\mathcal{F}$ is a parametric function of $\beta$ and $\zeta$,
of independent variable $x=(r-a)/H(a)$, and has the following form
\begin{eqnarray}
 \mathcal{F}(x,\beta,\zeta)&=&\left[p_{1}\,e^{-(x+p_{2})^{2}/p^{2}_{3}}
                           +p_{4}\,e^{-(x-p_{5})^{2}/p^{2}_{6}}\right]\nonumber\\%
                                         &  &\times\tanh{\left(p_{7}-p_{8}x\right)},
 \label{eq:F}
\end{eqnarray}
and $p_{i}=p_{i}(\beta,\zeta)$, with $i=1,\ldots,8$, are the parameters
resulting from the fit. This functional form \emph{is not} based on a 
physical model of resonant torques, but rather the result of searching 
among simple combinations of functions whose parameters have 
direct physical interpretations. So, in Equation~(\ref{eq:F}), 
parameters $p_{1}$ and $p_{4}$ are related to the amplitude of the 
positive and negative peaks, respectively, while parameters $p_{2}$ 
and $p_{5}$ are the distances of the peaks' positions from the 
planetary orbit. Parameters $p_{3}$ and $p_{6}$ regulate the width 
of the positive and negative portions of the function. 
The hyperbolic tangent affects the slope with which the profile 
transitions from positive to negative and the ratio $p_{7}/p_{8}$ 
gives the intercept of the function with the $x$-axis. Three
examples of the fit are displayed in Figure~\ref{fig:Tfit}, 
corresponding to gradients specified by pairs the 
$(\beta,\zeta)=(2,2)$ in the top panels, $(0.5,1)$ 
in the middle panels, and $(-2,-2)$ in the bottom 
panels. In the figure, the rescaled torque density distribution
$d\mathcal{T}(r)/dM$ (\textit{left panels}) and the cumulative
torque $\mathcal{T}(r)$ (\textit{right panel}s) from a simulations
(\textit{solid dots}) are compared against the fitting function
(\textit{solid lines}).

\begin{deluxetable*}{rcccccccc}
\tabletypesize{\footnotesize}
\tablecaption{Fit parameters for the function $\mathcal{F}$ in Equation~(\ref{eq:F}).\label{tbl:ppar}}
\tablewidth{0pt}
\tablehead{%
\colhead{$(\beta,\zeta)$}&\colhead{$p_{1}$}&\colhead{$p_{2}$}&\colhead{$p_{3}$}&\colhead{$p_{4}$}&\colhead{$p_{5}$}&\colhead{$p_{6}$}&\colhead{$p_{7}$}&\colhead{$p_{8}$}}
\startdata
        $(-2,-2)$    &$0.0501524$   &$0.849572$    &$0.920263$   &$0.0233445$   &$1.20039$         &$1.15790$     &$0.145358$     &$3.34263$\\
         $(0,-2)$    &$0.0419354$   &$0.984212$    &$0.861833$   &$0.0293424$   &$1.03610$         &$1.18903$     &$-0.481275$   &$4.38744$\\
         $(1,-2)$    &$0.0379796$   &$1.05644$      &$0.826359$   &$0.0328917$   &$0.945701$       &$1.21163$     &$-0.936489$    &$5.30156$\\
         $(2,-2)$    &$0.0315764$   &$1.15319$     &$0.803355$   &$0.0368841$   &$0.760423$       &$1.35752$     &$-0.989595$     &$4.54574$\\
         $(4,-2)$    &$0.0202990$    &$1.32162$     &$0.713699$   &$0.0464587$   &$0.557390$       &
$1.48935$     &$-1.11322$       &$3.58864$\\[0.6mm]
        \cline{1-9}\\[-1.5mm]
        $(-2,-1)$    &$0.0466666$    &$0.863280$   &$0.966481$   &$0.0264784$   &$1.13846$        &$1.12637$     &$0.300767$     &$3.29211$\\
         $(0,-1)$    &$0.0382860$    &$0.997974$   &$0.910000$    &$0.0328630$  &$1.00086$        &$1.15666$     &$-0.171042$    &$3.58431$\\
         $(1,-1)$    &$0.0349835$    &$1.05169$     &$0.902117$    &$0.0365165$  &$0.922390$      &$1.17324$    &$-0.403456$     &$3.31236$\\
         $(2,-1)$    &$0.0309523$    &$1.11777$     &$0.889403$    &$0.0405477$  &$0.840493$       &$1.22073$     &$-0.593193$    &$3.16310$\\
         $(4,-1)$    &$0.0228084$    &$1.34202$     &$0.789789$    &$0.0507717$  &$0.629326$       &
$1.32866$     &$-0.897774$    &$3.07277$\\[0.6mm]
        \cline{1-9}\\[-1.5mm]
        $(-2,0)$     &$0.0421255$    &$0.897070$   &$1.01784$      &$0.0304875$  &$1.02268$         &$1.14905$     &$0.428988$      &$2.89123$\\
         $(0,0)$     &$0.0341462$     &$1.04357$    &$0.958885$    &$0.0369814$  &$0.933042$        &$1.13253$    &$0.0358698$     &$3.23369$\\
         $(1,0)$     &$0.0303427$     &$1.11959$    &$0.928609$    &$0.0408822$  &$0.861233$        &$1.15412$    &$-0.181823$      &$3.07328$\\
         $(2,0)$     &$0.0287186$     &$1.21232$    &$0.849933$    &$0.0447468$  &$0.802171$        &$1.16904$    &$-0.466765$      &$3.59154$\\
         $(4,0)$     &$0.0223466$     &$1.32467$    &$0.785512$    &$0.0542731$  &$0.646047$        &    
$1.22339$    &$-0.713550$      &$2.69379$\\[0.6mm]
        \cline{1-9}\\[-1.5mm]
        $(-2,1)$     &$0.0342374$     &$0.956221$  &$1.11505$      &$0.0352915$  &$0.814786$        &$1.25951$    &$0.393368$        &$2.10245$\\
        $(0,1)$      &$0.0317793$      &$1.06971$   &$0.969339$    &$0.0401954$  &$0.931402$        &$1.04558$    &$0.221833$        &$4.50245$\\
         $(0.5,1)$  &$0.0297597$      &$1.09770$   &$0.938567$    &$0.0421186$  &$0.902328$        &$1.03579$    &$0.0981183$     &$4.68108$\\
         $(1.5,1)$  &$0.0270259$     &$1.22111$    &$0.891422$    &$0.0467822$  &$0.832732$        &$1.06184$    &$-0.226587$      &$4.07528$\\
         $(2,1)$     &$0.0255936$     &$1.27569$    &$0.867892$    &$0.0489470$  &$0.791497$        &$1.06753$    &$-0.358692$      &$3.69761$\\ 
         $(4,1)$     &$0.0178517$     &$1.39810$    &$0.912140$    &$0.0595500$  &$0.630116$        &$1.16562$    &$-0.487347$      &$1.89963$\\[0.6mm]
        \cline{1-9}\\[-1.5mm]
        $(-2,2)$     &$0.0347398$    &$0.864893$   &$1.21887$      &$0.0346082$  &$1.00993$         &$1.03900$     &$0.823430$      &$3.99184$\\
         $(0,2)$     &$0.0274646$    &$ 1.07401$    &$1.04454$      &$0.0435430$  &$0.924349$        &$0.975019$  &$0.538068$      &$5.37855$\\
         $(1,2)$     &$0.0244322$    &$1.20668$     &$0.986380$    &$0.0474291$  &$0.866602$        &$1.01086$    &$0.323656$      &$4.93965$\\
         $(2,2)$     &$0.0211401$    &$1.27928$     &$0.987525$    &$0.0518687$  &$0.785682$        &$1.05005$    &$-0.134179$     &$3.71919$\\
         $(4,2)$     &$0.0184703$    &$1.32365$     &$1.04665$      &$0.0666310$  &$0.671786$        &    
$1.01557$    &$-0.446362$     &$1.84684$
\enddata
\end{deluxetable*}
In Table~\ref{tbl:ppar}, we give parameters $p_{i}$ for a number
of representations of function $\mathcal{F}$, corresponding to 
various parameter pairs $(\beta,\zeta)$. We find that 
expression~(\ref{eq:F}) generally produces a very good or good 
fit to the distributions obtained from simulations. There are 
various ways to quantify how well Equation~(\ref{eq:F}) fits the 
data, one of the most stringent ways is to compare cumulative 
torques generated by the fitting functions against those 
determined from simulations. Three such comparisons are 
plotted in the right panels of Figure~\ref{fig:Tfit}, which have
a level of agreement ranging from a few per cent (\textit{top} 
and \textit{middle panels}) to $\approx 10$\% (\textit{bottom panel}) . 
The approximations of function $\mathcal{F}$ by means of 
Equation~(\ref{eq:F}), with parameters given in 
Table~\ref{tbl:ppar}, produce asymptotic cumulative torques 
that deviate from simulation data by typically less than $10$\%.

Parameters in Table~\ref{tbl:ppar} can be used to generate 
an approximation of function $\mathcal{F}$ for a given pair
$(\beta,\zeta)$. For example, one can proceed by selecting 
from the table four sets of parameters $p_{i}$ corresponding 
to $\beta_{1}$, $\beta_{2}$, $\zeta_{1}$, and $\zeta_{2}$, 
such that $\beta_{1}\le \beta < \beta_{2}$ and 
$\zeta_{1}\le \zeta < \zeta_{2}$. 
These parameters $p_{i}$ can then be interpolated at 
$(\beta,\zeta)$, and the interpolated values can be used 
in Equation~(\ref{eq:F}) to obtain function $\mathcal{F}$.
We use bi-linear interpolations to generate torque density 
distributions $d\mathcal{T}(r)/dM$ every
$\Delta\beta=\Delta\zeta=0.2$, which were then integrated 
according to Equation~(\ref{eq:dTdMdef}) to obtain the total 
torque as a function of the surface density and temperature 
gradients. By fitting these data, the coefficients of $\beta$ and 
$\zeta$ in Equation~(\ref{eq:Tdl}) can be recovered with good 
agreement, within a margin  of $10$\%.

\subsection{A Migration Test Case}
\label{sec:3Dtest}

%
\begin{figure}
\centering%
\resizebox{\linewidth}{!}{%
\includegraphics{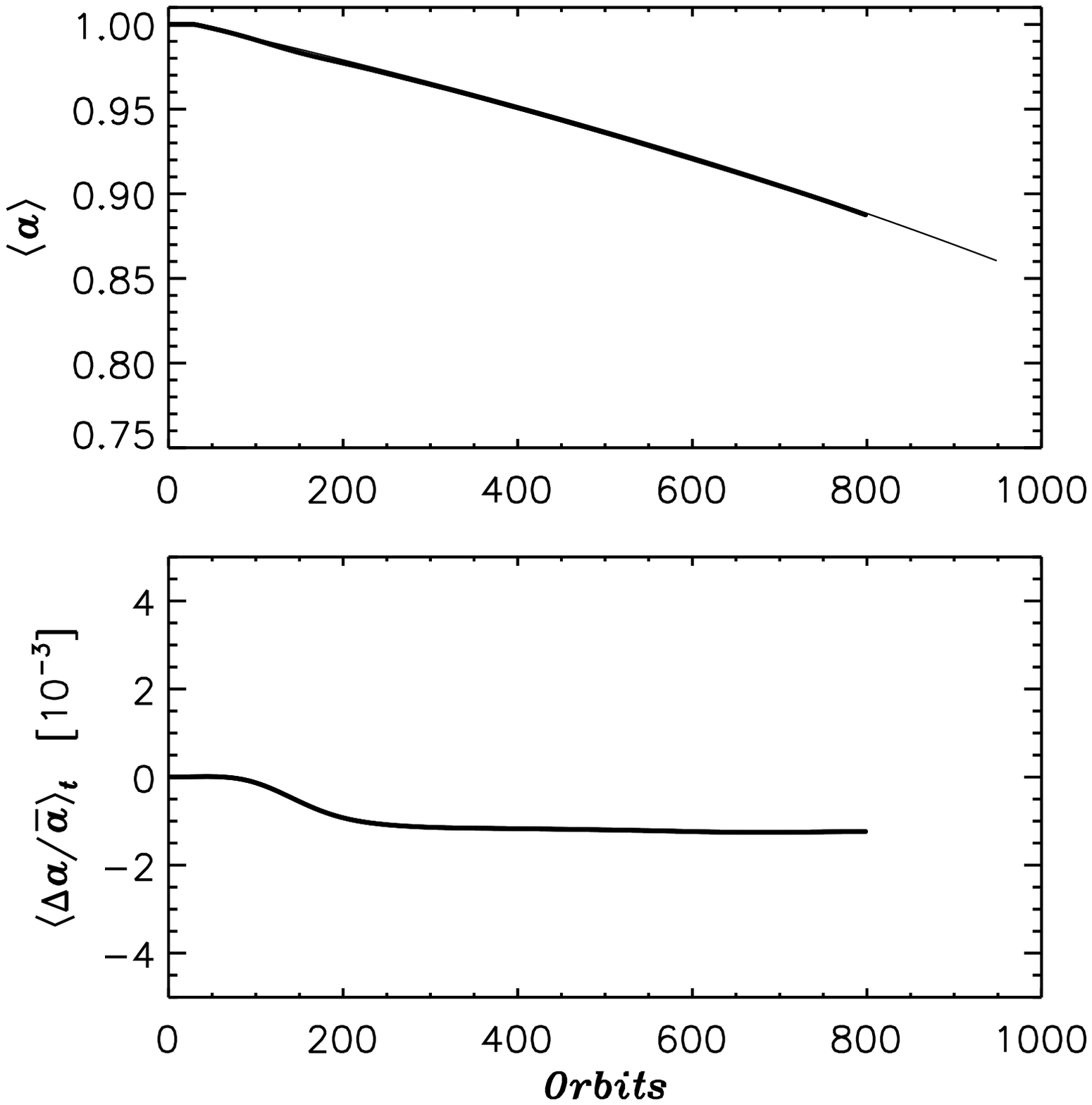}}
\caption{%
         \textit{Top panel}:
         a comparison between semi-major axis evolutions obtained
         from a three-dimensional model (\textit{thick line}, see the text for details)
         and from
         the integration of Equation~(\ref{eq:dadt})
         (\textit{thin line}). Function $\mathcal{F}$ uses parameters
         $p_{i}$ from Table~\ref{tbl:ppar}.
         \textit{Bottom panel}:
         running-time average of $\Delta a/\bar{a}$, defined
         as $(1/t)\int^{t}_{0}(\Delta a/\bar{a})dt'$, where
         $\Delta a$ is the difference and $\bar{a}$ is the mean
         value of the two curves in the top panel. Data are averaged
         every one (initial) orbital period.
         }
\label{fig:test3d}
\end{figure}
The formulae presented above in this section can be readily 
used to describe planet's migration, as indicated by the test case 
discussed here.
We set up a three-dimensional disk model with a planet whose mass is
$\Mp=10^{-5}\,\Ms$ (or $3\,\MEarth$ if $\Ms=1\,M_{\odot}$).
The disk is isothermal ($\zeta=0$) and has an initial surface density
corresponding to a parameter $\beta=2$. The slope of the azimuthally 
averaged $\Sigma(r)$ is maintained throughout the disk evolution 
by imposing a kinematic viscosity $\nu\propto r^{\beta}$, so that 
the disk is in an approximate steady state.

The planet is allowed to migrate under the action of disk torques
after $30$ orbital periods.
We adopt a treatment to approximately account for the 
axisymmetric effects of disk self-gravity by forcing the planet
to rotate at (approximately) the same rate as the disk
(see Appendix~C of \citetalias{gennaro2008}). The grid rotates
about its origin, $O$, at the same angular velocity as the planet
\citep[for details, see][]{gennaro2005}.
Therefore, the planet moves only radially through the grid
because of orbital migration.
Gravitational forces exerted on the planet are computed via
direct summation of elemental forces and the equations of motion
of the planet are integrated by means of a Bulirsch-Stoer
algorithm, which uses an adaptive timestep control to achieve
a relative accuracy of $10^{-5}$.

The evolution of the semi-major axis is illustrated in the top
panel of Figure~\ref{fig:test3d} as a thick solid line. 
We use coefficients $p_{i}$ in Table~\ref{tbl:ppar}, for the
pair $(\beta,\zeta)=(2,0)$, to obtain an approximation of function
$\mathcal{F}$ (Equation~(\ref{eq:F})) and then calculate the torque 
density distribution $d\mathcal{T}(r)/dM$ (Equation~(\ref{eq:dTdM})). 
The total torque $\mathcal{T}(a)$, as a function of the planet's 
semi-major axis $a$, is evaluated through the integral in 
Equation~(\ref{eq:dTdMdef}). Conservation of orbital angular 
momentum imposes that 
$da/dt=2\,\mathcal{T}(a)/(\Mp\,a\,\Omega_{\mathrm{K}})$, which
can be written more explicitly as
\begin{eqnarray}
 \frac{da}{dt}&=&4\pi\,\Omega_{\mathrm{K}}(a)\!%
 \left(\frac{a}{\Ms}\right)\!%
 \left(\frac{\Mp}{\Ms}\right)\!\left(\frac{a}{H}\right)^{4}\nonumber\\%
                    &  &\times\int^{r_{2}}_{r_{1}}{\!\!%
 \mathcal{F}\!\left(\frac{r-a}{H}, \beta, \zeta \right)%
                                 \!\Sigma(r)\, r\, dr}.
 \label{eq:dadt}
\end{eqnarray}
The integration limits in the above equation are such that
$r_{2}-a\gg H(a)$ and $a-r_{1}\gg H(a)$.

A numerical solution of Equation~(\ref{eq:dadt}) is shown as a 
thin solid line in the top panel of Figure~\ref{fig:test3d}. 
The average relative difference between the orbital evolution
obtained from the three-dimensional model and that simulated with the formulae
presented above is at the $10^{-3}$ level for $10$\% 
variations of the planet's semi-major axis (see 
Figure~\ref{fig:test3d}, \textit{bottom panel}).

\section{APPLICATIONS TO EVOLUTION TRACKS OF SUPER-EARTHS}
\label{sec:tracks}

Torque density distributions are often used to describe the
tidal interaction between a disk and embedded bodies.
In this section, we describe an application of the formulae
provided in \S~\ref{sec:dTdM} to protoplanetary disks.
Similar applications to other contexts involving astrophysical 
disks may employ an analogous strategy.

The synthesis of semi-major axis distributions of extrasolar 
planetary systems requires solving for the evolution of the
protoplanetary disk and, simultaneously, for the gravitational 
interaction occurring between the planet(s) and the disk. 
Since these calculations involve evolution timescales equal to 
typical lifetimes of protoplanetary disks (millions of years)
and length scales of hundreds of AU, they need to rely on one-dimensional 
models, as a consequence of the limitations of current 
computer capabilities.

The evolution of a thin, axisymmetric and viscous disk that 
interacts with one (or more) embedded planet(s) is described 
by the following equation \citep[e.g.,][]{lin1986b}
\begin{equation}
 \pi r\frac{\partial}{\partial t}(\Sigma+\Sigma_{\mathrm{pe}})=%
 \frac{\partial}{\partial r}\!\left\{%
 \frac{1}{r\,\Omega}\frac{\partial}{\partial r}\left[%
 \mathcal{T}_{\nu}(r)-\mathcal{T}(r)\right]\right\},
 \label{eq:devol0}
\end{equation}
where 
$\mathcal{T}_{\nu}(r)=-2\pi r^{3}\nu\Sigma\,d\Omega/dr%
=3\pi r^{2}\nu\Sigma\Omega$ \citep{lynden-bell1974} is the disk
viscous torque and $-\mathcal{T}(r)$ is the tidal torque exerted 
by the planet(s) at radius $r$. In Equation~(\ref{eq:devol0}), 
it is assumed that the disk's rotation rate is unperturbed, i.e.,
$\Omega=\Omega_{\mathrm{K}}$. 
Since one can write the second term in square brackets as
$\partial\mathcal{T}/\partial r=2\pi r\Sigma\,\partial\mathcal{T}/\partial M$,
Equation~(\ref{eq:devol0}) becomes
\begin{equation}
 r\frac{\partial}{\partial t}(\Sigma+\Sigma_{\mathrm{pe}})=%
 \frac{\partial}{\partial r}\!\left[%
 3\sqrt{r}\frac{\partial}{\partial r}\!\left(\nu\Sigma\sqrt{r}\right)%
-\frac{2\,\Sigma}{\Omega}\frac{\partial\mathcal{T}}{\partial M}(r)\right].
 \label{eq:devol1}
\end{equation}
In case of a multiple-planet system, the second term
on the right-hand side involves a
torque density distribution for each planet, and thus the summation
$\partial\mathcal{T}/\partial M=\sum_{n}{\partial\mathcal{T}_{n}/\partial M}$.

The left-hand side of Equations~(\ref{eq:devol0}) and (\ref{eq:devol1}) 
includes the term $\partial\Sigma_{\mathrm{pe}}/\partial t$, the mass 
loss flux from the disk. Here we assume a simple model of gas
removal due to photoevaporation by the central star that includes
effects of UV photons 
\begin{equation}
 \frac{\partial \Sigma_{\mathrm{pe}}}{\partial t}=
 3.7\times 10^{-13} \left(\frac{10\,\AU}{r}\right)^{5/2}\,%
 \left(\frac{M_{\odot}}{\AU^{2}\,\mathrm{yr}}\right),
 \label{eq:dSpedt}
\end{equation}
for $r> 10\,\AU$ and 
$\partial\Sigma_{\mathrm{pe}}/\partial t=0$ otherwise.
Equation~(\ref{eq:dSpedt}) applies to a disk surrounding a 
$1\,M_{\odot}$ central star that emits ionizing photons 
at a rate of $10^{41}\,\mathrm{s}^{-1}$
\citep[see, e.g.,][ and references therein]{martin2007}.

Equation~(\ref{eq:devol1}) is solved by means of an implicit scheme,
second-order accurate in time, 
over a grid that extends from $0.01\,\AU$ to $1850\,\AU$. The grid
is composed of $10000$ elements and uses a  constant logarithmic 
spacing. Torque density distributions are obtained from 
Equations~(\ref{eq:dTdM}) and (\ref{eq:F}), at each timestep,
according to the interpolation procedure outlined in the last paragraph 
of \S~\ref{sec:AATDD}. The parameter $\beta$ required to construct
function $\mathcal{F}$ is computed, every timestep, as an
average slope
\begin{equation}
 \bar{\beta}=-\frac{1}{(r_{2}-r_{1})}\int_{r_{1}}^{r_{2}}%
             \!\frac{r}{\Sigma}\frac{d\Sigma}{dr}\,dr,
 \label{eq:betapp}
\end{equation}
in which the integration limits are $r_{2,1}=a\pm 3H(a)$.
The orbital radius of the planet is computed by numerically solving
Equation~(\ref{eq:dadt}). 

The examples presented here are based on two disk configurations,
both intended to reproduce the initial mass distribution of gas in a
Minimum-Mass Solar Nebula. 
In the first,
the initial disk surface density is taken from \citet{davis2005}.
At $1\,\AU$, $\Sigma\approx 1100\,\mathrm{g\,cm}^{-2}$ at time
$t=0$.
Within $4\,\AU$, $\Sigma\propto 1/\sqrt{r}$, followed by an 
exponential decay at larger radii.
The disk contains a total mass of $\sim 0.02\,M_{\odot}$ inside 
of about $40\,\AU$. We assume that $H/r$ is constant (i.e., the
disk temperature $T\propto 1/r$) 
and that the kinematic viscosity is $\nu=\alpha\,H^2\Omega$, 
where $\alpha$ is also a constant (hence $\nu\propto \sqrt{r}$).  
In the second configuration, 
$\Sigma=1700\,\mathrm{g\,cm}^{-2} (1\,\AU/r)^{3/2}$ at time
$t=0$ \citep{hayashi1981}. An exponential decay is applied
beyond about $40\,\AU$ so that the disk mass within 
$\sim 70\,\AU$ is $\sim 0.02\,M_{\odot}$. 
The aspect-ratio of the disk 
is $H/r\propto r^{1/4}$ (and thus $T\propto 1/\sqrt{r}$). 
In this case, we assume that $\alpha\propto\sqrt{r}$, hence the
kinematic viscosity $\nu\propto r^{3/2}$. 

\begin{figure*}
\centering%
\resizebox{\linewidth}{!}{%
\includegraphics{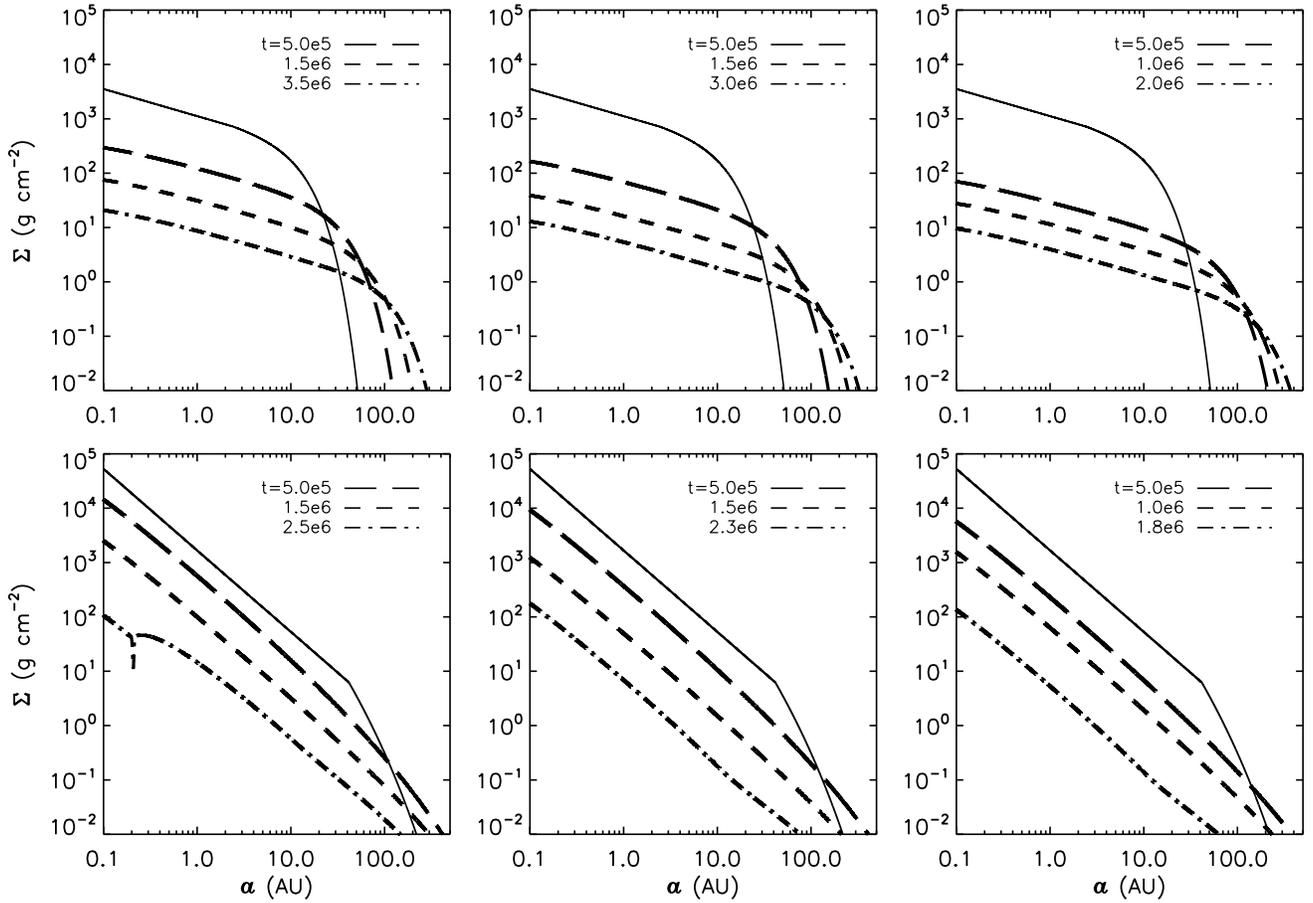}}
\caption{%
         Surface density at various times in years (see legends) of a
         protoplanetary disk that interacts with a low-mass
         planet ($\Mp=5\,\MEarth$ for $t\gtrsim 10^{6}$ years).
         The top panels refer to a configuration based on the 
         initial $\Sigma$ (\textit{solid lines}) from 
         \citet{davis2005}, constant $\alpha$, and temperature
         $T\propto 1/r$ (see the text for details). The bottom panels
         refer to a configuration based on the initial surface
         density from \citet{hayashi1981}, $\alpha\propto\sqrt{r}$,
         and $T\propto 1/\sqrt{r}$ . From left to right, the turbulence 
         parameter 
         $\alpha$ at $1\,\AU$ is $0.003$, $0.005$, and $0.01$ for
         models in the top panels, and $0.0035$, $0.005$, and 
         $0.007$ for models in the bottom panels.
         }
\label{fig:devol}
\end{figure*}
Figure~\ref{fig:devol} shows the surface density at various
times (see figure's caption for details). The top panels refer
to the first disk's configuration, whereas the bottom panels
refer to the second configuration. Each panel corresponds to
a given value of $\alpha$ at $1\,\AU$ (but recall that $\alpha$ 
is constant in the first configuration and only varies with 
radius in the second), 
which produces initial accretion rates on the star between
$10^{-8}\,M_{\odot}\,\mathrm{yr}^{-1}$
and $10^{-7}\,M_{\odot}\,\mathrm{yr}^{-1}$.

The bottom-left panel of Figure~\ref{fig:devol} indicates that
a density gap starts to form along the planet's orbit at late  
evolution times. 
This is due to the small viscosity and disk thickness at small 
radii. A condition for gap formation is that the symmetrized 
one-sided torque, i.e., the torque exerted by the planet on disk's 
gas interior and exterior of its orbit, exceeds the viscous torque
\citep[e.g.,][]{lin1986b}. 
The one-sided torque is of order $(a/H)|\mathcal{T}|$. This 
can be found by using Equation~(\ref{eq:dTdM}) and 
approximating the integral
$\int^{\infty}_{a}\!|\mathcal{F}_{S}(x)|\Sigma(r) r dr$ 
in Equation~(\ref{eq:dTdMdef}) as
$a H \Sigma(a)\!\int^{\infty}_{0}\!|\mathcal{F}_{S}(x)| dx$,
where $x=(r-a)/H(a)$ and 
$\mathcal{F}_{S}(x)=\left[\mathcal{F}(x)-\mathcal{F}(-x)\right]/2$
is a symmetrized form of function $\mathcal{F}(x)$.
Therefore, the condition for gap formation
becomes 
\begin{equation}
\xi\,\Sigma(a)\,\Omega^{2}(a) a^{4}\!%
\left(\frac{\Mp}{\Ms}\right)^{2}\left(\frac{a}{H}\right)^{3}%
\gtrsim\mathcal{T}_{\nu}(a), 
 \label{eq:gapcon0}
\end{equation}
with 
$\xi=2\pi\!\!\int^{\infty}_{0}\!|\mathcal{F}_{S}(x)| dx$.
Recalling the definition of viscous torque given above, one obtains 
the condition
\begin{equation}
\left(\frac{\Mp}{\Ms}\right)^{2}\gtrsim \frac{3\pi\alpha}{\xi}%
       \left(\frac{H}{a}\right)^{5},
 \label{eq:gapcon}
\end{equation}
where the factor $\xi$ is typically of order unity. 
In the second disk configuration, this 
condition is satisfied in disk regions interior of a few tenths 
of \AU, but it is never fulfilled in the first disk configuration.

\begin{figure*}
\centering%
\resizebox{\linewidth}{!}{%
\includegraphics{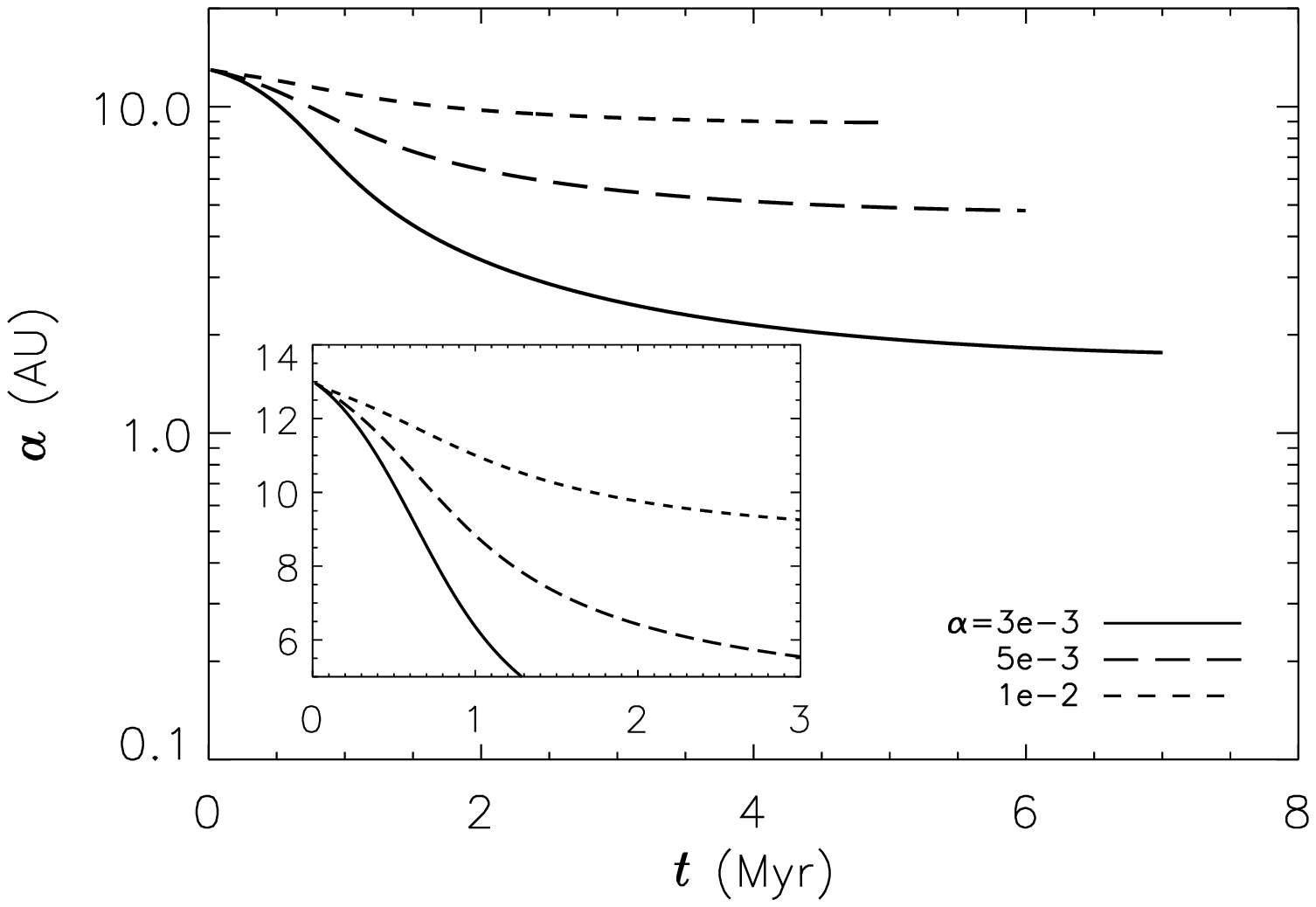}%
\includegraphics{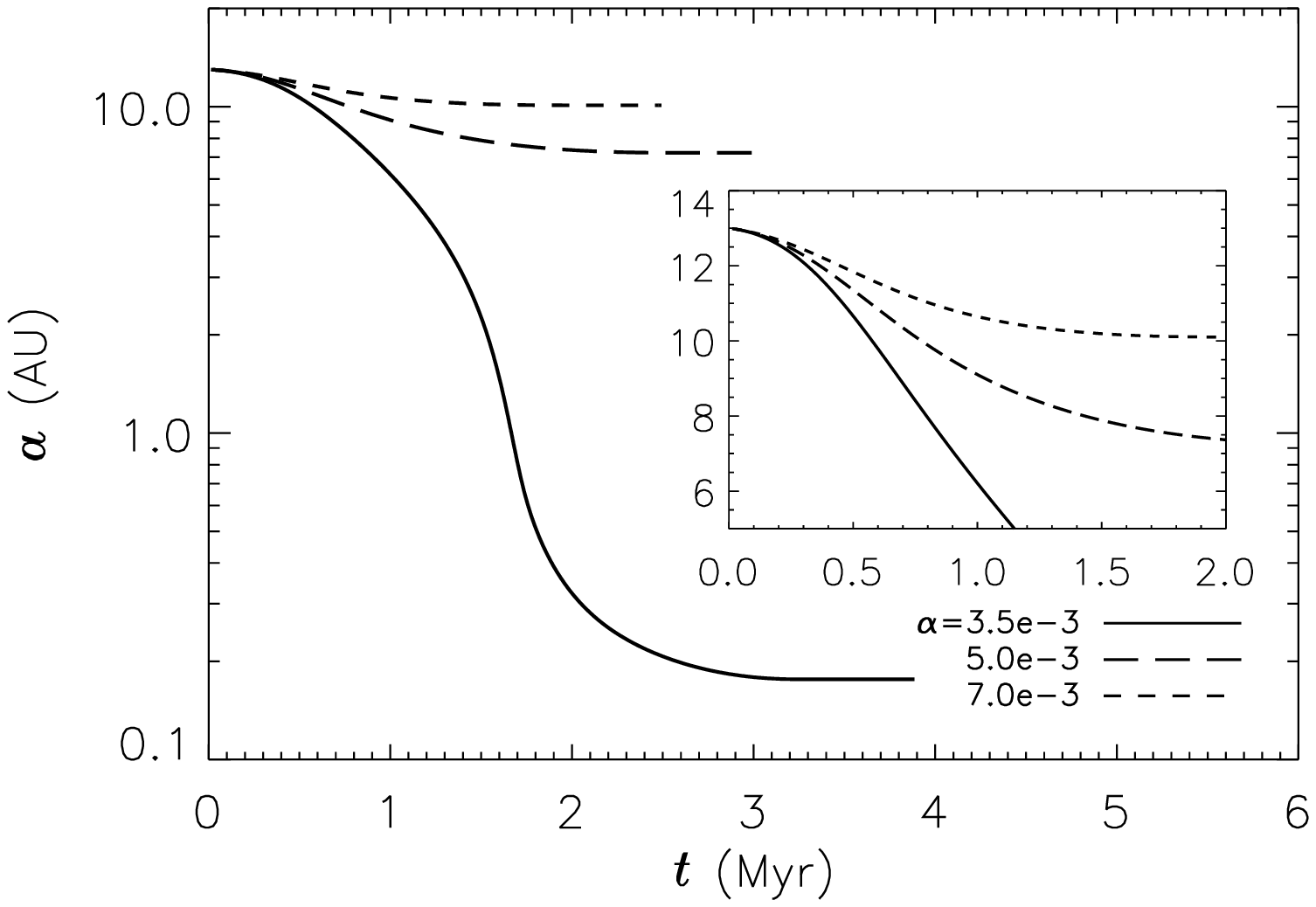}}
\caption{%
         Evolution of the semi-major axis of a planet that grows
         at an oligarchic rate ($d\Mp/dt\propto \Mp^{2/3}$), from 
         $0.1\,\MEarth$ to $5\,\MEarth$ in $\sim10^{6}$ years, and 
         interacts with a viscously-evolving and photoevaporating
         protoplanetary disk. The parameter characterizing disk
         viscosity at $1\,\AU$ is given in the bottom-right corner 
         of each panel. The left and right panels correspond, 
         respectively, to the first and second disk configuration 
         (see the text for details). The insets show details of the
         semi-major axis evolution over the first $3\times 10^{6}$
         (\textit{left}) and $2\times 10^{6}$ (\textit{right}) years.
         }
\label{fig:tracks}
\end{figure*}
The planet evolution tracks illustrated in Figure~\ref{fig:tracks}
assume that a planetary embryo grows at an oligarchic rate,
$d\Mp/dt\propto \Mp^{2/3}$, from $0.1\,\MEarth$ to $5\,\MEarth$
over a $\sim 10^{6}$ year period.
All tracks start at $13\,\AU$ and are computed until disk gas 
within tens of $\AU$ has almost entirely dissipated and 
$da/dt\approx 0$. 
The left and right panels of the figure refer, respectively, to 
the first and second disk configuration.

\begin{figure*}
\centering%
\resizebox{\linewidth}{!}{%
\includegraphics{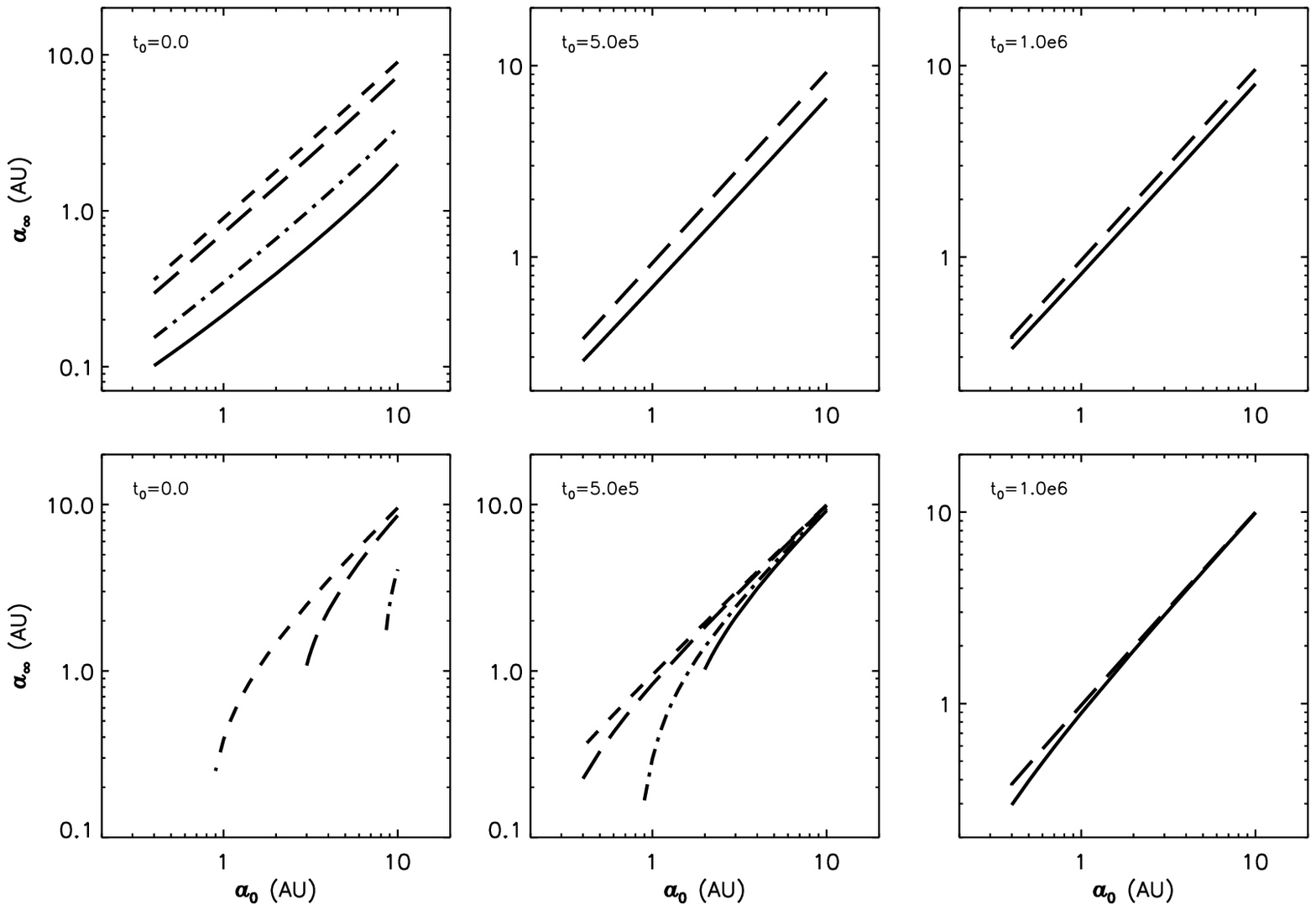}}
\caption{%
         Asymptotic orbital radius, $a_{\infty}$, as a function of the
         orbital radius at time $t_{0}$ (see the text), 
         for planets whose mass is
         $M_{p}=0.3\,\MEarth$ (\textit{short-dashed line}), $1\,\MEarth$
         (\textit{long-dashed line}), $3\,\MEarth$ (\textit{dot-dashed line}),
         and $5\,\MEarth$ (\textit{solid line}). Times $t_{0}$ are given 
         in the legends in units of years. In the top panels, the planet
         interacts with a protoplanetary disk whose initial configuration 
         is based on the surface density from \citet{davis2005}, 
         a constant turbulence parameter $\alpha=0.01$, and a 
         temperature $T\propto 1/r$. In the bottom panels, the initial
         disk configuration is based on the surface density from 
         \citet{hayashi1981}, $\alpha=0.01\sqrt{r/1\,\AU}$,
         and $T\propto 1/\sqrt{r}$. Notice that some panels only report
         two mass cases to improve legibility.
         }
\label{fig:a0ai}
\end{figure*}

Observations show that super-Earths, i.e., planets of several Earth 
masses, are also found within a few tenths of an $\AU$ from their 
stars, a likely result of orbital migration. This occurrence is 
predicted in the case of the second disk configuration, as 
indicated by the solid-line track in the right panel of 
Figure~\ref{fig:tracks}.
The first disk configuration would also predict a similar outcome
for a somewhat smaller turbulence parameter 
($\alpha\approx 0.001$) or for different initial conditions of
planet's mass and orbital radius (see Figure~\ref{fig:a0ai}).

The disk configurations introduced above are also used to produce the
results illustrated in Figure~\ref{fig:a0ai}. In both cases, a turbulent viscosity 
parameter $\alpha=0.01$, at $1\,\AU$, is applied. The disk is allowed
to evolve without any planet until time $t_{0}$. At time $t_{0}$, we assume 
that a planet of mass $\Mp$ is located at the orbital radius $a_{0}$. 
The disk-planet system is then allowed to evolve until negligible amounts 
of gas remain in the disk and $da/dt$ drops virtually to zero. The orbital radius 
attained by the planet at the end of the calculation is indicated as $a_{\infty}$, 
its asymptotic value. 
Figure~\ref{fig:a0ai} shows $a_{\infty}$ as a function of $a_{0}$ for cases 
in which $t_{0}=0$ 
(\textit{left panels}), $5\times 10^{5}$ years (\textit{middle panels}), and 
$10^{6}$ years (\textit{right panels}). Top and bottom panels refer,
respectively, to the first and second disk's initial configuration.
In each panel, different lines styles correspond to models with different 
planet masses (which is taken to be constant): 
$M_{p}=0.3\,\MEarth$ (\textit{short-dashed line}), 
$1\,\MEarth$ (\textit{long-dashed line}), $3\,\MEarth$ 
(\textit{dot-dashed line}), and $5\,\MEarth$ (\textit{solid line}). 
Clearly, initial conditions play a key role in the determination of 
semi-major axis distributions of planet populations.

\section{EFFECTS OF GRADIENT VARIATIONS ON TORQUE DENSITY DISTRIBUTIONS}
\label{sec:dTdMvars}

As discussed in \S~1, a number of studies have looked 
into the possibility of stopping migration by trapping the planet at disk 
locations where the radial gradients of surface density and/or temperature 
may be strong and vary over short radial distances, as it may occur 
due to rapid changes in disk opacity or turbulent viscosity.

The results developed in \S~\ref{sec:dTdM} were derived under 
the condition that surface density and temperature slopes, $\beta$ 
and $\zeta$, are nearly constant over a radial region of about $3 H(a)$ 
on either side of the planet's orbit, where most of the torque is produced.
Here, we wish to show that the same formulae are applicable when 
disk radial gradients vary somewhat over this region. This is because
torque density distributions respond smoothly to mildly varying gradients
and can be approximated in terms of mean slopes.

\begin{figure*}
\centering%
\resizebox{\linewidth}{!}{%
\includegraphics{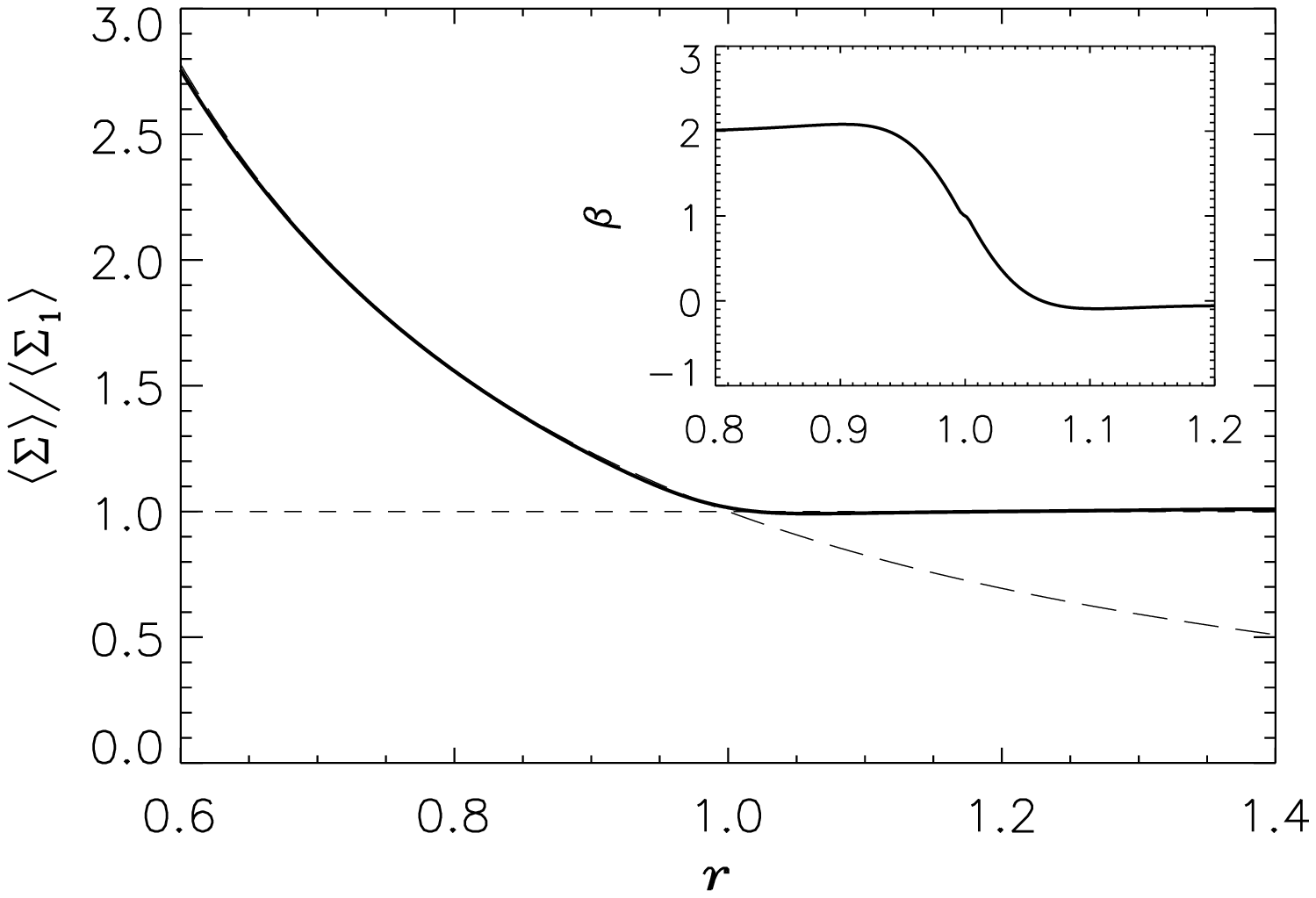}%
\includegraphics{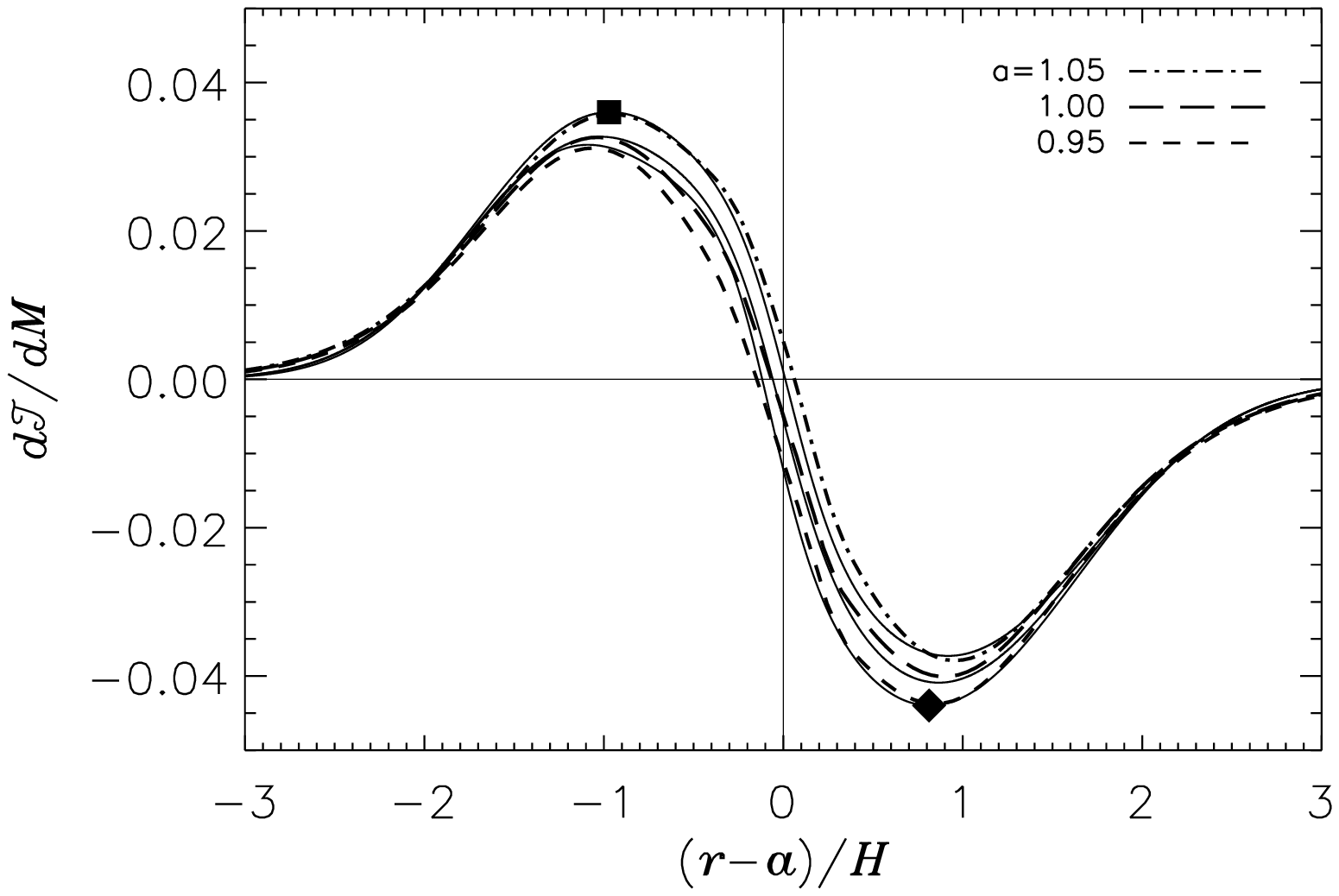}}
\caption{%
         Surface density across a disk transition region (\textit{left}) obtained 
         from a three-dimensional calculation. The long-dashed line is the function $1/r^{2}$.
         The transition is produced by choosing an appropriate kinematic
         viscosity $\nu(r)$. The inset shows details of the transition in 
         terms of the slope $\beta$.
         The distributions $d\mathcal{T}(r)/dM$ (\textit{right}) are
         scaled by $\Omega^{2}\,a^{2}\,(\Mp/\Ms)^{2}\,(a/H)^{4}$. 
         The thin solid curves are representations of function $\mathcal{F}$
         (Equation~(\ref{eq:F})) with parameters $p_{i}$ (Table~\ref{tbl:ppar}) 
         interpolated at $\beta\approx 0.2$ (profile marked by a filled square) 
         and $\approx 1.8$ (profile marked by a filled diamond). 
         The third thin solid curve has a slope parameter $\beta=1$.
         All slopes are mean values computed with Equation~(\ref{eq:betapp}).
         }
\label{fig:bzvar}
\end{figure*}
We set up a three-dimensional model with a surface density $\Sigma$ whose azimuthal 
average is illustrated in the left panel of Figure~\ref{fig:bzvar}. The profile 
is such that $\Sigma\propto 1/r^{2}$ at $r\ll 1$ and constant at $r\gg 1$, 
as also indicated by the long-dashed and short-dashed lines, respectively.
This setup mimics the conditions of the opacity jumps in 
\citetalias{menou2004} for a locally isothermal disk. Details of the slope 
$\beta$ as a function of radius, in the transition region, are shown in the inset.
This profile is maintained stable in time by imposing a kinematic viscosity 
$\nu\propto 1/\Sigma$. The temperature is constant throughout. 
In the right panel, rescaled torque density distributions $d\mathcal{T}(r)/dM$
are plotted when the planet ($\Mp=10^{-5}\,\Ms$) is located at three different 
radial positions, indicated in the legend.

The torque density distributions in Figure~\ref{fig:bzvar} exhibit a behavior 
reminiscent of that shown in the top-left panel of Figure~\ref{fig:Tvsbz} 
where different, but radially constant, slopes $\beta$ are imposed.
This behavior is also seen by comparing the dot-dashed and 
short-dashed curves in the right panel with the thin solid curves 
marked by a filled square and diamond, respectively. 
These are torque densities per unit disk mass obtained from 
Equation~(\ref{eq:F}) and parameters $p_{i}$ (from Table~\ref{tbl:ppar}) 
interpolated at slopes $\bar{\beta}\approx 0.2$ (\textit{filled square}) and 
$\bar{\beta}\approx 1.8$ (\textit{filled diamond}). Both slopes are obtained 
as mean values according to Equation~(\ref{eq:betapp}). The torque density 
distribution produced when the planet is located at radius $r=1$ 
(\textit{long-dashed line}) has an intermediate behavior and can be 
approximated by using a mean slope $\bar{\beta}=1$, again resulting
from Equation~(\ref{eq:betapp}).

Therefore, if a disk has radial gradients that are mildly varying over 
a region of a few times $H$, one can still apply the expressions derived 
in \S~\ref{sec:dTdM} after having introduced appropriate mean slopes.
One important aspect to appreciate is the small variation of the 
torque density distribution $d\mathcal{T}(r)/dM$ as the density 
gradient changes. Even though $\beta$ varies by factors of order unity,
the peaks of $d\mathcal{T}(r)/dM$ are only affected at $\sim 10$\% 
levels (see also Figure~\ref{fig:Tvsbz}, \textit{top-left panel}).
The results do not show any indication of a migration trapping.

\begin{figure}
\centering%
\resizebox{\linewidth}{!}{%
\includegraphics{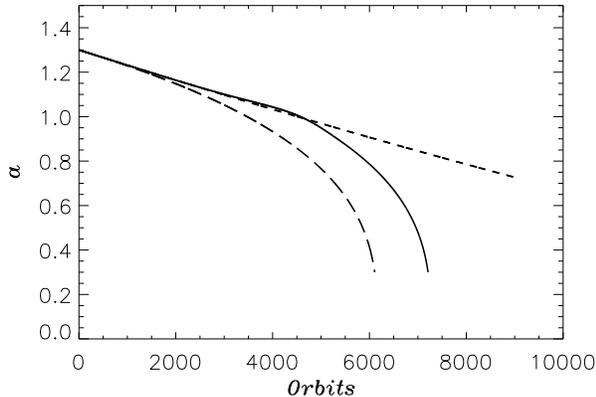}}
\caption{%
              Migration track of an $\Mp=10^{-5}\,\Ms$ planet 
              (\textit{solid line}) that moves across the surface density 
              transition region illustrated in Figure~\ref{fig:bzvar}.
              The track was obtained by numerically integrating
              Equation~(\ref{eq:dadt}). The surface density is that
              displayed in Figure~\ref{fig:bzvar} and the mean slope 
              $\bar{\beta}$ (to compute function $\mathcal{F}$) is 
              given by Equation~(\ref{eq:betapp}).
              Also shown are migration tracks through an isothermal 
              disk with constant surface density (\textit{short dashed line})
              and with $\Sigma\propto 1/r^{2}$ (\textit{long dashed line}).
         }
\label{fig:migtran}
\end{figure}
Figure~\ref{fig:migtran} shows, as a solid line, the migration track 
of an $\Mp=10^{-5}\,\Ms$ planet across the transition region of 
Figure~\ref{fig:bzvar}. The migration track is obtained by integrating 
Equation~(\ref{eq:dadt}) ($\Sigma$ is that illustrated in Figure~\ref{fig:bzvar}) 
and using Equation~(\ref{eq:betapp}) to evaluate the mean slope 
$\bar{\beta}$ that characterizes function $\mathcal{F}$. 
For comparison purposes, the figure also shows the tracks that 
the same planet would follow if $\Sigma$ was constant 
(\textit{short dashed line}) and $\Sigma\propto 1/r^{2}$ 
(\textit{long dashed line}). 
The solid and short dashed lines are similar until the
density gradient $\bar{\beta}$ becomes greater than zero.
Incidentally, the slope of the long dashed line is similar to the 
other two initially since the reduced density value compensates 
for the larger density gradient (see Equation~(\ref{eq:Tdl})).

The rapid decay at small radii of the planet's semi-major axis 
seen in the solid and long dashed curves of Figure~\ref{fig:migtran} 
is what one would expect from, e.g., Equation~(\ref{eq:Tdl}) applied
to a stationary disk: $|\mathcal{T}(a)|\propto a^{(\zeta-\beta)}$, hence
\begin{equation}
 \left|\frac{da}{dt}\right|\propto a^{(\zeta-\beta+1/2)}.
 \label{eq:adotvsa}
\end{equation}
The solid curve is therefore expected to steepen ($|da/dt|$ diverges) 
as the semi-major axis decreases, whereas the short dashed curve is 
expected to flatten ($|da/dt|$ tends to zero) as $a$ decreases.

Density transitions occurring at the edges of a ``dead zone'' may be 
quite abrupt and the slope $\beta$ may change by large amounts over
radial distances of order $H$ \citep[e.g.,][]{matsumura2007}. 
The analysis provided here may not apply to those sharp transitions,
whose effects on torques would require a more detailed analysis.

\section{TORQUE DENSITY DISTRIBUTIONS FOR GAP-OPENING PLANETS}
\label{sec:dTdMgap}

The torque density distributions given in \S~\ref{sec:dTdM}
were derived in a regime in which the mass density distribution
along the planet's orbit remains basically unperturbed. 
In general, this condition ceases to hold  when tidal torques
exceed viscous torques (see Equation~(\ref{eq:gapcon})), 
which is roughly expressed by the inequality (see \S~\ref{sec:tracks})
\begin{equation}
 \left(\frac{\xi}{3\pi\alpha}\right)^{1/2}\left(\frac{\Mp}{\Ms}\right)%
 \left(\frac{a}{H}\right)^{5/2}\gtrsim 1.
 \label{eq:gcon}
\end{equation}

\begin{figure}
\centering%
\resizebox{\linewidth}{!}{%
\includegraphics{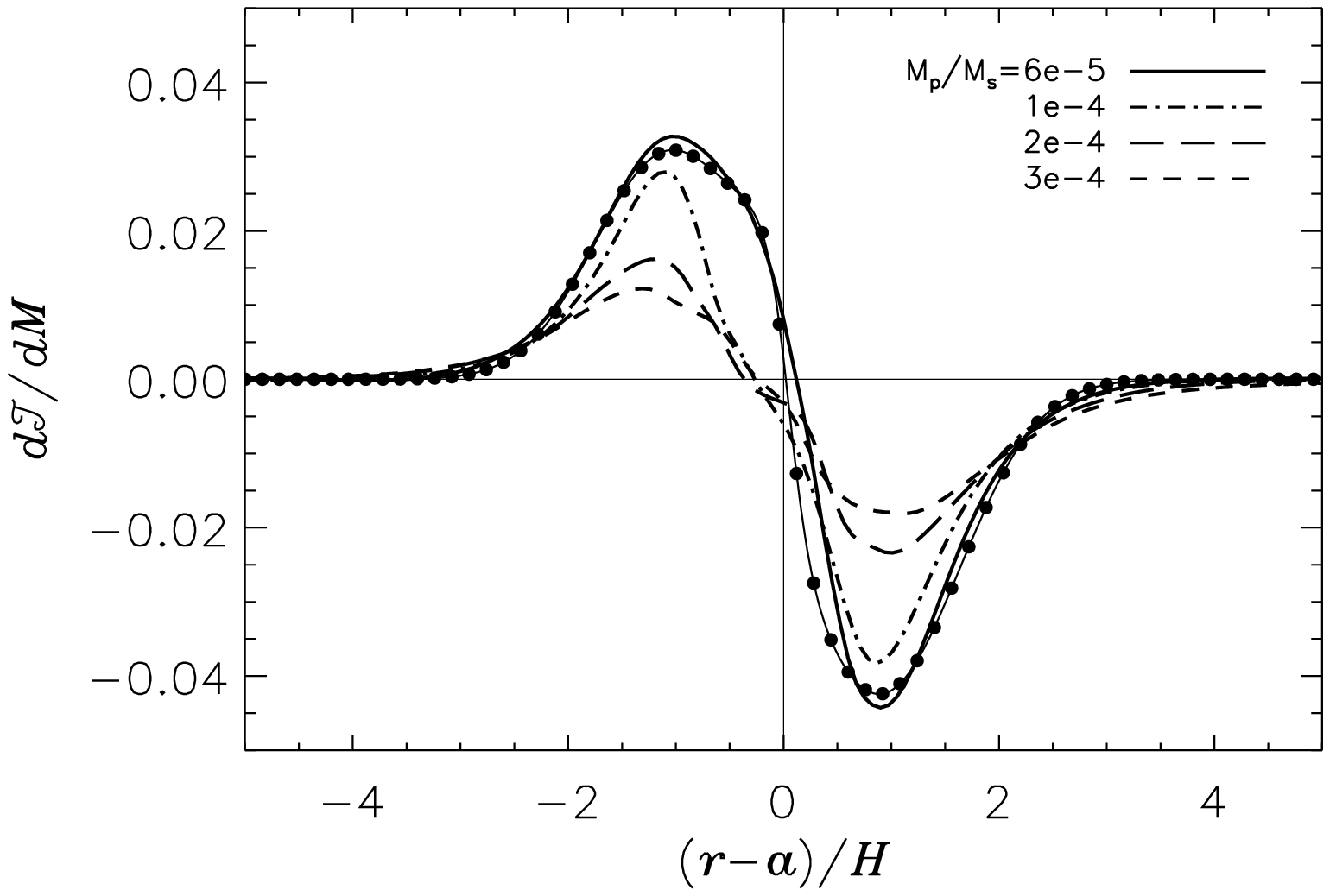}}
\resizebox{\linewidth}{!}{%
\includegraphics{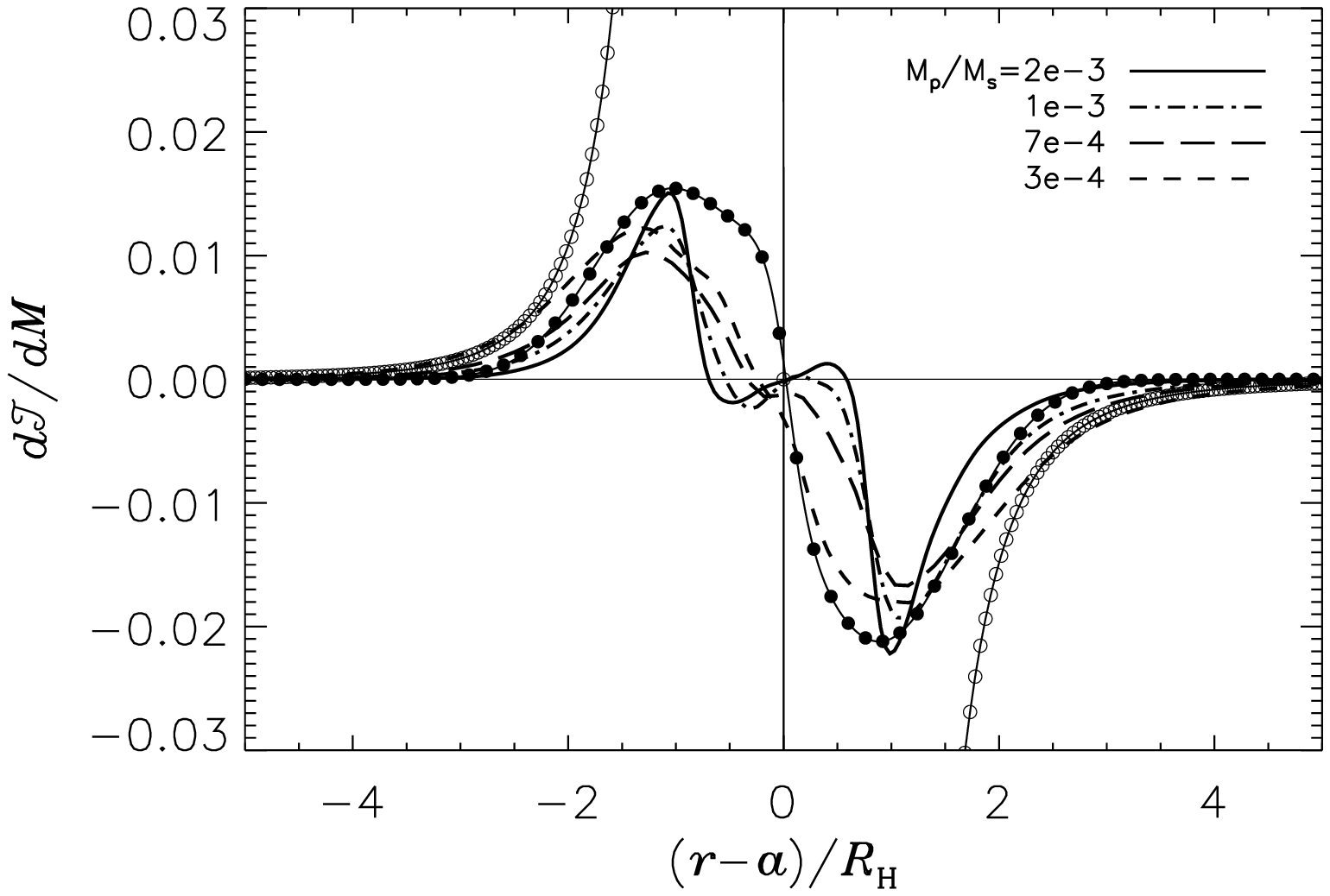}}
\caption{%
         Torque density distributions in a locally isothermal disk
         for various values of the planet-to-star mass ratio $\Mp/\Ms$, 
         indicated in the legends (see the text for further details).
         The disk has parameters $H/a= 0.05$ 
         and $\alpha=0.004$. In both panels, the torque is 
         scaled by $\Omega^{2}\,a^{2}\,(\Mp/\Ms)^{2}\,(a/H)^{4}$.
         The distributions $d\mathcal{T}(r)/dM$ are shown for cases 
         where $\Rhill<H$ (\textit{top}) and $\Rhill>H$ (\textit{bottom}).
         The short-dashed curve in the two panels refers to the same 
         model. See the text for further details.
         }
\label{fig:dtdm_gap}
\end{figure}
For standard disk parameters $H/a\approx 0.05$, $\alpha$ of a few 
times $10^{-3}$, and $\sqrt{\xi}$ of order unity, Equation~(\ref{eq:gcon}) 
requires a planet-to-star mass ratio $\Mp/\Ms\gtrsim 10^{-4}$. 
Gap formation can induce average variations in the surface density, 
relative to the unperturbed state, of $2$ orders of magnitudes when 
the planet mass grows from one Earth mass to one Jupiter mass.
However, the corresponding variation in the strength of the peaks of  
the rescaled $d\mathcal{T}(r)/dM$ is only a factor between $2$ and $3$, 
as shown in Figure~\ref{fig:dtdm_gap}. This figure displays torque density 
distributions produced by planets of various masses, for cases in which 
$\Rhill<H$ (\textit{top panel}) and $\Rhill>H$ (\textit{bottom panel}). 
Notice that the radial positions of the peaks are located at 
$r-a\approx \pm \Rhill$, when $\Rhill>H$ \citepalias[see also][]{gennaro2008}. 

The thin solid curve with filled circles in the top panel of 
Figure~\ref{fig:dtdm_gap} is function $\mathcal{F}$ (Equation~(\ref{eq:F})) 
of variable $x=(r-a)/H(a)$ and parameters $p_{i}$ given in 
Table~\ref{tbl:ppar}, for disk gradients corresponding to $\beta=0.5$ and 
$\zeta=1$. 
The same function, but divided by $2$ and of variable $x=(r-a)/\Rhill$, is 
plotted in the bottom panel. 
The peak strength of the rescaled $d\mathcal{T}(r)/dM$ shows only 
small changes for $\Rhill>H$.
(For $\Rhill \gg H$, the torque density scalings can be different from
those in Equation~(\ref{eq:dTdM})).
The factor $2$ is given by the left-hand side of Equation~(\ref{eq:gcon})
for $\Rhill\approx H$ and a more accurate determination of $\sqrt{\xi}$
(see definition in \S~\ref{sec:tracks}).

For a gap-opening planet, the details of the torque density distribution 
well within the gap region ($|r-a|\lesssim \Rhill$) are not very important 
for computing torques in the framework of one-dimensional disk models  (of the type 
discussed in \S~\ref{sec:tracks}) since those portions of 
$d\mathcal{T}(r)/dM$ are multiplied by a surface density that is relatively 
small (because of mass depletion). For reference purposes, in the bottom 
panel of Figure~\ref{fig:dtdm_gap} we also plot as a thin solid curve with 
empty circles the torque density per unit disk mass that results from the 
impulse approximation, as implemented by \citet{veras2004}.

\section{SUMMARY AND DISCUSSION}
\label{sec:summary}

We obtained an expression, Equation~(\ref{eq:Tdl}), for the disk-planet 
torque in three dimensions for a locally isothermal disk that contains radial density and 
temperature variations, based on numerical simulations. This expression was
derived for density and temperature gradients that are fairly constant over 
the torque producing region of the disk, which extends roughly over a radial 
distance of $3 H$ inside and outside the orbit of the planet.
(see, e.g., Figures~\ref{fig:dtdm_h}, \ref{fig:dtdm_q}, and \ref{fig:dtdm_alpha}).
In the case of a fully isothermal disk, the torque expression agrees well with
previous analytic, linear results obtained by \citetalias{tanaka2002}
(Equation~(\ref{eq:Tt})).
The analysis presented by \citetalias{tanaka2002} cannot be used to deduce 
the dependence of the torque on the temperature gradient (see discussion in 
\S~\ref{sec:3Disot}), which we provided in Equation~(\ref{eq:Tdl}).
We tested the torque expression for a range of disk densities, temperatures,
viscosities, and planet masses and found good agreement. The results agree 
well with the expected scalings of linear theory with planet mass and gas 
sound speed (see Figures~\ref{fig:dtdm_h} and \ref{fig:dtdm_q} for which 
linear theory predicts the curves overlap in each figure). Although these 
torque scalings agree with linear theory, we cannot claim that 
Equation~(\ref{eq:Tdl}) with a nonzero temperature gradient agrees with 
linear theory because, as noted above, there is no expression available
for the linear torque in a three-dimensional locally isothermal disk. These results suggest 
that substantially slowing or halting planet migration by means of smoothly 
varying but large negative density gradients and small temperature gradients 
in a locally isothermal disk is difficult. In addition, considerably reducing or 
stopping migration in a locally isothermal disk by means of opacity variations 
(e.g., \citetalias{menou2004}) or shadowing effects 
\citep[e.g.,][]{jang-condell2005} appears unlikely (see \S~\ref{sec:dTdMvars}).

We also obtained an expression for the torque density per unit disk mass 
(Equations~(\ref{eq:dTdM}) and (\ref{eq:F})). We tested the range of 
applicability of this expression for a range of disk temperatures, planet masses, 
and viscosities and found good agreement with the expected scaling predicted 
by linear theory. This expression can be applied to one-dimensional simulations of disk-planet 
interactions, such as, but not limited to, the cases of super-Earth evolution 
discussed in \S~\ref{sec:tracks}. The torque density expression commonly 
used is based on the impulse approximation and contains a long power-law 
tail that extends far from the planet. Ad hoc modifications have been made 
to make it more physically reasonable \citep[e.g.,][]{veras2004}. 
The expression we obtained is more accurate and overcomes that difficulty.
The formalism presented in \S~\ref{sec:dTdM} can also be employed for 
computing torques in the presence of mildly varying disk gradients  (see 
Figure~\ref{fig:bzvar}) and of gap-opening planets (see Figure~\ref{fig:dtdm_gap}).
Moreover, applications to other contexts involving tidal interactions between
astrophysical disks and embedded objects may also be possible.

The Lindblad-only torque formula of \citetalias{menou2004}, Equation~(\ref{eq:Tmg}), 
based on the torque density of \citet{ward1997} modified for a planet potential 
smoothed over a distance $H$, results in a positive torque (outward migration) 
for a steep enough and radially decreasing  surface density. We notice that, in the fully
isothermal case, there is disagreement between the two-dimensional (no softened potential) 
formula of \citetalias{tanaka2002} (for Lindblad torques only) and the results of 
\citet{ward1997}.
At the end of \S~\ref{sec:3Dvs2Dsoft}, we discussed some possible reasons for the
discrepancy. But, we cannot be certain at this point what the reason is.
In addition, we found that corotation torques play a significant role and, therefore,
the use of a Lindblad-only torque formula may not be appropriate in the
regimes that we investigated.

For zero radial density and temperature gradients in the disk, the torques 
resulting from nonlinear calculations of a two-dimensional disk, with a planet potential 
smoothed over a distance $H$, are in fair agreement with the three-dimensional results. 
However, there is substantial quantitative disagreement in the effects of these 
gradients on the torques (compare Equations~(\ref{eq:Tdl}) and (\ref{eq:T2d})). 
One possible source of this disagreement may be due to the effects of the 
softened two-dimensional potential in modifying the torques in the presence of disk 
gradients. 

The analysis presented here is limited by the range of the parameter space 
we considered, involving a turbulent viscosity $5\times 10^{-4} \la \alpha \la 0.05$.
In addition, it is limited by the assumption that the disk thermodynamics behaves 
in a locally isothermal manner. The results may not hold under conditions 
very different from those we assumed. For example, if the gas behaves 
somewhat adiabatically, the torque can be modified and may even result 
in outward migration \citep[e.g.,][]{baruteau2008,paardekooper2008,kley2009}.
Under such conditions, nonlinear effects may be more important than 
they are in the locally isothermal case.

In weakly turbulent disk regions ($\alpha \la 10^{-4}$), such as in ``dead zones'',
we expect the disk response to a planet to be dominated by the highly nonlinear 
effects of waves launched at Lindblad resonances resulting in a complex behavior, 
including the formation of vortices  \citep{li2009,yu2010}. Such effects produce 
strong changes in the underlying gas distribution near a low-mass ($\sim 5\,\MEarth$) 
planet that in turn affect its migration rate. Type~I planet migration can cease 
and a vortex dominated, jumpy form of migration can prevail. The linear torque 
formulae presented above are inapplicable in such a regime. 
For $\alpha \ga 5\times 10^{-4}$, the viscosity regime we considered, the nonlinear 
feedback effects of shocks on migration subside due to turbulent diffusion
that washes out these sharp features in the gas density \citep{li2009}. 
However, these simulations were performed in two dimensions and it would be useful 
to understand how three-dimensional effects could alter this form of migration. In particular, 
the formation of vortices at gap edges may operate differently in three dimensions.

In highly turbulent disk regions ($\alpha \ga 0.1$), there may be other types of
changes in the properties of the flow that can potentially affect disk-planet 
interactions.
Preliminary results suggest that the shape of the torque density distribution
may also be altered. Therefore, in such disk regions, the torque acting on low 
mass planets may not be represented by the formulae presented here.

\citet{paardekooper2009} performed linear and nonlinear two-dimensional calculations of
torques acting on low-mass planets in fully isothermal disks and found 
that nonlinearities can develop, depending on the level of viscosity. For a 
$\Mp\sim 10^{-5}\,\Ms$ planet embedded in a disk with $H/a=0.05$, they 
found agreement between linear and nonlinear calculations only at very high
viscosity, beyond $\alpha\sim 0.1$. Those two-dimensional results, however, are highly sensitive 
to the value of the smoothing length applied in the planet's potential.
As discussed above, in the fully isothermal case, our three-dimensional torques are in good 
agreement, including the effects of density gradients, with the linear theory of 
\citetalias{tanaka2002} at much smaller viscosity levels, $\alpha\sim 0.001$.
In the more general locally isothermal case, we found that the overall scalings 
of the torque with planet mass and gas sound speed agree with the predictions 
of linear theory. 
Nonlinear feedback effects due to partial saturation of the corotation torque 
would in general give rise to different scalings with planet mass 
\citep[e.g.,][]{ogilvie2003}. Such effects are not expected to be present in the 
parameter space we investigated (see Equation~(\ref{eq:alcr})), as we 
also verified in simulations (see Figures~\ref{fig:satu} and \ref{fig:dtdm_t}). 
Other nonlinear effects are possible, especially in other regimes as mentioned 
above, but we did not see any evidence for them to have a significant impact
in these three-dimensional simulations.

The analysis provided here may not be applicable to cases where substantially
large variations of density and temperature gradients occur within the disk 
zone where torques are produced, i.e., a few times $H$ or less in radius.
Such situations may arise near the edges of ``dead zones'' 
\citep[e.g.,][]{matsumura2007}, although there are limits on the slope of 
these edges imposed by disk instabilities \citep[e.g.,][]{yang2010}.
It would be desirable to explore such situations along the lines of the 
approach taken in this paper.

\acknowledgments

We benefitted from discussions with Fr{\'e}d{\'e}ric Masset and Kristen 
Menou at the KITP program entitled ``The Theory and Observation of Exoplanets''.
We are grateful to Gordon Ogilvie and Jim Pringle for helpful discussions and 
useful feedback on this work. We thank Hui Li and Shengtai Li for running 
some two-dimensional models and for informative discussions.
We acknowledge support from NASA Origins of Solar Systems
Program grants NNX08AH82G (G.D.) and NNX07AI72G (S.L.\ and G.D.).
G.D.\ was also supported in part by the National Science Foundation under 
grant no.\ NSF PHY05-51164.
S.L.\ acknowledges visitor support from the IoA and Newton Institute at 
Cambridge University and many beneficial discussions at the Newton 
Institute programme  ``Dynamics of Discs and Planets''.
Resources supporting this work were provided by the NASA High-End 
Computing (HEC) Program through the NASA Advanced Supercomputing 
(NAS) Division at Ames Research Center.






\appendix

\section{SENSITIVITY OF RESULTS TO  IMPOSED PARAMETERS}
\label{sec:INP}

In numerical simulations of the kind discussed in this paper, there
is often much emphasis placed on the form of the gravitational potential
of the planet. In a context in which disk self-gravity is neglected
and the ``effective'' radius of the planet is much smaller than the
smallest length scale resolved numerically, this potential is that of 
a dimensionless mass $\Mp$, i.e., $\Phi_{p}=-G\,\Mp/S$, 
where $S$ is the distance 
from said mass. Since this potential is singular at the position of the
planet and the planet lies on the computational grid, the singularity
could generate numerical instabilities because the source terms of
the momenta equations could involve the calculation of a diverging 
gravitational force. (Notice that there is no problem with employing
a point-mass potential for the gravity field generated by the star, 
when it is located outside of the computational grid.)
Therefore, the singularity is usually removed by introducing a 
modified potential as in Equation~(\ref{eq:Phip}). 
In this case, the potential is softened over a given length, which is 
either a fraction of the planet's Hill radius, $\Rhill$, or on the order 
of the local (hydrostatic) disk scale-height, $H(a)$. 
The second option is meaningful only in a two-dimensional ($r$--$\phi$) geometry, 
where the smoothing length is used, in effect, as a proxy for the 
third (vertical) dimension of the disk. We choose the first
option and adopt a softening length that is a tenth of $\Rhill$.

\begin{figure*}
\centering%
\resizebox{\linewidth}{!}{%
\includegraphics{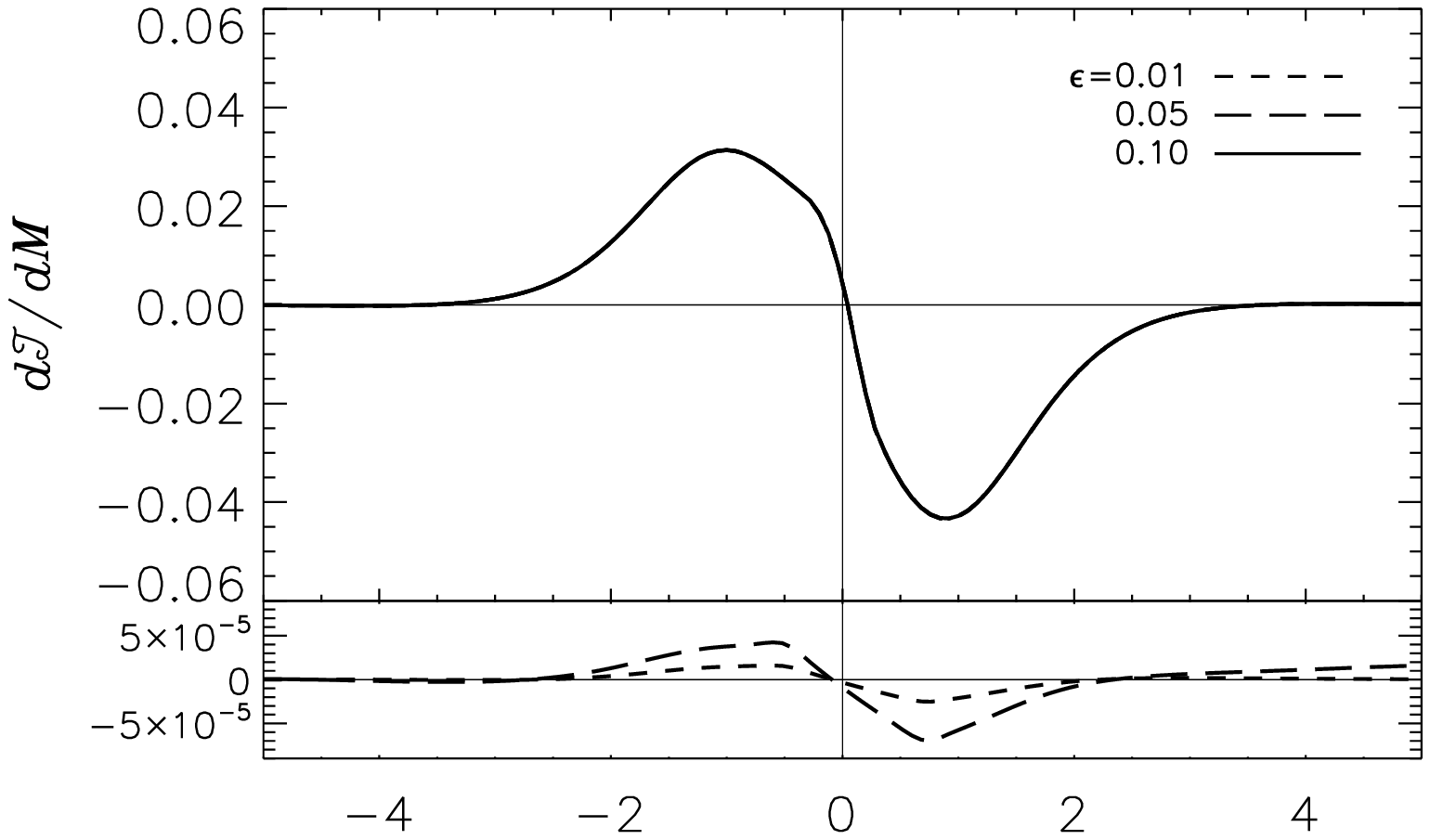}%
\includegraphics{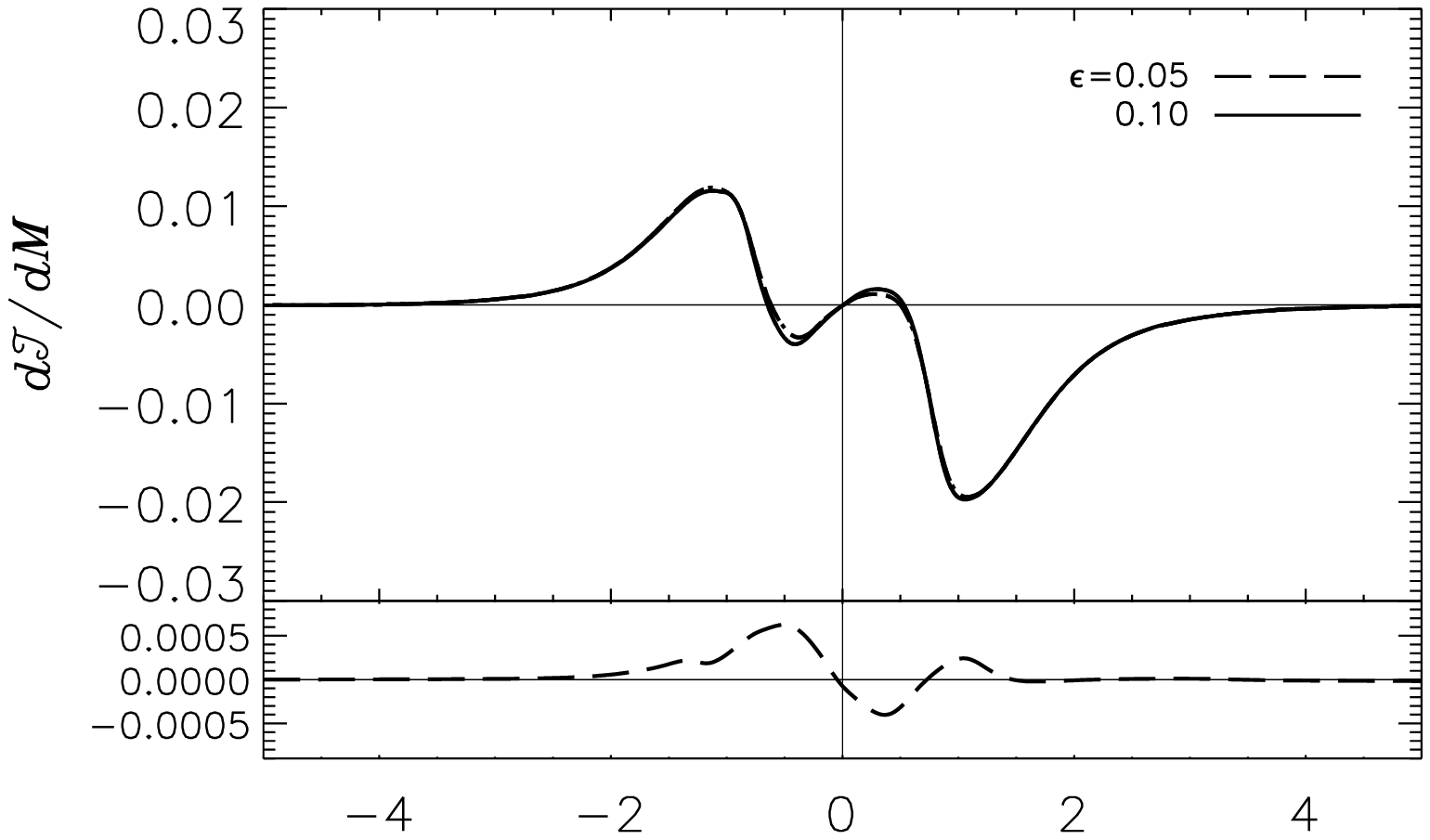}}
\resizebox{\linewidth}{!}{%
\includegraphics{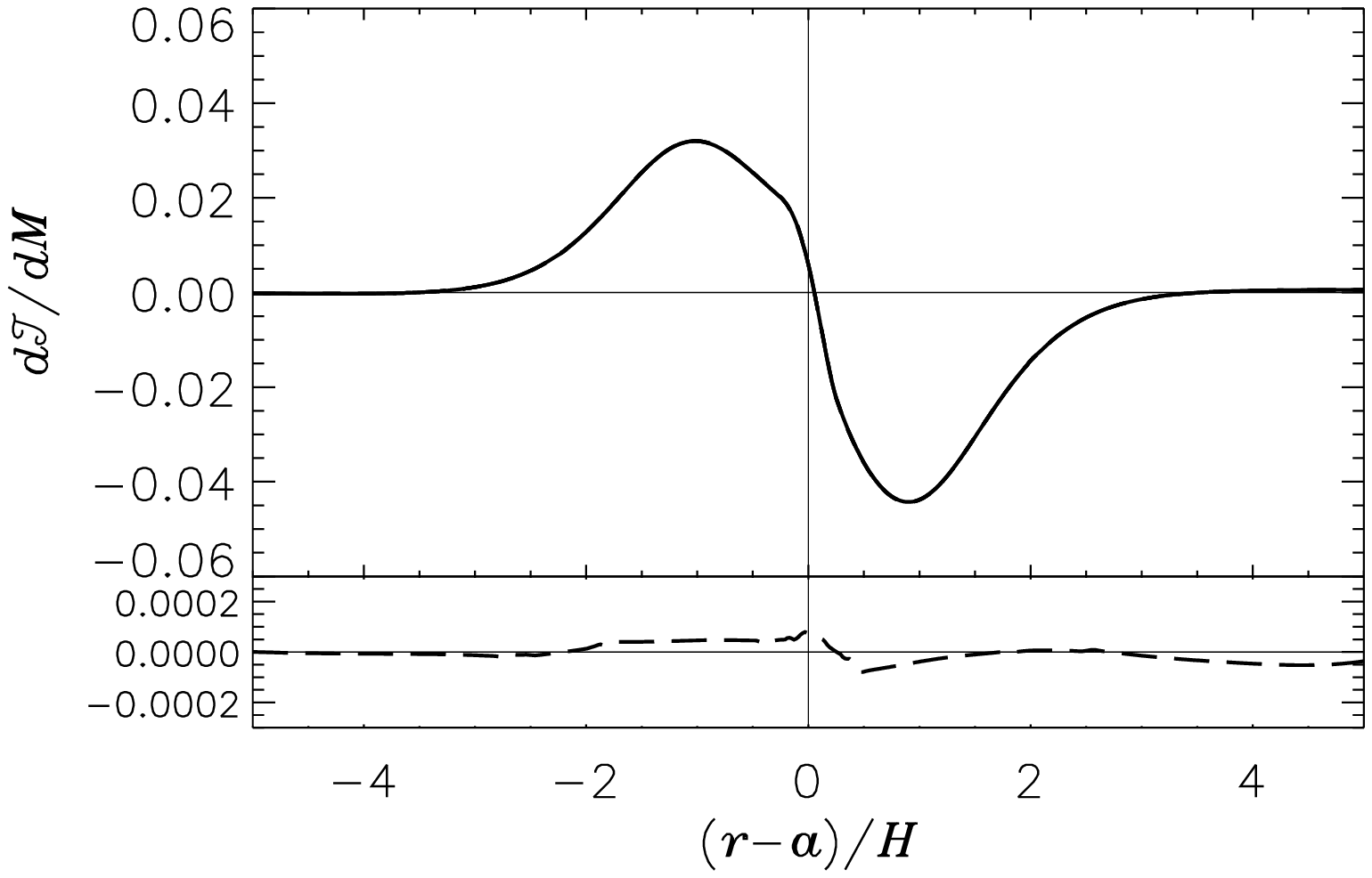}%
\includegraphics{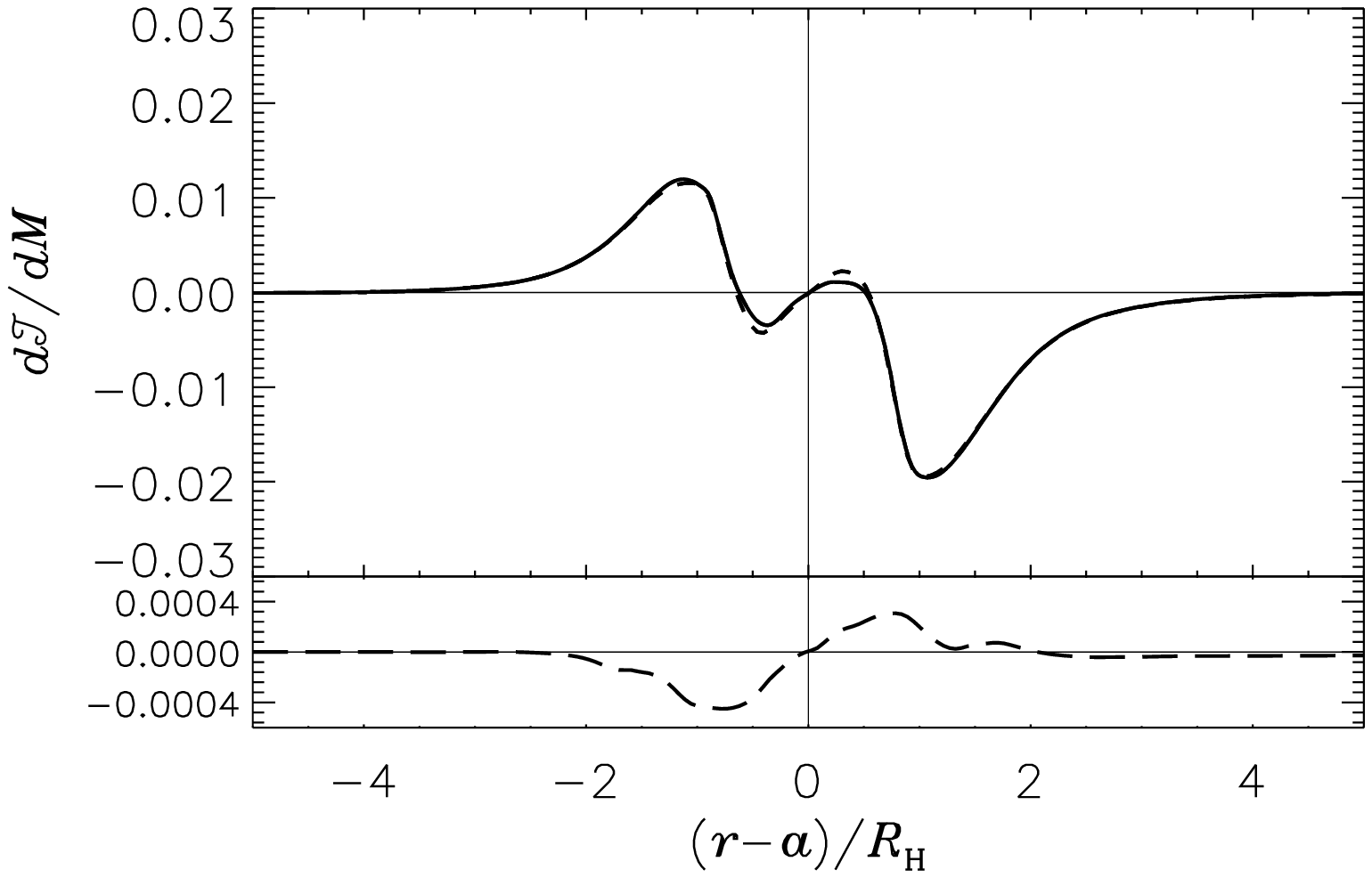}}
\caption{%
         \textit{Top}:
         torque density distributions in a locally isothermal disk, 
         interacting with an $\Mp=5\times 10^{-6}\,\Ms$ planet
         (\textit{left}) and an $\Mp=10^{-3}\,\Ms$ planet 
         (\textit{right}), for different values of the softening 
         parameter $\varepsilon$ (see Equation~(\ref{eq:Phip})),
         as indicated in the legends. Torques are rescaled by
         $\Omega^{2}\,a^{2}\,(\Mp/\Ms)^{2}\,(a/H)^{4}$,
         where $\Omega=\Omega(a)$ and $H=H(a)$.
         The curves overlap each other and total torques differ
         by $\sim 1$\% (or less for the lower mass planet).
         The differences of $d\mathcal{T}(r)/dM$ are shown beneath
         the torque density distributions. On the left, the differences
         are between cases which use $\varepsilon=0.05$ 
         and $\varepsilon=0.1$ (\textit{long-dashed line}) and cases 
         which use $\varepsilon=0.01$ and $\varepsilon=0.05$
         (\textit{short-dashed line}).
         \textit{Bottom}:
         convergence study of $d\mathcal{T}(r)/dM$ for the
         models in the top panels, for cases that use 
         average linear resolutions of $2\times 10^{-3}$ and 
         $10^{-3}$ (\textit{left}) and of $4\times 10^{-3}$ and 
         $2\times 10^{-3}$ (\textit{right}).
         The agreement is again at the $\sim 1$\% level or better.
         }
\label{fig:dtdmvssm}
\end{figure*}
The deviation of the softened potential in Equation~(\ref{eq:Phip}) 
from the point-mass potential, normalized to the point-mass potential,
is $\approx (\varepsilon^{2}/2) (\Rhill/S)^2$, which is $\le 12$\% for 
$S/\Rhill \ge 2\,\varepsilon$ and $\le 2$\% for $S/\Rhill \ge 5\,\varepsilon$. 
We check that our results are largely insensitive to the choice of 
$\varepsilon=0.1$ by performing calculations for a 
$\Mp=5\times 10^{-6}\,\Ms$ planet in a disk, whose viscosity parameter 
is $\alpha=0.005$, and using softening lengths ($\varepsilon \Rhill$) 
with $\varepsilon=0.05$ and $0.01$. 
The resulting net torques agree at about the $10^{-3}$ level, 
implying that they are basically converged with respect to the 
parameter $\varepsilon$.
We perform a similar experiment with an $\Mp=10^{-3}\,\Ms$ planet,
using $\varepsilon=0.05$, and found discrepancies at the
$10^{-2}$ level.
Torque density distributions are plotted in Figure~\ref{fig:dtdmvssm}
(\textit{top panels}) for the $\Mp=5\times 10^{-6}\,\Ms$ planet (\textit{left})
and the $\Mp=10^{-3}\,\Ms$ planet (\textit{right}), with values of 
$\varepsilon$ indicated in the legends. Given the level of agreement, 
profiles corresponding to different softening length cannot be 
distinguished from one another, or barely so. The differences of 
$d\mathcal{T}(r)/dM$, taking cases with closest values of $\varepsilon$
(see figure's caption for details), are shown beneath the torque density 
distributions.

The outcome illustrated in the top panels of Figure~\ref{fig:dtdmvssm} 
can be explained with the property of the relative distance between the 
planet and the locations where momenta and gravitational potential are 
computed. The planet position is fixed on the (rotating) grid at a ``corner'' 
of a volume element of the grid (where eight such elements are in contact). 
Momenta are computed at the geometrical center of each of the six
faces of a volume element, whereas the gravitational potential is 
computed at the geometrical center of the volume element. Therefore, 
there is always a buffer distance of $\sqrt{3}L/2$ (neglecting curvature), 
where $L$ is the average grid spacing, which prevents $\Phi_{p}$ 
from diverging, even if $\varepsilon=0$. 
This argument also applies when the planet migrates since, by virtue 
of the time-varying rotation rate of the reference frame 
(see \S~\ref{sec:3Dtest}), it only moves 
along the radius, $R$, but not along the azimuth, $\phi$. Hence, there 
still exists a minimum buffer distance of $\sqrt{2}L/2$ between the
 planet position and the center of a volume element.
(Notice that, on a two-dimensional grid, these buffer distances become shorter and
equal to $\sqrt{2}L/2$ and $L/2$, respectively.)

Numerical resolution is another potential issue that may have an
impact on simulation results. In the bottom panels of 
Figure~\ref{fig:dtdmvssm}, we test for convergence of torque density 
distributions with respect to numerical resolution. As mentioned in 
\S~\ref{sec:NumericalMethod}, the average linear resolution in the 
disk region of interest ($|R-a|\le 4\,H$) applied in this study ranges 
from $2\times 10^{-3}\,a$ to $5\times 10^{-3}\,a$. These grid systems 
contain typically from $10$ to over $40$ millions of grid elements.
The torque density distributions shown in Figure~\ref{fig:dtdmvssm}
(\textit{bottom panels}) are calculated using average linear resolutions 
of $2\times 10^{-3}$ and $10^{-3}$ (\textit{left}) and
of $4\times 10^{-3}$ and $2\times 10^{-3}$ (\textit{right}). 
The higher resolution cases in these plots are based on grid
systems that have nearly $110$ (\textit{left}) and $90$ (\textit{right}) million 
grid elements.
The profiles lie on top of each other and, again, can be hardly 
distinguished. The differences are displayed beneath the plots
of $d\mathcal{T}(r)/dM$, showing agreement at a $\sim 10^{-2}$
level or better.



\end{document}